\documentclass[ALICE,manyauthors]{cernphprep}
\usepackage[comma,square,numbers,sort&compress]{natbib}
\usepackage{hyperref}
\usepackage{lineno}
\usepackage{xspace}
\usepackage{subcaption}
\usepackage{xcolor}
\usepackage{float}
\usepackage{physics}
\usepackage{graphicx} 
\usepackage{subfigure}
\usepackage[T1]{fontenc}
\usepackage{orcidlink} 

\usepackage[nameinlink]{cleveref}
\Crefname{figure}{Fig.}{Figs.}

\begin{document}
%

\newcommand{\pp}           {pp\xspace}
\newcommand{\ppbar}        {\mbox{$\mathrm {p\overline{p}}$}\xspace}
\newcommand{\XeXe}         {\mbox{Xe--Xe}\xspace}
\newcommand{\PbPb}         {\mbox{Pb--Pb}\xspace}
\newcommand{\pA}           {\mbox{pA}\xspace}
\newcommand{\pPb}          {\mbox{p--Pb}\xspace}
\newcommand{\AuAu}         {\mbox{Au--Au}\xspace}
\newcommand{\dAu}          {\mbox{d--Au}\xspace}

\newcommand{\s}            {\ensuremath{\sqrt{s}}\xspace}
\newcommand{\snn}          {\ensuremath{\sqrt{s_{\mathrm{NN}}}}\xspace}
\newcommand{\pt}           {\ensuremath{p_{\rm T}}\xspace}
\newcommand{\meanpt}       {$\langle p_{\mathrm{T}}\rangle$\xspace}
\newcommand{\ycms}         {\ensuremath{y_{\rm CMS}}\xspace}
\newcommand{\ylab}         {\ensuremath{y_{\rm lab}}\xspace}
\newcommand{\etarange}[1]  {\mbox{$\left | \eta \right |~<~#1$}}
\newcommand{\yrange}[1]    {\mbox{$\left | y \right |~<~#1$}}
\newcommand{\dndy}         {\ensuremath{\mathrm{d}N_\mathrm{ch}/\mathrm{d}y}\xspace}
\newcommand{\dndeta}       {\ensuremath{\mathrm{d}N_\mathrm{ch}/\mathrm{d}\eta}\xspace}
\newcommand{\avdndeta}     {\ensuremath{\langle\dndeta\rangle}\xspace}
\newcommand{\dNdy}         {\ensuremath{\mathrm{d}N_\mathrm{ch}/\mathrm{d}y}\xspace}
\newcommand{\Npart}        {\ensuremath{N_\mathrm{part}}\xspace}
\newcommand{\Ncoll}        {\ensuremath{N_\mathrm{coll}}\xspace}
\newcommand{\dEdx}         {\ensuremath{\textrm{d}E/\textrm{d}x}\xspace}
\newcommand{\RpPb}         {\ensuremath{R_{\rm pPb}}\xspace}

\newcommand{\nineH}        {$\sqrt{s}~=~0.9$~Te\kern-.1emV\xspace}
\newcommand{\seven}        {$\sqrt{s}~=~7$~Te\kern-.1emV\xspace}
\newcommand{\twoH}         {$\sqrt{s}~=~0.2$~Te\kern-.1emV\xspace}
\newcommand{\twosevensix}  {$\sqrt{s}~=~2.76$~Te\kern-.1emV\xspace}
\newcommand{\five}         {$\sqrt{s}~=~5.02$~Te\kern-.1emV\xspace}
\newcommand{\twosevensixnn}{$\sqrt{s_{\mathrm{NN}}}~=~2.76$~Te\kern-.1emV\xspace}
\newcommand{\fivenn}       {$\sqrt{s_{\mathrm{NN}}}~=~5.02$~Te\kern-.1emV\xspace}
\newcommand{\LT}           {L{\'e}vy-Tsallis\xspace}
\newcommand{\GeVc}         {Ge\kern-.1emV/$c$\xspace}
\newcommand{\MeVc}         {Me\kern-.1emV/$c$\xspace}
\newcommand{\TeV}          {Te\kern-.1emV\xspace}
\newcommand{\GeV}          {Ge\kern-.1emV\xspace}
\newcommand{\MeV}          {Me\kern-.1emV\xspace}
\newcommand{\GeVmass}      {Ge\kern-.2emV/$c^2$\xspace}
\newcommand{\MeVmass}      {Me\kern-.2emV/$c^2$\xspace}
\newcommand{\lumi}         {\ensuremath{\mathcal{L}}\xspace}

\newcommand{\ITS}          {\rm{ITS}\xspace}
\newcommand{\TOF}          {\rm{TOF}\xspace}
\newcommand{\ZDC}          {\rm{ZDC}\xspace}
\newcommand{\ZDCs}         {\rm{ZDCs}\xspace}
\newcommand{\ZNA}          {\rm{ZNA}\xspace}
\newcommand{\ZNC}          {\rm{ZNC}\xspace}
\newcommand{\SPD}          {\rm{SPD}\xspace}
\newcommand{\SDD}          {\rm{SDD}\xspace}
\newcommand{\SSD}          {\rm{SSD}\xspace}
\newcommand{\TPC}          {\rm{TPC}\xspace}
\newcommand{\TRD}          {\rm{TRD}\xspace}
\newcommand{\VZERO}        {\rm{V0}\xspace}
\newcommand{\VZEROA}       {\rm{V0A}\xspace}
\newcommand{\VZEROC}       {\rm{V0C}\xspace}
\newcommand{\Vdecay} 	   {\ensuremath{V^{0}}\xspace}

\newcommand{\ee}           {\ensuremath{e^{+}e^{-}}} 
\newcommand{\pip}          {\ensuremath{\pi^{+}}\xspace}
\newcommand{\pim}          {\ensuremath{\pi^{-}}\xspace}
\newcommand{\kap}          {\ensuremath{\rm{K}^{+}}\xspace}
\newcommand{\kam}          {\ensuremath{\rm{K}^{-}}\xspace}
\newcommand{\pbar}         {\ensuremath{\rm\overline{p}}\xspace}
\newcommand{\kzero}        {\ensuremath{{\rm K}^{0}_{\rm{S}}}\xspace}
\newcommand{\lmb}          {\ensuremath{\Lambda}\xspace}
\newcommand{\almb}         {\ensuremath{\overline{\Lambda}}\xspace}
\newcommand{\Om}           {\ensuremath{\Omega^-}\xspace}
\newcommand{\Mo}           {\ensuremath{\overline{\Omega}^+}\xspace}
\newcommand{\X}            {\ensuremath{\Xi^-}\xspace}
\newcommand{\Ix}           {\ensuremath{\overline{\Xi}^+}\xspace}
\newcommand{\Xis}          {\ensuremath{\Xi^{\pm}}\xspace}
\newcommand{\Oms}          {\ensuremath{\Omega^{\pm}}\xspace}
\newcommand{\degree}       {\ensuremath{^{\rm o}}\xspace}

\begin{titlepage}
\PHyear{2025}       
\PHnumber{253}      
\PHdate{04 November}  

\title{Multiplicity dependence of two-particle angular correlations of identified particles in pp collisions at $\mathbf{\sqrt{s} = 13}$ TeV}
\ShortTitle{Two-particle angular correlations in pp collisions at $\sqrt{s} = 13$ TeV}   

\Collaboration{ALICE Collaboration\thanks{See Appendix~\ref{app:collab} for the list of collaboration members}}
\ShortAuthor{ALICE Collaboration} 

\begin{abstract}
Two-particle angular correlations probe particle production mechanisms and the underlying event-wide phenomena present in hadronic collisions. The correlations are examined as a function of rapidity and azimuthal-angle differences ($\Delta y, \Delta \varphi$) for pairs of like- and unlike-sign pions, kaons, and (anti-)protons produced in pp collisions at $\sqrt{s}$ = 13 TeV, measured by the ALICE experiment. Two-particle correlation functions are provided together with $\Delta y$ and $\Delta\varphi$ projections and compared to Monte Carlo (MC) model predictions. For the first time, the measurement is performed as a function of the event's charged-particle density. Previous studies conducted for pp collisions at $\sqrt{s}$ = 7 TeV at ALICE revealed a near-side anticorrelation for baryon--baryon and antibaryon--antibaryon pairs, whose origin remains unresolved. Here, an additional approach is introduced to study the multiplicity dependence and the expected inverse multiplicity scaling of the correlation function. This method highlights qualitative differences in the underlying sources of correlations, such as quantum-statistics effects, final-state interactions, and resonance decays. The puzzling near-side anticorrelation in baryon–baryon measurements is observed across all multiplicity classes and continues to challenge current particle-production models. 
Furthermore, the multiplicity dependence of the correlations between mesons provides an independent probe of the sensitivity of current MC models to soft-QCD effects and hadronization dynamics. The presented measurements, together with the baryon results, enrich the experimental picture of two-particle correlations in pp collisions and serve as valuable input for ongoing theoretical developments.
\end{abstract}
\end{titlepage}

\setcounter{page}{2} 


\section{Introduction} 
The study of particle production mechanisms in ultrarelativistic proton--proton (pp) collisions at the LHC is a cornerstone of high-energy physics research. These mechanisms are typically explored through inclusive single-particle distributions and various correlation observables~\cite{Pruneau_2017}. Among these, two-particle angular analysis has emerged as a powerful tool for investigating the underlying dynamics of particle production~\cite{ALICEInsight,ACPhysRevC.80.064912,AC1PhysRevLett.104.062301, AC7ATLAS:2012cix, AC0Alver_2010,AC32010, AC5ATLAS:2016yzd, AC6CMS:2011cqy, AC9CMS:2012xss}.
Such studies are typically performed as a function of the difference in azimuthal angle ($\Delta\varphi$) and pseudorapidity ($\Delta \eta$) or rapidity ($\Delta y$). These correlations provide insights into diverse physical phenomena, including contributions from (mini)jets arising from parton fragmentation, quantum-statistical effects such as Bose--Einstein and Fermi--Dirac, resonance decays, and event-wide effects like the conservation of energy, momentum, electric charge, baryon number, and others. Each of these mechanisms imprints distinct signatures on the correlation function, enabling their characterization and further description using theoretical models.

Research into particle production mechanisms in elementary collisions dates back to the work of R.~Feynman and R. Field, who introduced a model to explain the creation of jets~\cite{Field:1977fa,Field:1976ve}. It contained rules for particle production, energy distribution, and the constraints imposed by conservation laws.
Key elements of this framework are still used today in widely adopted fragmentation models. Aligning these theoretical implementations with experimental observations remains essential. 
Such investigations have primarily been conducted in $e^+e^-$ collisions, typically at lower energies and with significantly smaller datasets than available today~\cite{TPCTwoGamma:1986yqu,TASSO:1984ajj,DELPHI:1997fmy,OPAL:1993wav}. For example, Ref.~\cite{TPCTwoGamma:1986yqu} examined two-particle rapidity correlations in $e^+e^-$ collisions at $\sqrt{s}=29$~GeV. The study found strong correlations between particle pairs with opposite baryon numbers and significant anticorrelations for pairs with the same baryon number. Additionally, it highlighted the importance of mechanisms ensuring local baryon number conservation during parton fragmentation. Models based on the Lund framework~\cite{SJOSTRAND1986347}, such as the Lund string-fragmentation model~\cite{TPCTwoGamma:1986yqu}, which incorporates local baryon number compensation, were able to accurately reproduce the observed correlations and anticorrelations in $e^+e^-$ data.


Measurements performed in 2017 by the ALICE Collaboration in pp collisions at $\sqrt{s} = 7$ TeV~\cite{ALICEInsight} have revealed differences in the shape of global correlations between various types of particles. In particular, correlation functions for both the baryon--baryon and antibaryon--antibaryon pairs exhibit a depletion around $(\Delta\eta,\Delta\varphi) \sim (0,0)$, which is qualitatively similar to the effect observed earlier in $e^+e^-$ collisions~\cite{TPCTwoGamma:1986yqu}. To investigate this phenomenon, numerous studies have been conducted, focusing on various ranges of transverse momentum ($p_{\rm{T}}$) and energy, different baryonic species, and the influence of the Coulomb interaction and Fermi--Dirac quantum-statistics~\cite{ALICEInsight}. Such behavior was also observed by the STAR Collaboration at RHIC in Au--Au collisions across multiple collision energies in the Beam Energy Scan program~\cite{,STARbeamenergy}. These investigations consistently revealed an anticorrelation for particle pairs with the same baryon number. Interestingly, theoretical models which did describe the anticorrelation effect at lower energies, such as the Lund model, fail to do so for the RHIC and LHC energy regime. These results challenge our current understanding of baryon production.

Since 2017, both experimental and theoretical advances have shed new light on the observed anticorrelation.
A recent study by the ALICE Collaboration utilized per-trigger yields for $\Xi$ baryons in more detail~\cite{alicecollaboration2023studyingstrangenessbaryonproduction}. This work provided critical insights into baryon-number production, conservation, and dispersion.
Furthermore, the study revealed that near-side suppression for same-baryon-number correlations and near-side peaks for unlike-sign baryon-number correlations were consistently observed in the $\Xi-\Lambda$ and $\Xi-\Xi$ correlation functions.

On the theoretical side, several advances have been made. An investigation of the origin of the near-side suppression in two-baryon correlations was performed within the context of the AMPT model~\cite{ZHANG2022137063}. That study suggested that the baryon--baryon and antibaryon--antibaryon anticorrelations in pp collisions arise from a combination of initial parton-level correlations, quark coalescence, and finite parton-system expansion, pointing to the formation of partonic matter with strong space-momentum correlations. In~\cite{SJOSTRAND201943} it was emphasized that the Lund string model used in the PYTHIA event generator lacks a fundamental mechanism for baryon production. Recently, advances within the PYTHIA framework introduced two ad-hoc modifications to address the baryon anticorrelation feature~\cite{Demazure:2022gfl}. These modifications, termed the “one-baryon” (a string breaking can produce at most one baryon) and “always-baryon” (a string breaking always has to produce a baryon) policies, qualitatively reproduce the observed anticorrelations. The one-baryon mechanism prevents two baryons from originating from the same string, removing forward correlations but reducing the total baryon count. The always-baryon policy compensates for this by increasing the overall number of baryons, yielding better agreement with ALICE experimental data~\cite{ALICEInsight}. These findings also suggest that missing Pauli-principle corrections at the quark level may underlie the discrepancies between experimental results and existing string-based models. Another study within the context of the PYTHIA model was done in~\cite{Lonnblad:2023kft}, where the authors state that the positive baryon--baryon correlation observed in PYTHIA might be related to the so-called gluon kinks, i.e., changes in the direction of the color string caused by gluon emissions during fragmentation. Suppressing this mechanism reduces the (mini-)jet like correlation present in the simulation.

Together, these experimental and theoretical advances highlight the complexity of baryon production and correlations in pp collisions, motivating further exploration into the interplay of initial-state effects, hadronization mechanisms, and the possible formation of partonic matter in small collision systems, for example, in proton–proton and proton–nucleus interactions.

This paper focuses on a detailed analysis of angular correlation functions in the ($\Delta y, \Delta\varphi$) space across different multiplicity classes, aiming to shed light on the mechanisms of particle production and enhance our understanding of the nature of anticorrelations. The study was performed in 13 multiplicity classes, including the class of triggered high-multiplicity collisions, for pions, kaons, and protons, including both like- and unlike-sign charge combinations. 
Understanding how the correlations change with multiplicity is essential to improve the models, as high-multiplicity events typically involve more multi-parton interactions (MPIs), creating a complex environment where subleading color effects from QCD, such as color reconnection~\cite{ColourReconnectionFromSoftGluons,AngelesMartinez:2018cfz}, become more prominent.
In these events, higher multiplicity implies higher phase-space density, which allows for more available color states in the same patch of phase space and in particular increasing (or even allowing) the formation of baryonic clusters and strings.

In the previous study by ALICE~\cite{ALICEInsight},  a definition of the correlation function was employed that inherently includes a $1/N$ scaling factor, where $N$ denotes the number of particles per event in each multiplicity class. 
In the present study, to explore the underlying physics beyond the multiplicity scaling behavior, an alternative formulation of the correlation function~\cite{Pruneau_2017,rescaledcumulant} is adopted, where this dependence is removed, allowing for a more detailed examination of additional effects.

\section{Angular correlations}
In this study, the correlation function is examined in the angular space defined by $\Delta y$ and $\Delta\varphi$. Rapidity is preferred for studying angular correlations between particles in relativistic collisions due to its precise incorporation of particle kinematics. This approach is made possible by the outstanding particle-identification (PID) capabilities of the ALICE experiment, which, together with precise tracking of transverse momentum and dip angle, enables an accurate calculation of rapidity.
Experimentally, the correlation function is constructed in the ($\Delta y, \Delta \varphi$) space as follows:
\begin{equation}
    C_{\rm{P}}(\Delta y,\Delta\varphi) = \frac{S(\Delta y,\Delta\varphi)}{B(\Delta y,\Delta\varphi)},
\label{eq:CP}
\end{equation}
where $C_{\rm{P}}$ is referred to as the probability ratio correlation function.
The term represents a ratio of conditional probability of production of two particles with given relative kinematics to the product of single-particle probabilities~\cite{ALICEInsight}.
The $S(\Delta y,\Delta\varphi)$ represents the distribution of correlated pairs constructed from particles originating from the same event
\begin{equation}
    S(\Delta y,\Delta\varphi) = \frac{1}{N_{\rm{pairs}}^{\rm{signal}}} \frac{{\rm{d}}^{2}N_{\rm{pairs}}^{\rm{signal}}}{{\rm{d}}\Delta y\rm{d}\Delta\varphi},
\end{equation}
where \( N_{\rm{pairs}}^{\rm{signal}} \) is the total number of same-event pairs.  
The signal distribution typically shows a peak around \( (\Delta y, \Delta \varphi) \approx (0,0) \). This peak is overlaid by a triangular-shaped background in \( \Delta y \), resulting from the convolution of two uniform rapidity distributions expected in a Lorentz boost-invariant system.
To isolate genuine physical correlations, $S(\Delta y,\Delta\varphi)$ is divided by the reference distribution $B(\Delta y,\Delta\varphi)$ constructed using the event-mixing technique, which accounts for non-physical correlations arising from the complexity of single-particle acceptance.
The mixed-event distribution is defined as:  \begin{equation}
    B(\Delta y,\Delta\varphi) = \frac{1}{N_{\rm{pairs}}^{\rm{mixed}}} \frac{{\rm{d}}^{2}N_{\rm{pairs}}^{\rm{mixed}}}{{\rm{d}}\Delta y {\rm{d}}\Delta \varphi},
\end{equation}

where \( N_{\rm{pairs}}^{\rm{mixed}} \) is the total number of mixed-event pairs across all events. To ensure that the mixed-event background closely matches the acceptance of the signal distribution, events are selected within bins of similar primary vertex position (within 5 cm) and multiplicity (within 30 tracks). This binning reduces systematic differences between mixed and same-event pairs, improving the accuracy of background estimation.
Both the signal and background distributions are normalized by the respective number of pairs in each distribution. This results in correlations being expressed as deviations from unity.

The $C_{\rm P}$ definition encounters limitations when analyzing correlations across different multiplicity classes. This is primarily due to its normalization factor, which introduces a trivial scaling by $1/N$, where $N$ represents the multiplicity. Consequently, correlation functions exhibit an artificial increase in lower multiplicity classes, resulting from the convolution of trivial scaling effects with underlying physical phenomena, making it difficult to disentangle the two. To overcome this challenge, the rescaled two-particle correlation function is introduced~\cite{rescaledcumulant}:
\begin{equation}
    C_{\rm R}(\Delta y,\Delta\varphi) = \frac{1}{2\pi}\left\langle \dv{N_a}{y}\right\rangle (C_{\rm{P}} - 1).
\end{equation}
Here, $\frac{1}{2\pi} \left\langle \dv{N_a}{y}\right\rangle$ represents the multiplicity of charged particles of type $\it{a}$ (where $\it{a} = \pi,\, K,\, \text{or}\, p$) calculated within $|y|<0.5$ and corrected for efficiency. It can be noted that, in this definition, the correlations are expressed as deviations from zero rather than from unity.

\section{Data sample and experimental setup}
\label{settings}
The study utilized a data sample of pp collisions collected by the ALICE experiment~\cite{ALICE:2008ngc} in 2016, 2017, and 2018. 
Particle-track reconstruction is performed using the Inner Tracking System (ITS)~\cite{ALICE:ITS}, and the Time Projection Chamber (TPC)~\cite{ALICE:TPC}, while the PID is done using the information from TPC and the Time-of-Flight (TOF)~\cite{ALICE:TOF} detectors. The charged-particle multiplicity estimation is carried out using the V0 detector~\cite{ALICE:V0}.
The ITS serves as a high-precision tracking and vertexing system, consisting of six cylindrical layers of coordinate-sensitive detectors, covering a pseudorapidity range of $|\eta| < 0.9$. The two innermost layers contain the Silicon Pixel Detector (SPD)~\cite{ALICE:ITS2}, which also serves as a trigger detector. The four outer layers include the Silicon Drift Detector (SDD) and Silicon Strip Detector (SSD)~\cite{ALICE:ITS2}, particularly effective for reconstructing particles with low transverse momentum.

The TPC is the primary tracking detector in ALICE's central barrel, structured as a cylindrical gaseous chamber with radial dimensions from 85 cm to 247 cm from the beam and longitudinal dimensions spanning from $-250$~cm to 250~cm. Using drift-time measurements, the TPC reconstructs three-dimensional space points for charged-particle tracks within the pseudorapidity range of $|\eta| < 0.9$. Charged-particle identification in the TPC is based on specific energy-loss measurements.

The TOF detector, located 3.7 meters from the beamline, has a cylindrical geometry with a polar acceptance of $|\eta| < 0.9$, providing full azimuthal coverage in the $\varphi$ angle. Equipped with Multigap Resistive Plate Chamber (MRPC) strips, it measures the time-of-flight between the collision event and particle detection, facilitating the identification of charged particles, particularly in the intermediate momentum range ($p_{\rm T} < 5$ GeV/$c$).

The V0 detector is composed of two arrays containing 32 scintillator counters each, placed on opposite sides of the ALICE interaction point. V0A is situated at $z = 3.4$ m, covering the pseudorapidity range $2.8 < \eta < 5.1$, while V0C is positioned at $z = -0.9$ m, covering the pseudorapidity range $-3.7 < \eta < -1.7$. Its capabilities include minimum bias triggers, centrality determination, event-plane determination, and acting as a luminometer.

\section{Event selection and track reconstruction}
\label{eventandtrackreco}
The analyzed data were recorded using both minimum bias (MB) and high-multiplicity (HM) triggers~\cite{Alice:triggerpaper, ALICE:trigger2}, corresponding to approximately $1.3 \times 10^9$ MB events and $2.5 \times 10^9$ HM events.
The MB trigger is designed to select inelastic collisions within the detector acceptance. It is typically implemented using the condition that requires at least one hit in the SPD or in either of the V0A and V0C detectors, synchronized with the LHC bunch crossing. The HM trigger builds on the MB condition by requiring that the total signal amplitude from the V0 detectors (V0M) exceeds a threshold, usually set to about five times the average V0M value in MB events. This selection enhances the sample with high-multiplicity events and corresponds to approximately the top 0.17\% of the inelastic event multiplicity distribution for events with at least one charged particle within $|\eta| < 1$.
The HM trigger is used in pp collisions to study rare phenomena, such as charged or neutral jets, along with additional signals.
In addition, high-multiplicity pp collisions are particularly interesting as they exhibit multiplicities comparable to those found in peripheral heavy-ion collisions.
The primary vertex was reconstructed using global (ITS+TPC) tracking, requiring a distance of $|z_{\rm {vtx}}|<10 $ cm along the beam axis from the center of the TPC. Pile-up events were filtered out by rejecting multiple vertices within the readout time of the SPD. An additional criterion in this analysis is that each event must contain at least one particle of each required type after all selection criteria have been applied. Events that do not meet this requirement are excluded.
The events were analyzed across twelve multiplicity classes, which included 0--1\%, 1--5\%, 5--10\%, 10--20\%, 20--30\%, 30--40\%, 40--50\%, 50---60\%, 60--70\%, 70--80\%, 80--90\%, and 90--100\% of the total interaction cross section, as measured by the V0 detector using the V0M amplitude. Additionally, HM events were considered in this analysis as a special multiplicity class corresponding to 0--0.17\%.

The analysis uses tracks reconstructed within the kinematic acceptance set by $|\eta|<$ 0.8, with the additional rapidity cut of $|y|<$ 0.5, within transverse-momentum ($p_{\rm{T}}$) ranges 0.2--2.5 GeV/$c$ for pions and 0.5--2.5 GeV/$c$ for kaons and protons.
Particle identification is carried out for each track with the $N_{\rm{\sigma}}$ method, where $\sigma$ denotes the standard deviation derived from the expected specific energy loss in the TPC or time of flight in the TOF detector.
The selection of particles is based on the difference between the measured signal for a given track, denoted by ``a"  to represent one of the three particle type hypotheses, and the expected signals for pions, kaons, or protons within both the TPC and TOF detectors. Specifically, only particle tracks satisfying $N_{\rm{\sigma}} < $ 2 for a given species hypothesis are selected.
For $p_{\rm{T}} < 0.5$~GeV$/c$, only information from the TPC is utilized, resulting in $N_{\rm{\sigma,PID}}^{\rm{a}} = N_{\rm{\sigma,TPC}}^{\rm{a}}$. 
For $p_{\rm{T}} > 0.5$~GeV$/c$, the selection criteria are determined by the combined information from both the TPC and TOF detectors, with ${N_{\rm{\sigma,PID}}^{\rm{a}}}^{\rm{2}} = {N_{\rm{\sigma,TOF}}^{\rm{a}}}^{\rm{2}} + {N_{\rm{\sigma,TPC}}^{\rm{a}}}^{\rm{2}}$. Tracks fulfilling the condition $N_{\rm{\sigma,PID}}^{\rm{a}} < 3$ for more than one particle species hypothesis are rejected. 

The application of these criteria achieves exceptional purity levels, exceeding 99\% for both pions and protons and exceeding 98\% for kaons. Additionally, the contamination from secondary particles, which originate from weak decays or interactions with the detector material, was computed using MC simulation. This calculation considers the distance of the closest approach of the track to the primary vertex in the transverse plane perpendicular to the beam axis ($\rm{DCA}_{xy}$) and longitudinal ($\rm{DCA}_{z}$) directions which is required to be less than $0.0105 + 0.0350 /(p_{\rm{T}}/(\rm{GeV}/\it{c})^{\rm{-1.1}}$ cm and $3.2$ cm, respectively.

\section{Systematic uncertainties}
To assess the systematic uncertainty associated with the measurement, modifications were made to the particle-selection and reconstruction criteria. The various distributions were compared using both the probability ratio ($C_{\rm{P}}$) and the rescaled two-particle correlation function ($C_{\rm{R}}$) definitions. An initial comparison was conducted between the three data sets from the periods 2016, 2017, and 2018. To account for any impurities, the default particle-identification method was modified by keeping the same acceptance criteria while varying the rejection thresholds. Specifically, if a track was identified as belonging to more than one particle species, it was excluded from the analysis, discarding the particle within a $2\sigma$ range. The influence of secondary contamination was verified by varying the strict default DCA selection with $DCA_{\rm {xy}} < 2.4$ cm and $DCA_{\rm {z}} <3.2$ cm. Another variation was obtained by excluding the electron-rejection cut ($N_\sigma^{\mathrm{electron}} < 1$). Event mixing was also considered as a source of systematic uncertainties, and the impact was estimated by varying the number of z-vertex bins, with 30 bins as default and 10 bins as the alternative. Additionally, the effect of pile-up on the events was assessed by comparing the default analysis with one that did not include pile-up rejection.

All described systematic variations were estimated for both the $C_{\rm{P}}$ and $C_{\rm{R}}$ definitions independently. In all the figures, the relative uncertainty was used for the $C_{\rm{P}}$, while the absolute uncertainty was applied to the $C_{\rm{R}}$ functions. The following tables present the values obtained for all multiplicities and particle species analyzed. Table~\ref{unc12multprob} summarizes the final systematic uncertainty values for the $C_{\rm{P}}$ definition, while Table~\ref{unc12multcum} presents the uncertainty values for the $C_{\rm{R}}$, for clarity expressed as a percentage of the uncertainty affecting the peak of the correlation functions.

\begin{table}[!h]
\caption{Summary of all systematic-uncertainty studies performed for all pairs of particles involved and for all multiplicity classes using the $C_{\rm{P}}$ definition. 
In the table, for each variation listed, the range of uncertainties from the highest to the lowest multiplicity class is shown.}
\resizebox{\textwidth}{!}{
\begin{tabular}{lcccccc}
\hline
source              &  $\pi^{+}\pi^{+}+\pi^{-}\pi^{-}$  & $\pi^{+}\pi^{-}$ & K$^{+}$K$^{+}$+K$^{-}$K$^{-}$  & K$^{+}$K$^{-}$ & pp$+\bar{\rm{p}}\bar{\rm{p}}$ & p$\bar{\rm{p}}$ \\ \hline
Track selection          & {[}0.08--0.5{]}\%    & {[}0.05--0.9{]}\%    & {[}0.5--3.25{]}\%  & {[}0.3--2.1{]}\%   & {[}1.3--5.2{]}\% & {[}1.2--1.5{]}\%  \\
Electron rejection  & {[}0.0015--0.005{]}\% & {[}0.002--0.004{]}\% & {[}0.04--0.2{]}\% & {[}0.03--0.15{]}\% & {[}0.05--0.4{]}\% & {[}0.05--0.3{]}\% \\
Event mixing buffer & {[}0.009--0.025{]}\%  & {[}0.009--0.02{]}\%  & {[}0.1--0.6{]}\%   & {[}0.1--1{]}\%    & {[}0.2--0.7{]}\%  & {[}0.25--1.6{]}\% \\
Pile-up             & {[}0.01--0.02{]}\%   & {[}0.01--0.03{]}\%   & {[}0.08--0.6{]}\%  & {[}0.09--0.7{]}\%  & {[}0.15--1{]}\%   & {[}0.1--1{]}\%   \\
Particle identification               & {[}0.008--0.01{]}\%    & {[}0.02--0.02{]}\%  & {[}0.07--0.4{]}\%   & {[}0.06--0.5{]}\%  & {[}0.2--0.9{]}\%  & {[}0.2--0.6{]}\%  \\
Dataset comparison       & {[}0.07--0.2{]}\%     & {[}0.08--0.25{]}\%   & {[}0.9--7{]}\%     & {[}1.1--5.6{]}\%  & {[}1.7--15{]}\%   & {[}1.2--7.4{]}\% \\ \hline
\textbf{Total} & \textbf{[0.11--0.54]\%} & \textbf{[0.10--0.94]\%} & \textbf{[1.0--7.8]\%} & \textbf{[1.2--6.1]\%} & \textbf{[2.2--15.9]\%} & \textbf{[1.7--7.8]\%} \\ \hline
\end{tabular}
}
\label{unc12multprob}
\end{table}

\begin{table}[!h]
\caption{Summary of all systematic-uncertainty studies performed for all pairs of particles involved and for all multiplicity classes using the $C_{\rm{R}}$ definition. 
In the table, for each variation listed, the range of absolute error influence on the peak is estimated from the highest to the lowest multiplicity class.}
\resizebox{\textwidth}{!}{
\begin{tabular}{lcccccc}
\hline
source              &  $\pi^{+}\pi^{+}+\pi^{-}\pi^{-}$  & $\pi^{+}\pi^{-}$ & K$^{+}$K$^{+}$+K$^{-}$K$^{-}$  & K$^{+}$K$^{-}$ & pp$+\bar{\rm{p}}\bar{\rm{p}}$ & p$\bar{\rm{p}}$ \\ \hline
Track selection          & {[}0.4--3.6{]}\%  & {[}1.5--2.9{]}\%  & {[}7.9--9.7{]}\%   & {[}0.7--1.2{]}\% & {[}5.7--7.8{]}\%   & {[}2.1--2.6{]}\% \\
Electron rejection  & {[}0.1--0.2{]}\%  & {[}0.2--1.3{]}\%  & {[}4.3--6.1{]}\%  & {[}0.3--0.7{]}\%  & {[}4--5.6{]}\%     & {[}1.3--2.2{]}\%  \\
Event mixing buffer & {[}0.15--0.2{]}\%  & {[}0.2--1.4{]}\%  & {[}4.7--6.7{]}\%   & {[}0.4--0.8{]}\%  & {[}4--5.7{]}\%     & {[}1.5--2.7{]}\% \\
Pile-up             & {[}0.3--0.2{]}\% & {[}0.4--1.4{]}\% & {[}4.7--6.6{]}\%   & {[}0.4--0.8{]}\% & {[}4--6.2{]}\%     & {[}1.4--2.4{]}\%  \\
Particle identification   & {[}6--8.8{]}\%     & {[}8.7--11.3{]}\% & {[}9.3--12.3{]}\%  & {[}4.6--5.7{]}\%   & {[}15.2--11{]}\%   & {[}9--7{]}\%      \\
Dataset comparison        & {[}0.7--0.8{]}\%  & {[}1.3--7.9{]}\%  & {[}13.7--22.7{]}\% & {[}1.9--3.5{]}\%   & {[}11.7--22.6{]}\% & {[}4.6--8{]}\%  \\ \hline 
\textbf{Total} & \textbf{[6.1--9.5]\%} & \textbf{[8.9--14.1]\%} & \textbf{[21.3--29.1]\%} & \textbf{[5.1--6.9]\%} & \textbf{[22.1--28.8]\%} & \textbf{[10.9--12.4]\%} \\ \hline
\end{tabular}
}
\label{unc12multcum}
\end{table}
The total systematic uncertainties for the $C_{\rm{P}}$ exhibit a noticeable increase as the event multiplicity decreases. This behavior can be attributed to the reduced number of particle pairs available in low-multiplicity events, which amplifies the relative impact of statistical fluctuations on the estimation of systematic variations. In contrast, for the $C_{\rm{R}}$, the absolute uncertainties remain approximately constant across the full multiplicity range. This uniformity suggests that the sources of systematic uncertainty in $C_{\rm{R}}$ are less dependent on event multiplicity and are likely dominated by analysis selections that affect all multiplicity classes similarly.

\section{Results}
\begin{figure}[!h]
 \centering
 \subfigure[]{
   \includegraphics[width=4.9cm]{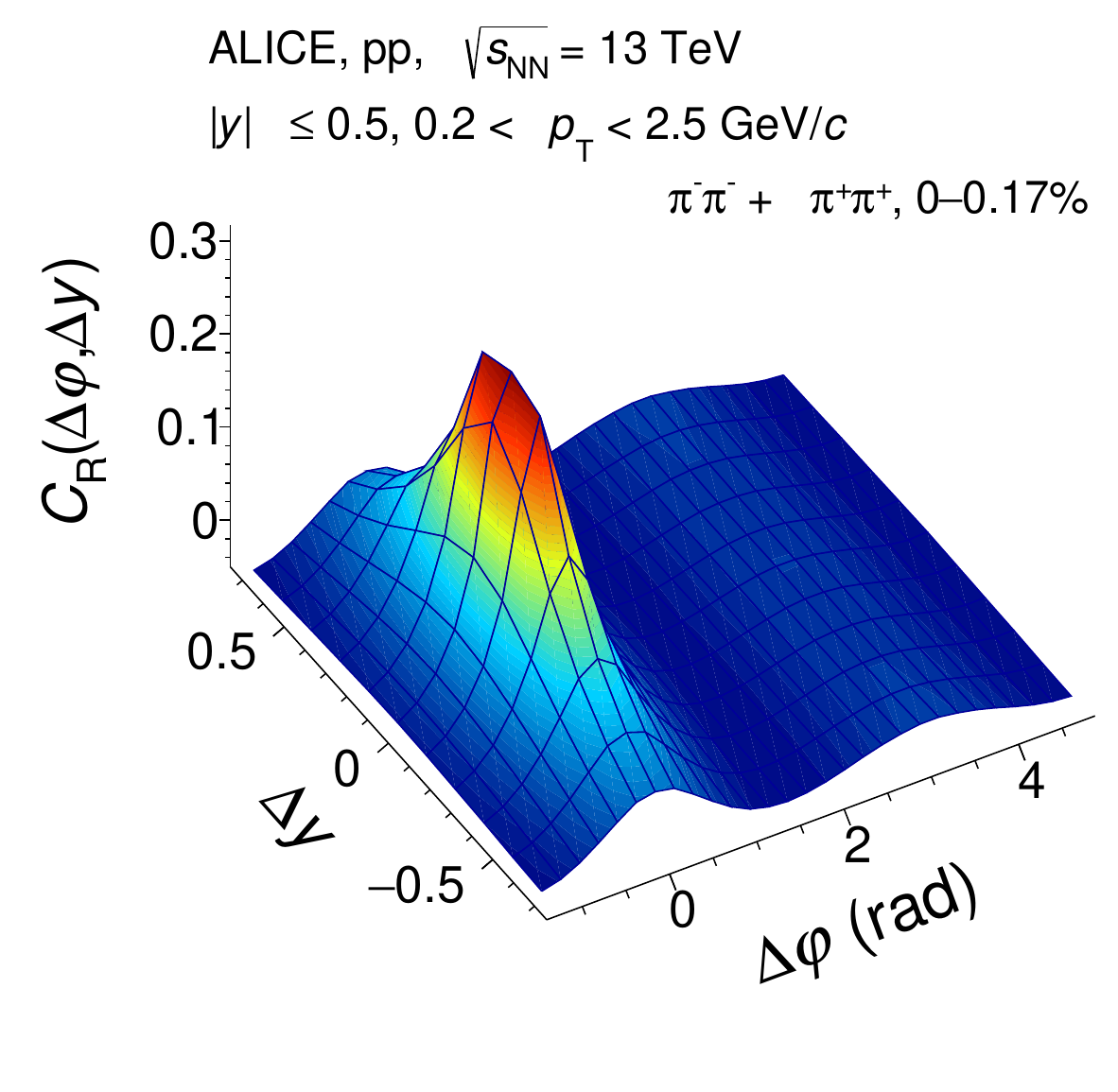} 
   \label{fig-a}
 }
 \subfigure[]{
   \includegraphics[width=4.9cm]{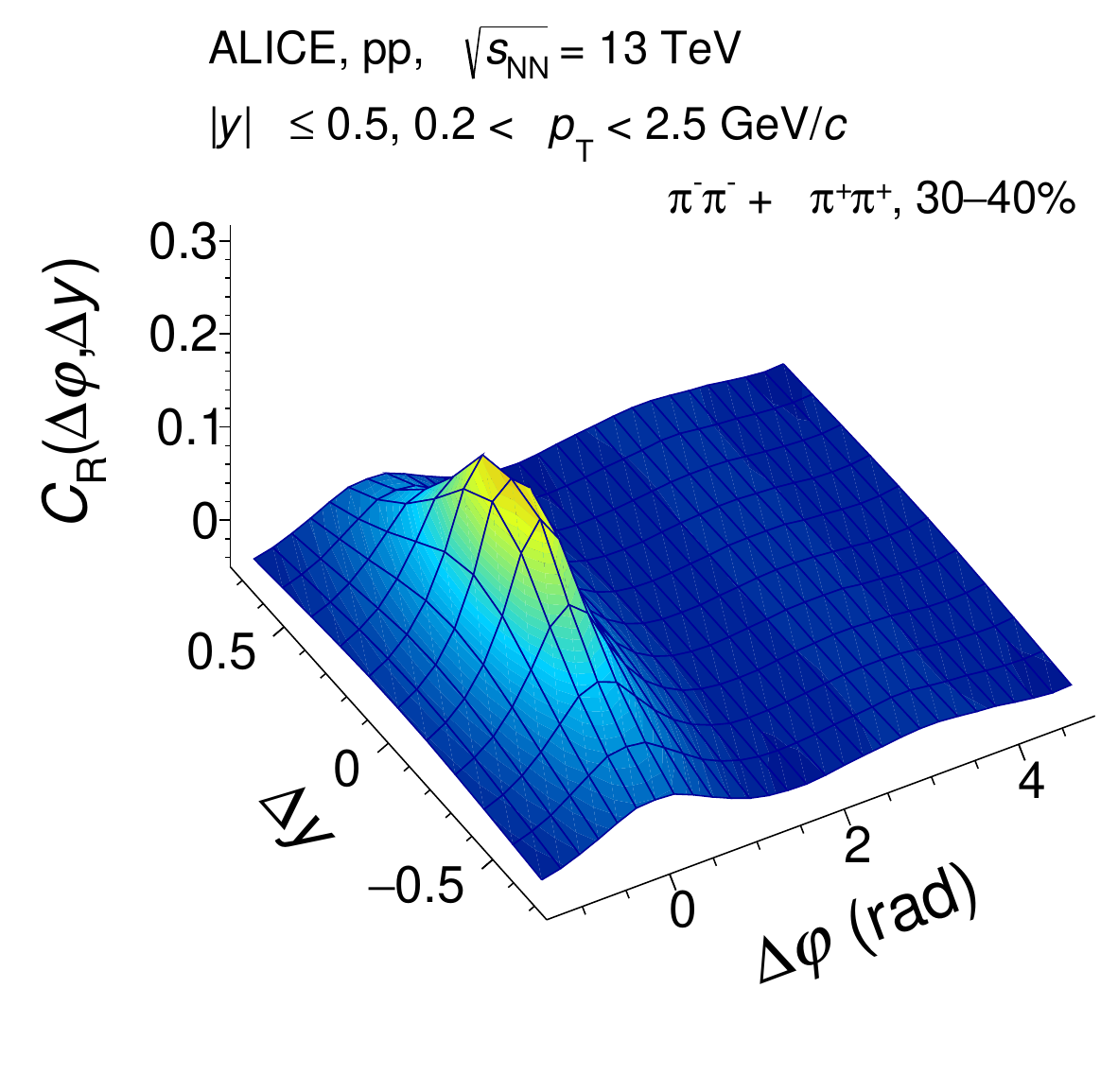} 
   \label{fig-b}
 }
 \subfigure[]{
   \includegraphics[width=4.9cm]{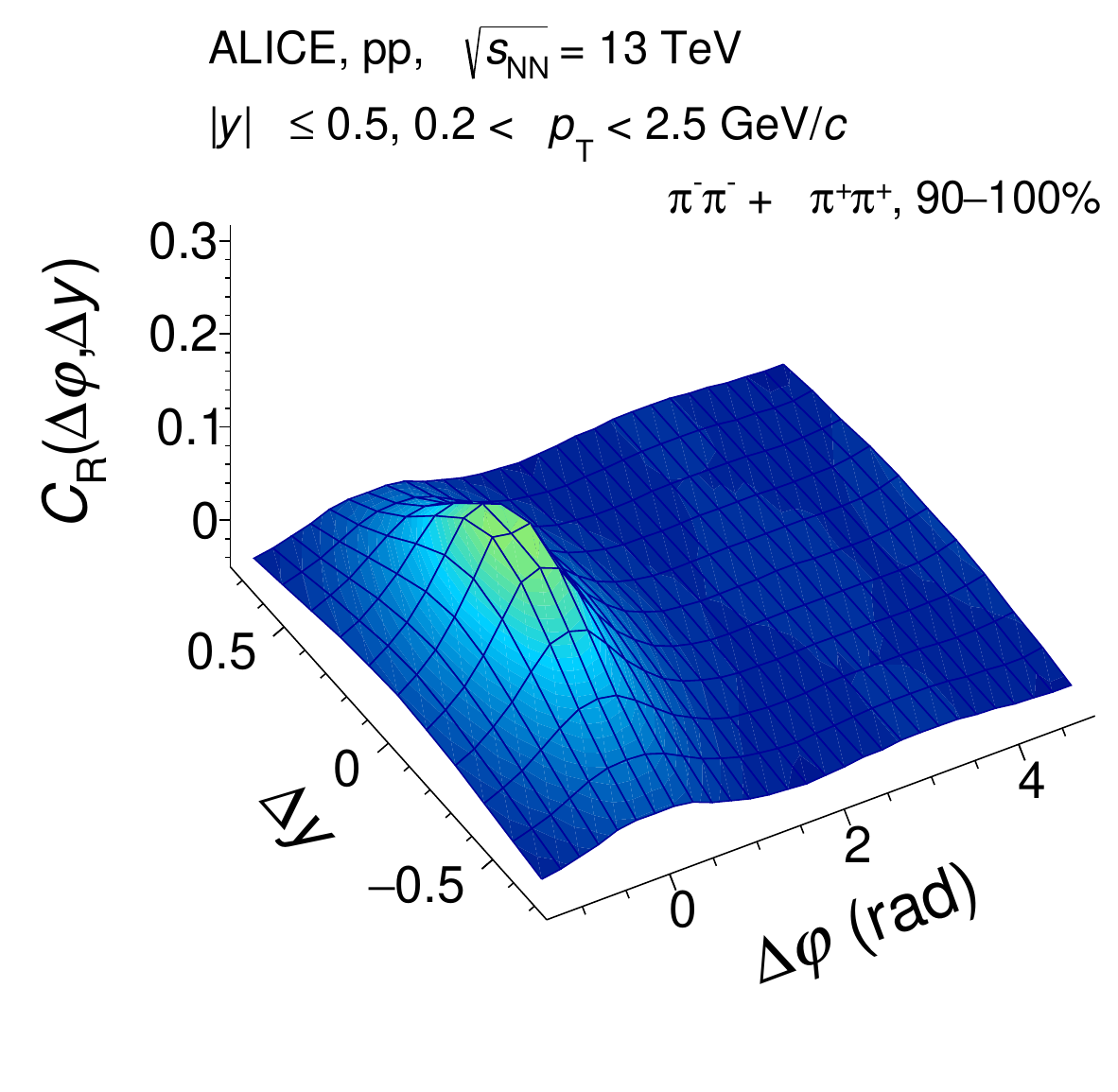}
   \label{fig-c}
}
\subfigure[]{
   \includegraphics[width=4.9cm]{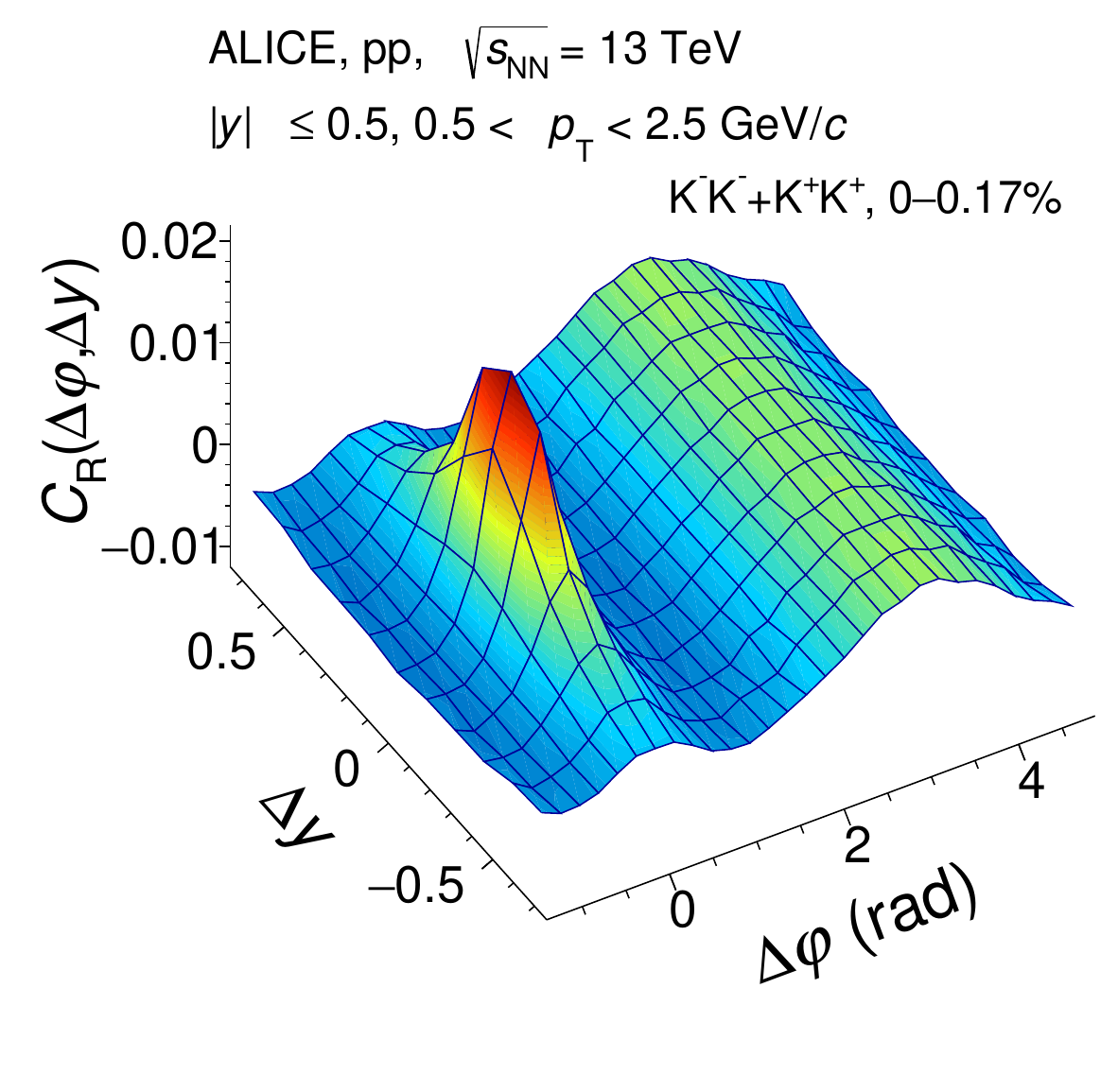}
   \label{fig-d}
 }
 \subfigure[]{
   \includegraphics[width=4.9cm]{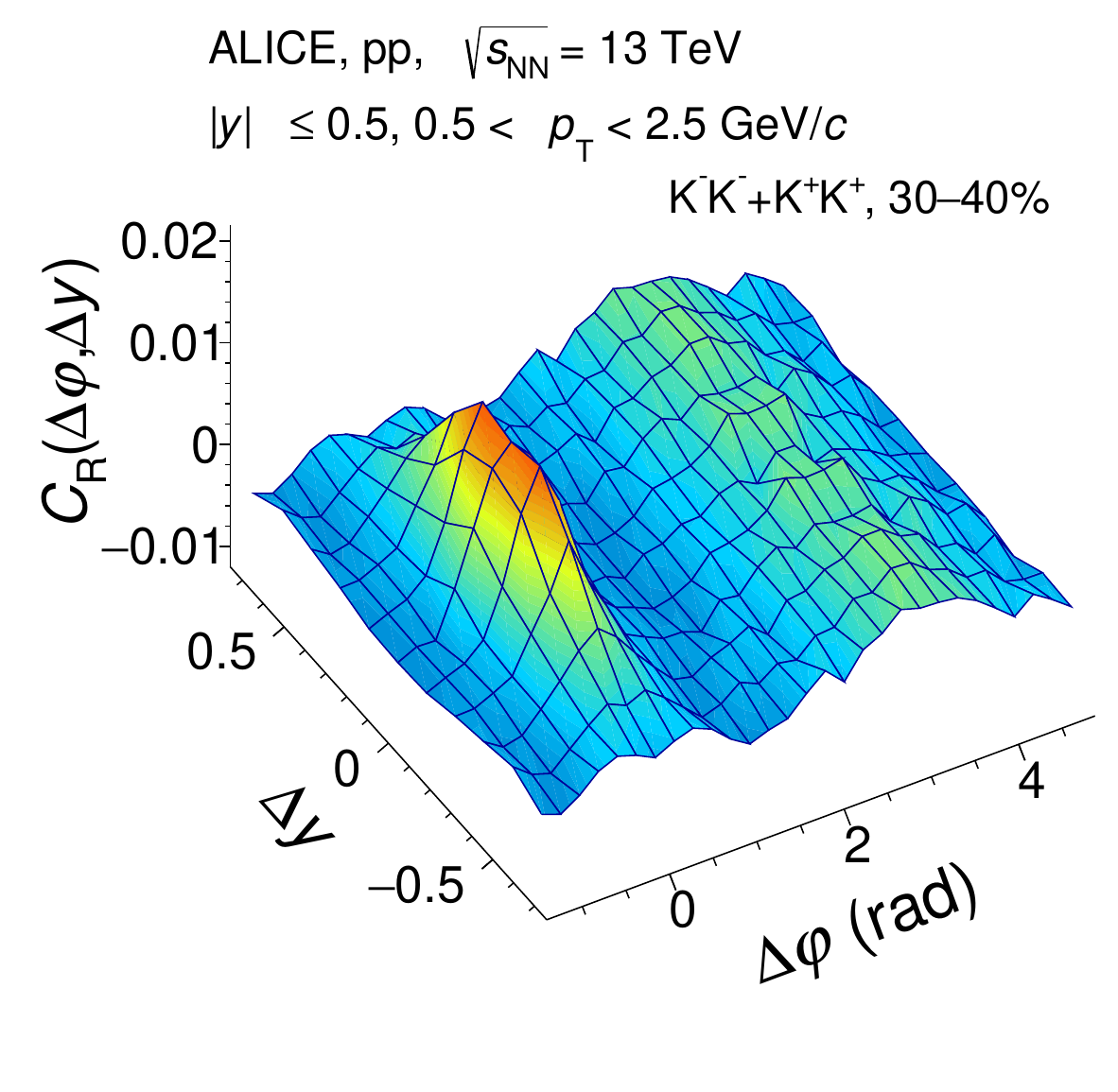}
   \label{fig-e}
 }
 \subfigure[]{
   \includegraphics[width=4.9cm]{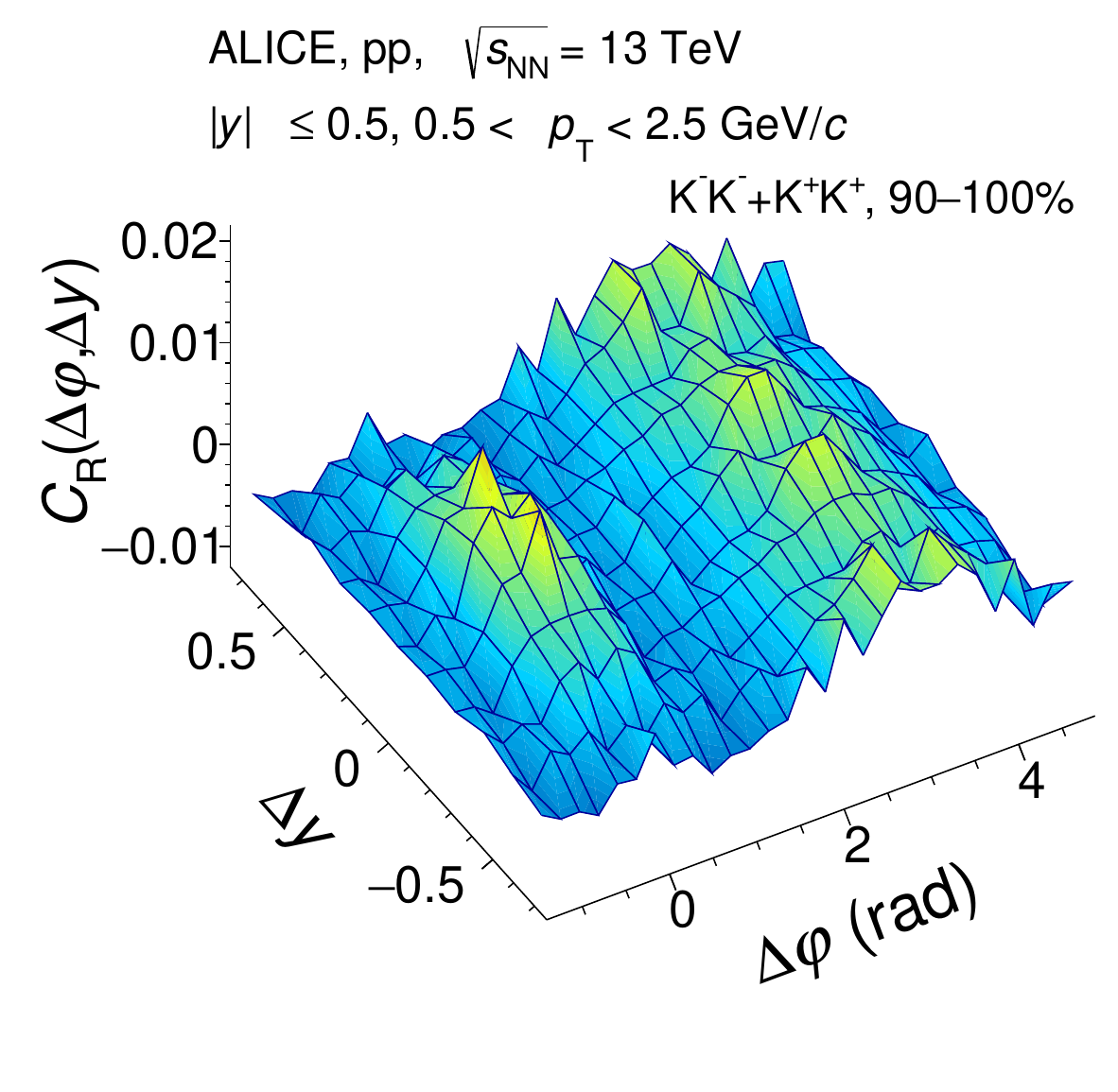}
   \label{fig-f}
}
\subfigure[]{
   \includegraphics[width=4.9cm]{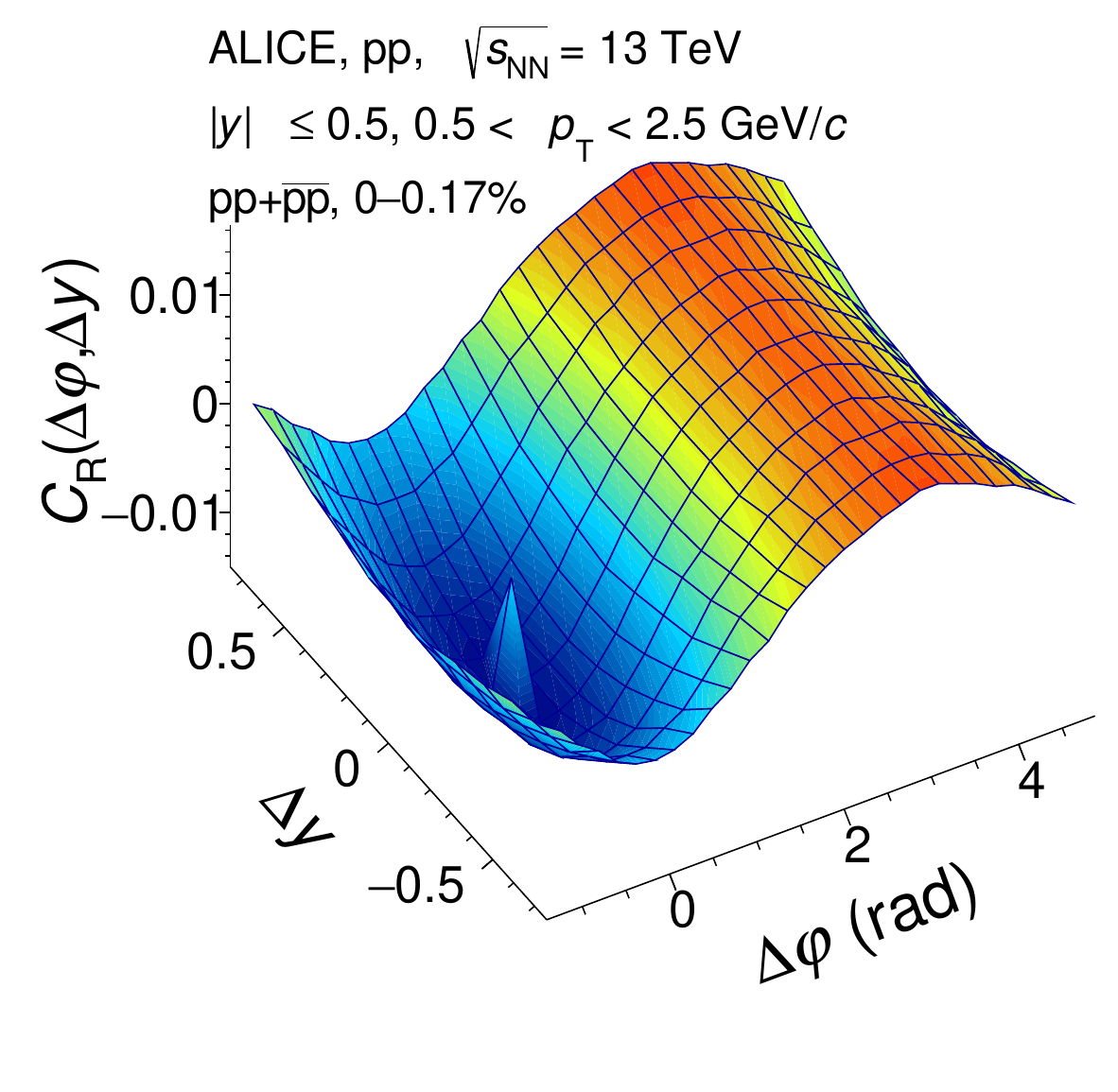}
   \label{fig-g}
 }
 \subfigure[]{
   \includegraphics[width=4.9cm]{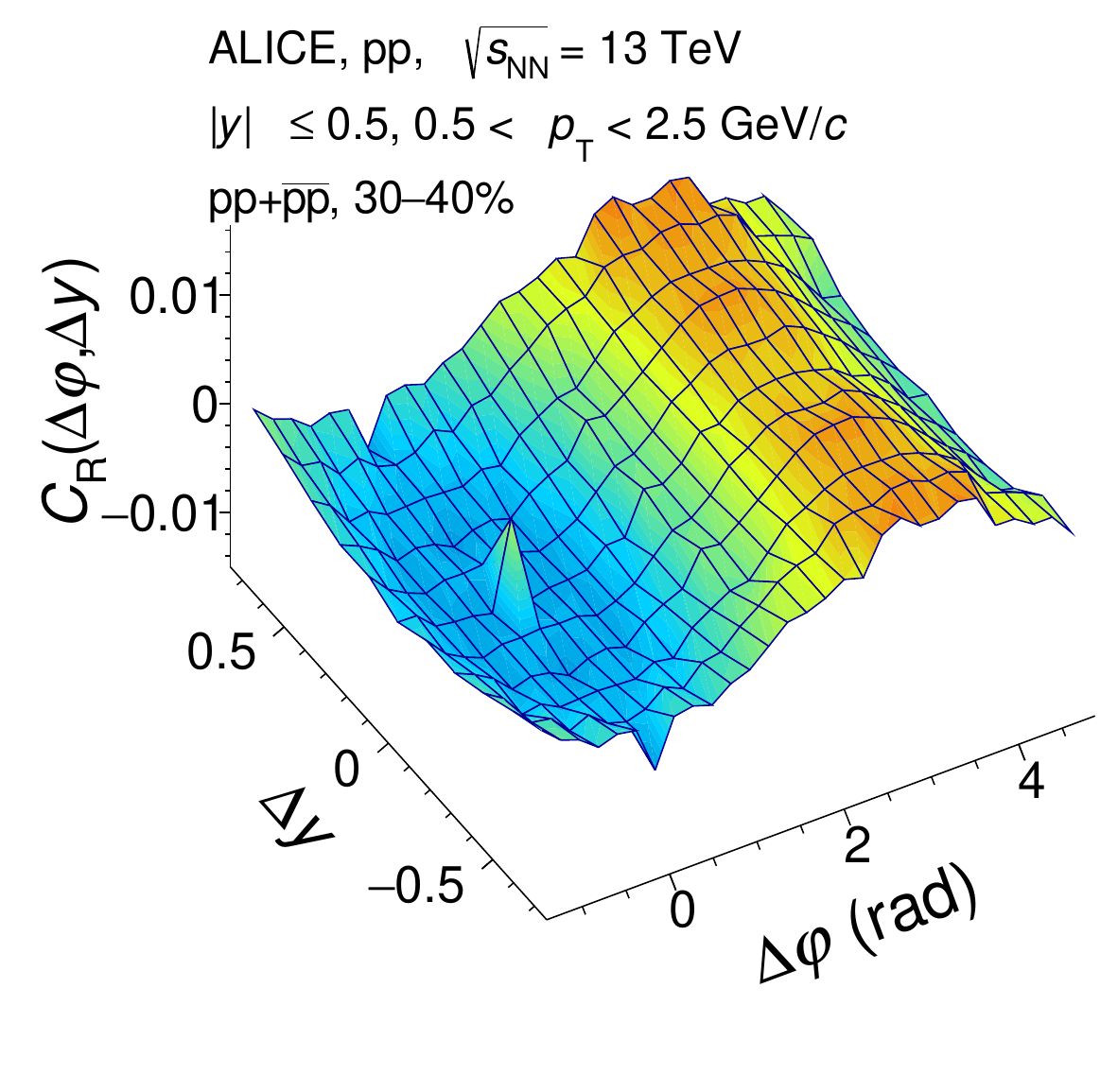}
   \label{fig-h}
 }
 \subfigure[]{
   \includegraphics[width=4.9cm]{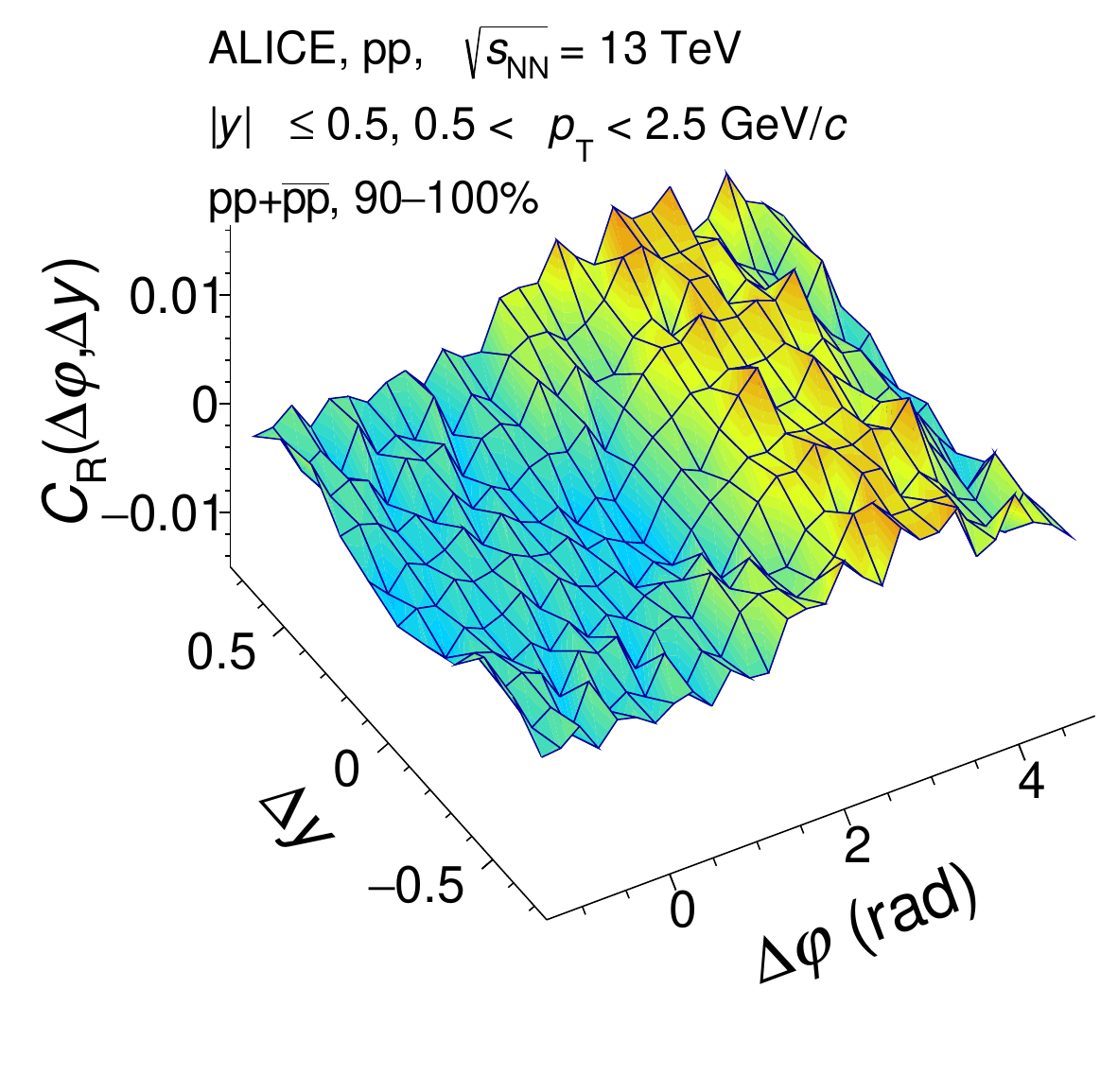}
   \label{fig-i}
}
\caption{The ($\Delta y,\Delta\varphi$) correlation functions for like-sign pions, kaons, and protons pairs (top-to-bottom) in the highest (0--0.17\%), the middle (30--40\%), and lowest (90--100\%) multiplicity classes (left-to-right) of pp collisions at $\sqrt{s} = 13$~TeV using the $C_{\rm{R}}$ definition.}
\label{corrfunc2D_like-sign}
\end{figure}

\begin{figure}[!h]
   \centering
 \subfigure[]{
   \includegraphics[width=4.9cm]{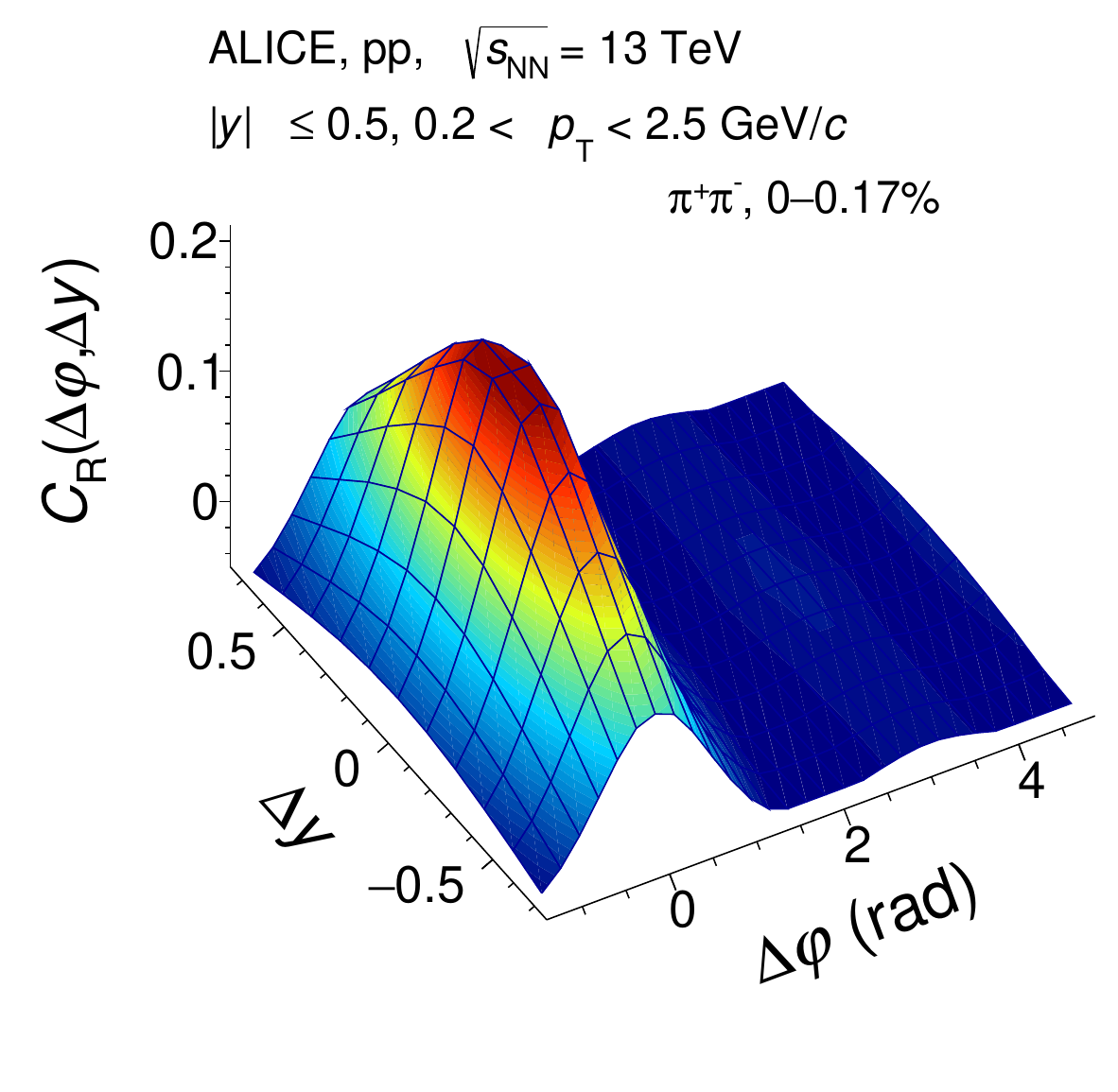} 
   \label{fig2-a}}
    \subfigure[]{
   \includegraphics[width=4.9cm]{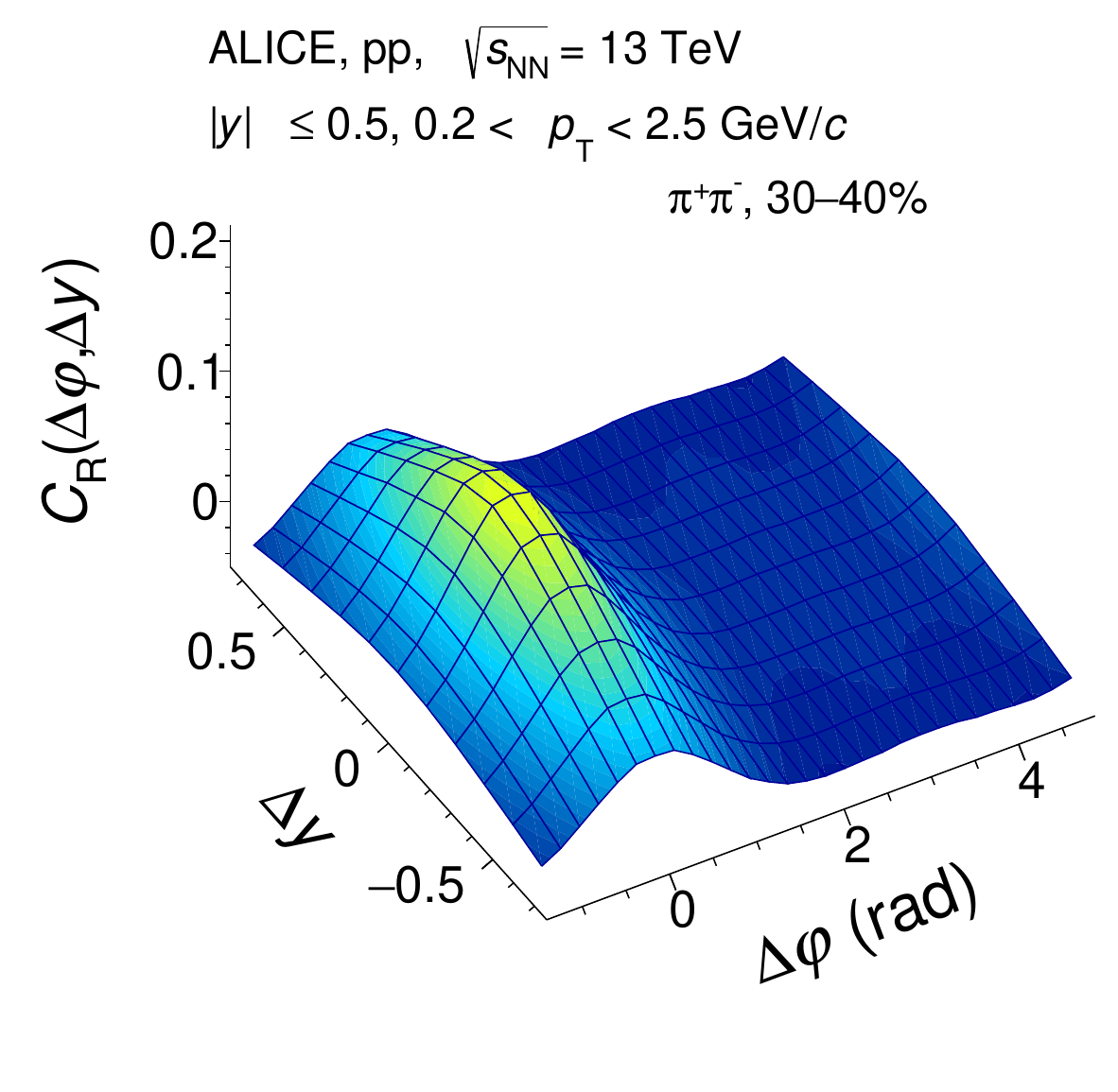} 
   \label{fig2-b}}
 \subfigure[]{
   \includegraphics[width=4.9cm]{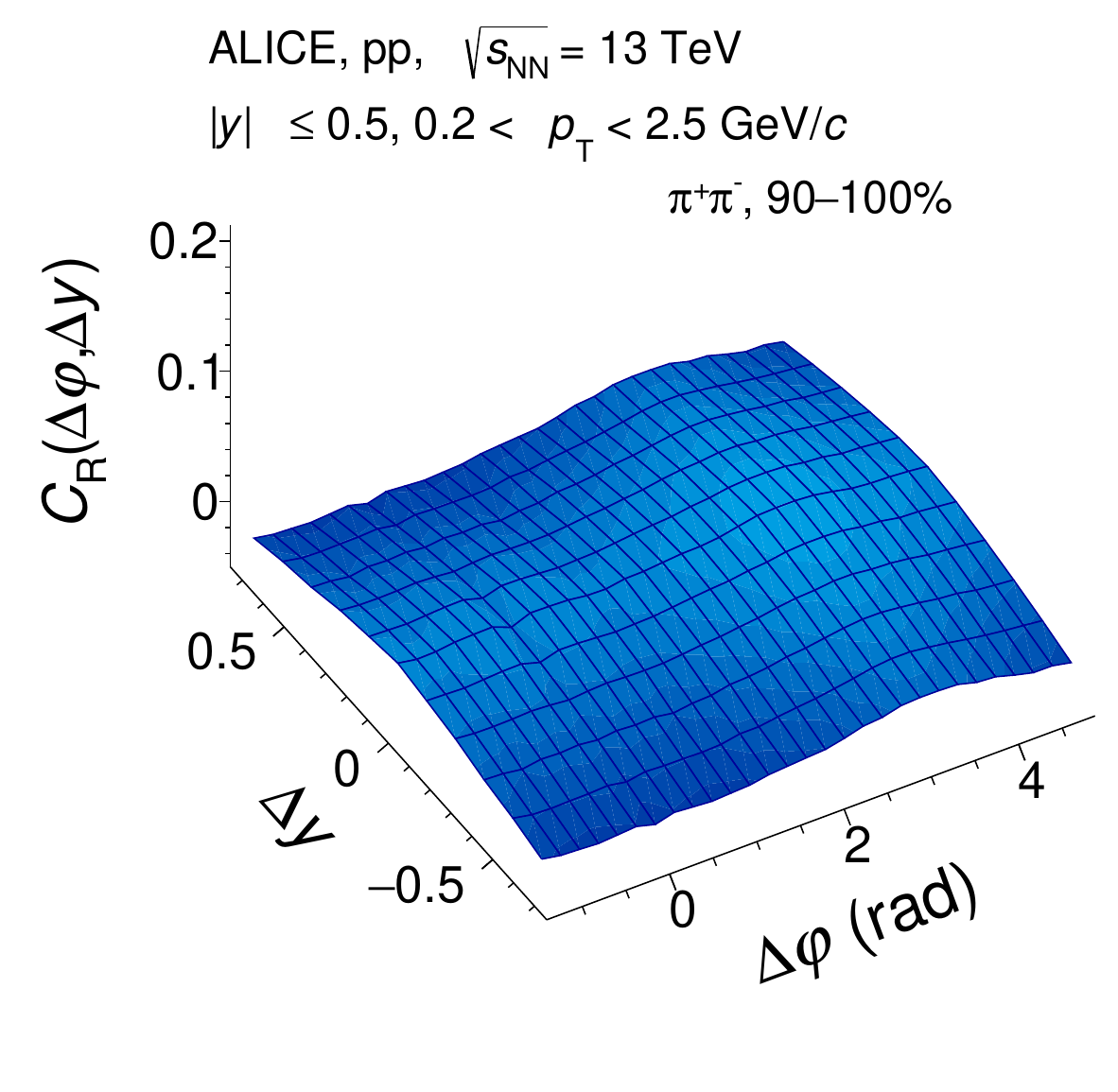}
   \label{fig2-c}}
\subfigure[]{
   \includegraphics[width=4.9cm]{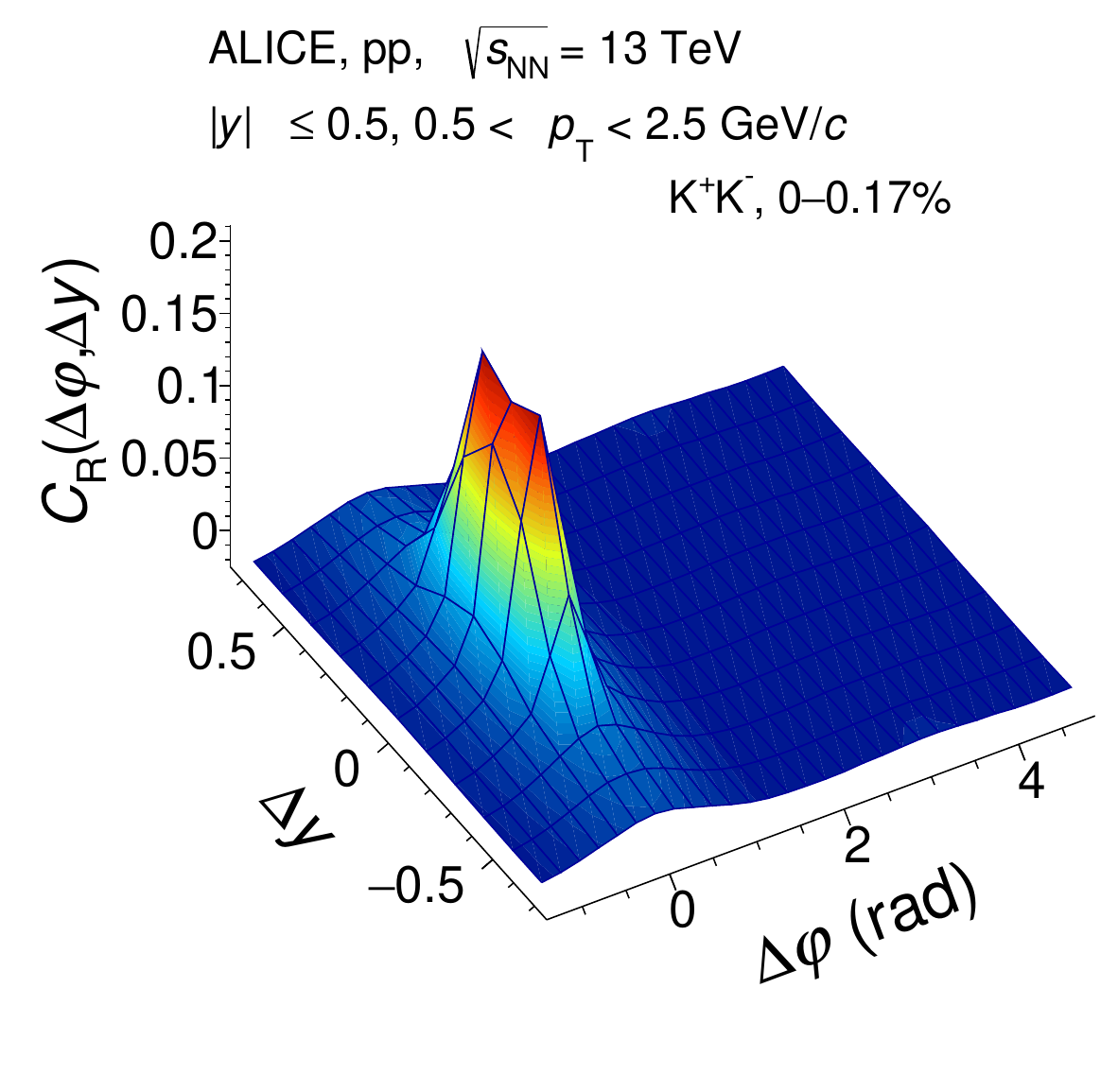}
   \label{fig2-d}}
   \subfigure[]{
   \includegraphics[width=4.9cm]{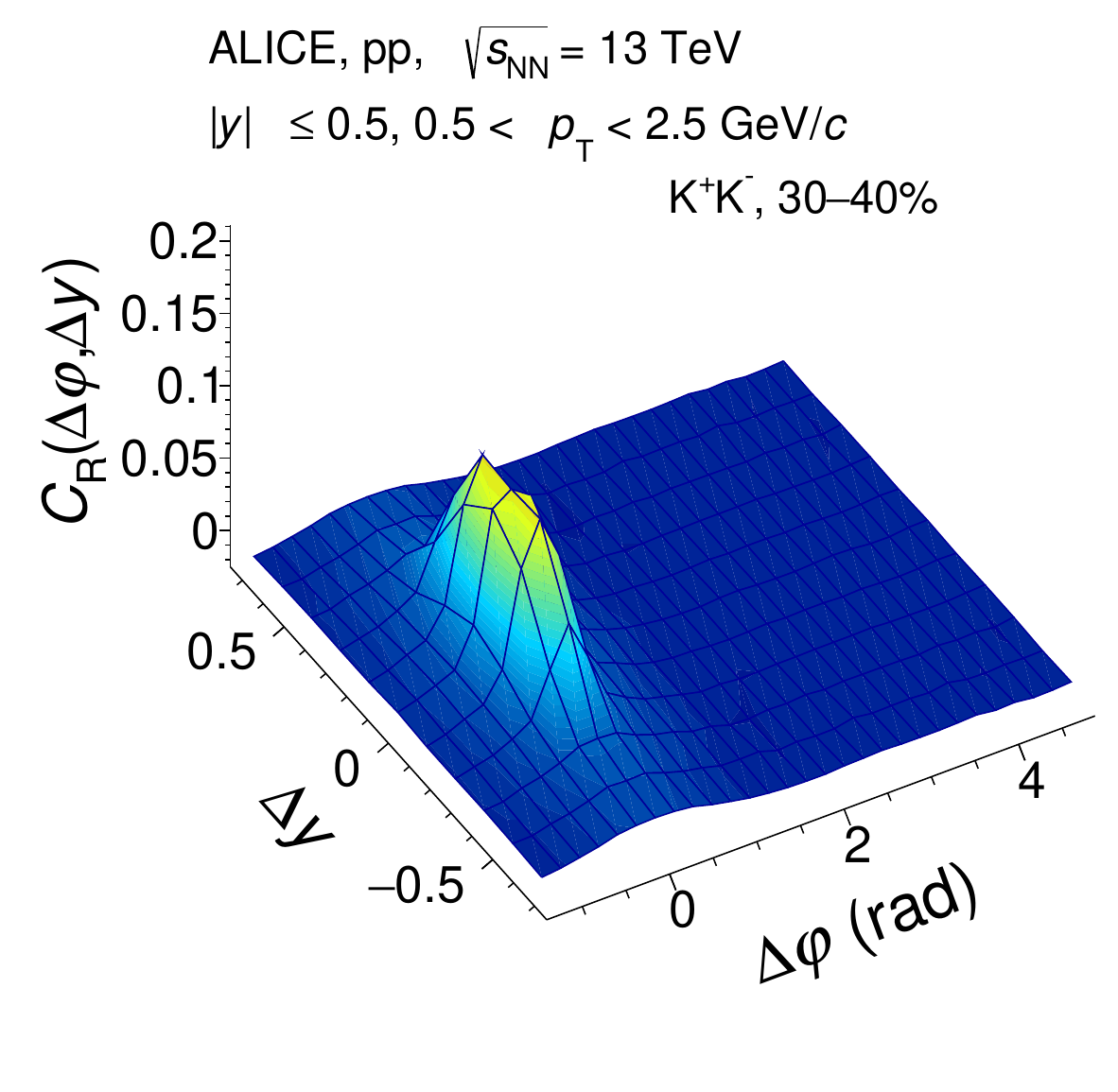}
   \label{fig2-e}}
 \subfigure[]{
   \includegraphics[width=4.9cm]{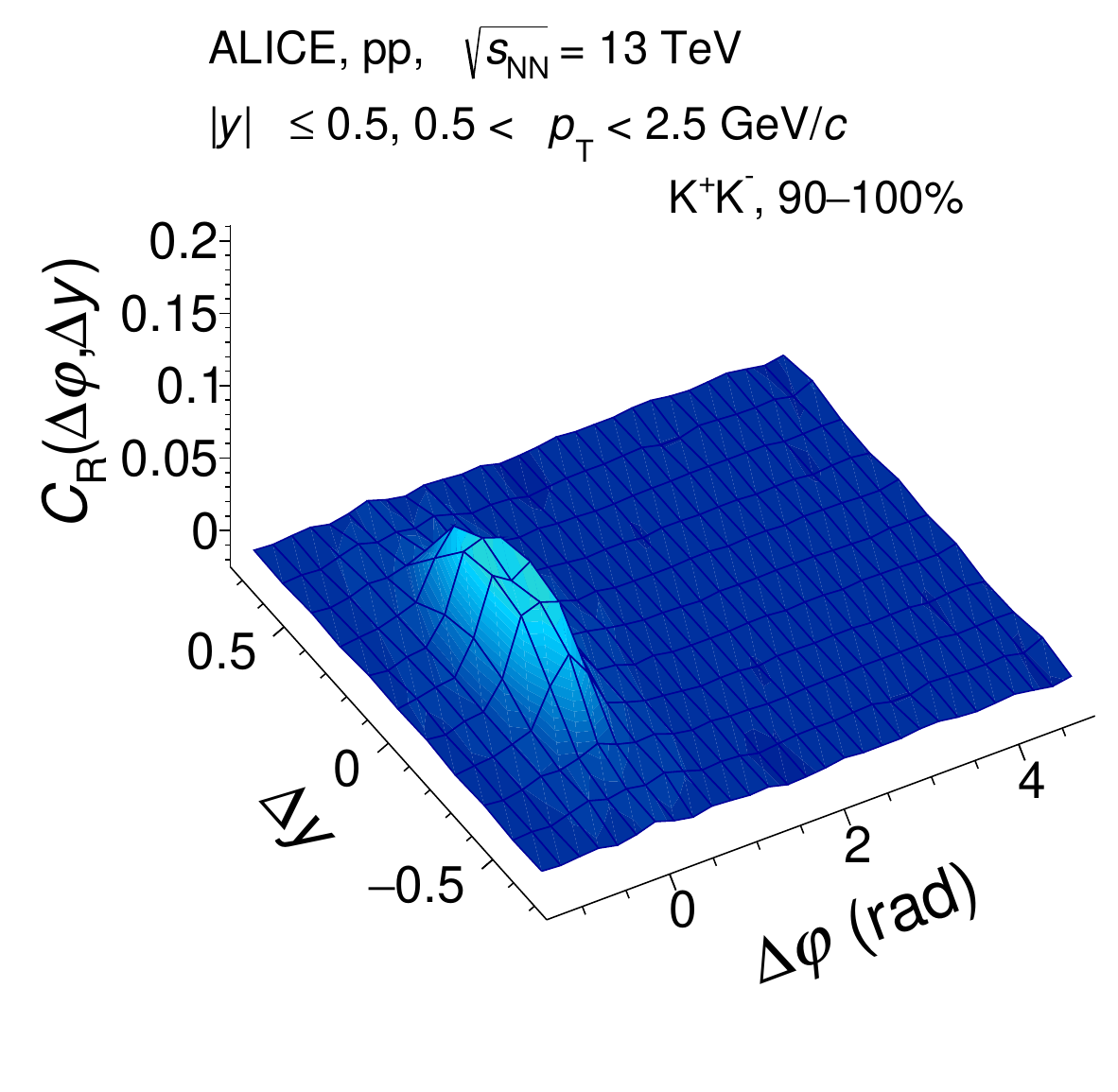}
   \label{fig2-f}}
\subfigure[]{
   \includegraphics[width=4.9cm]{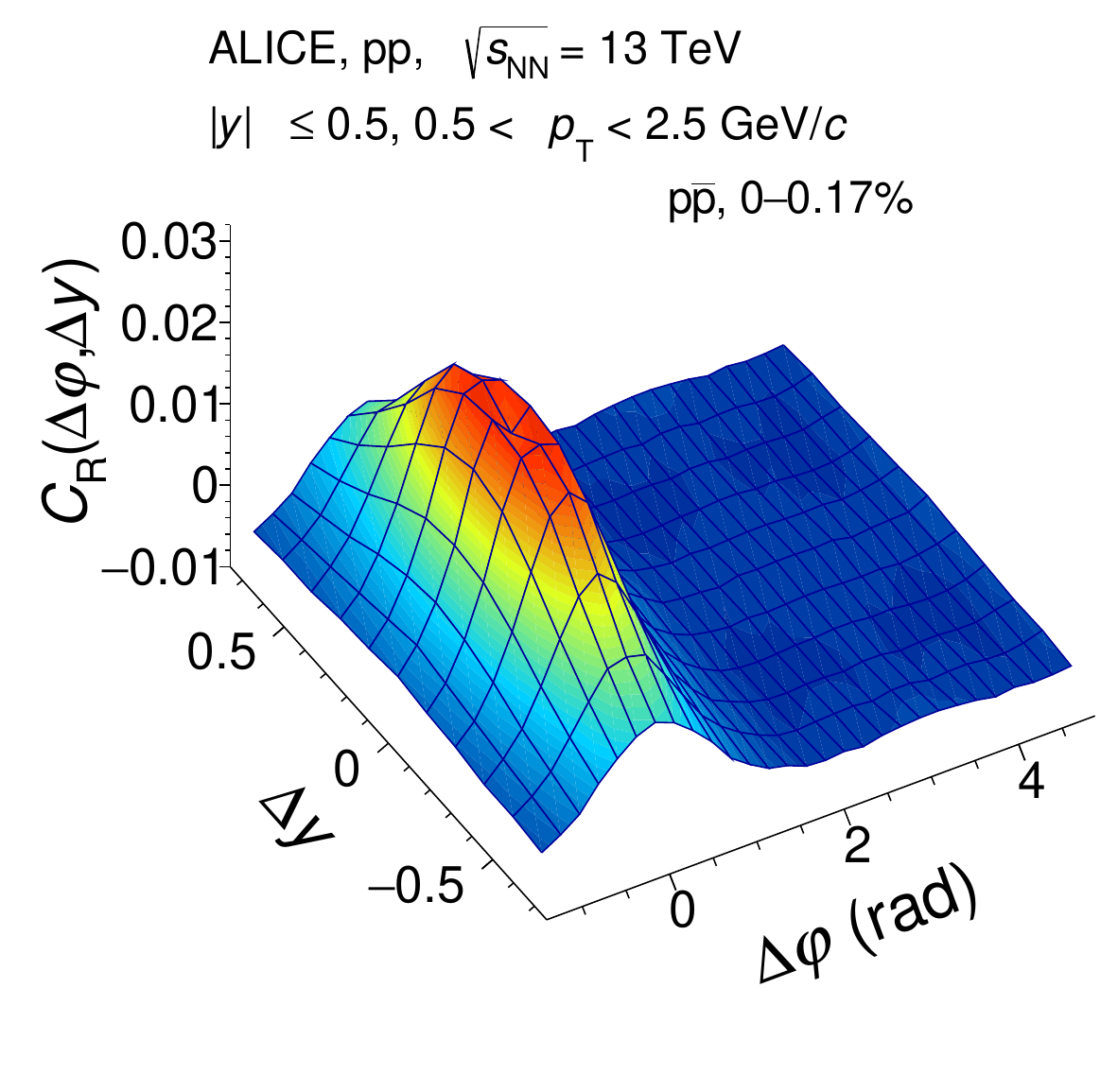}
   \label{fig2-g}}
   \subfigure[]{
   \includegraphics[width=4.9cm]{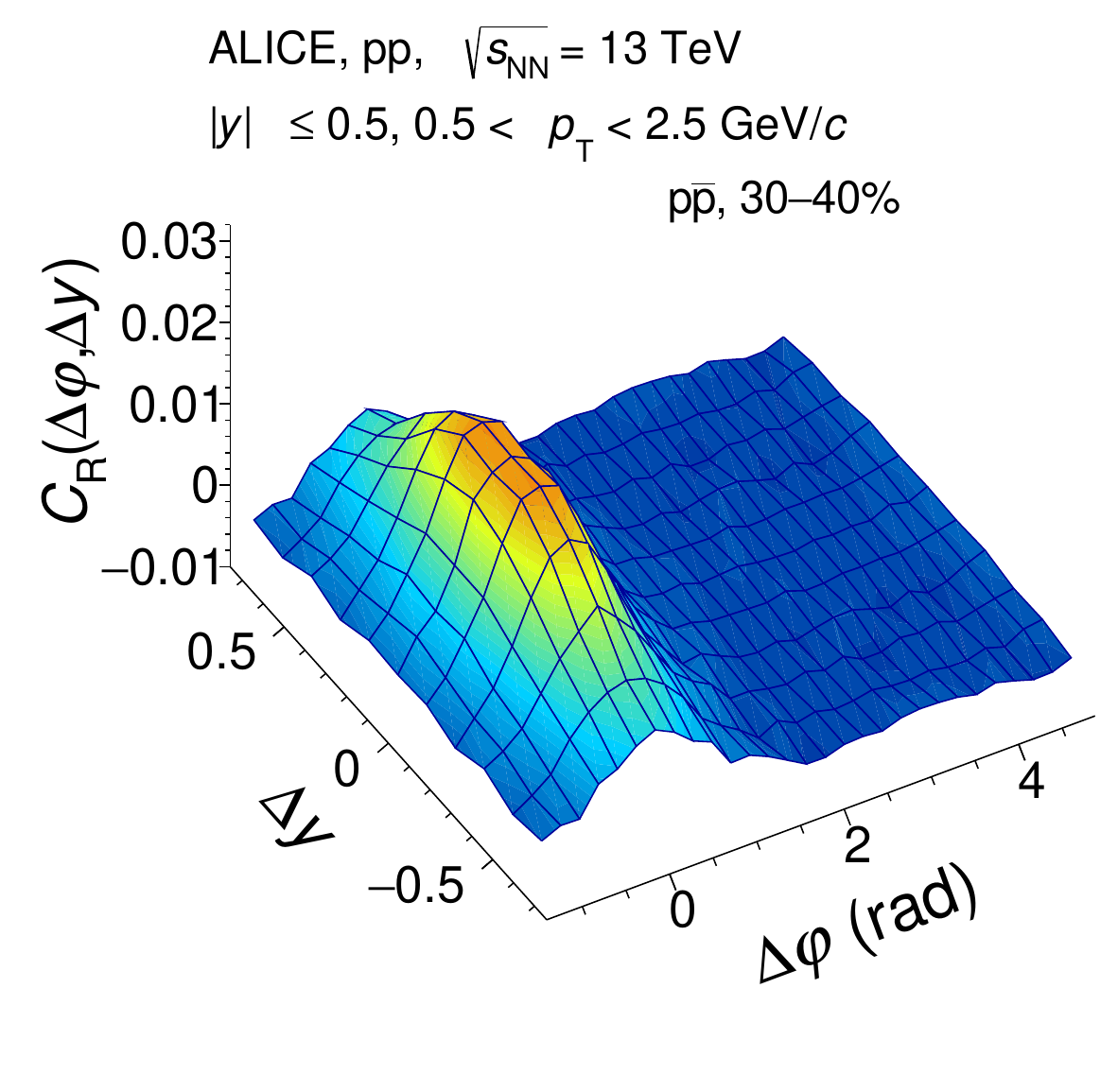}
   \label{fig2-h}}
 \subfigure[]{
   \includegraphics[width=4.9cm]{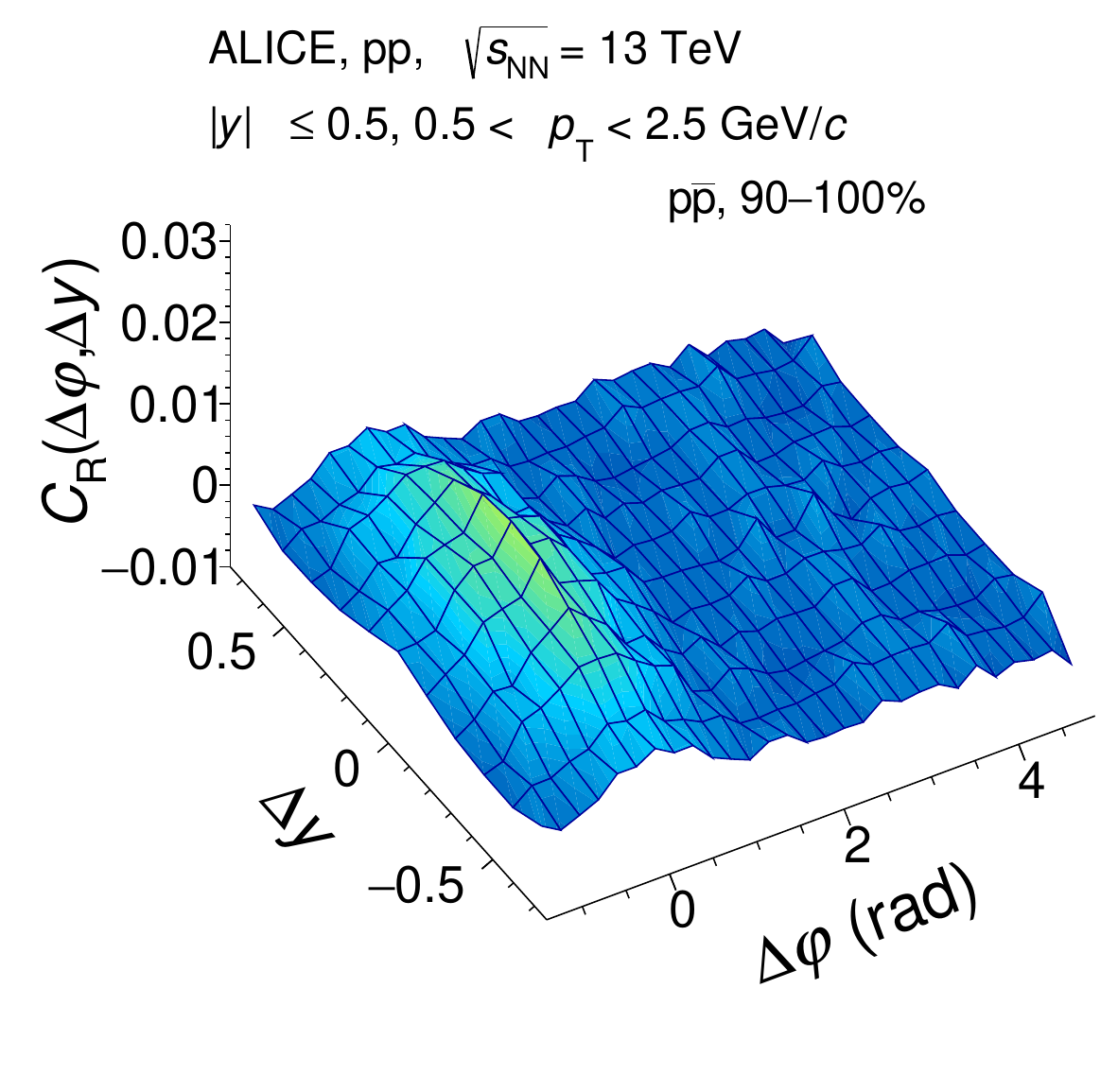}
   \label{fig2-i}}
\caption{The ($\Delta y,\Delta\varphi$) correlation functions for unlike-sign pions, kaons, and protons pairs (top-to-bottom) in the highest (0--0.17\%), the middle (30--40\%), and lowest (90--100\%) multiplicity classes (left-to-right) of pp collisions at $\sqrt{s} = 13$~TeV using the $C_{\rm{R}}$ definition.}
\label{corrfunc2D_unlike-sign}
\end{figure}

The $\Delta y\Delta\varphi$ correlation functions for pions, kaons and protons for like and unlike-sign pairs are shown in \Cref{corrfunc2D_like-sign,corrfunc2D_unlike-sign} for the highest (0--0.17\%), middle (30--40\%), and lowest (90--100\%) multiplicity class, respectively. 
Due to statistical fluctuations at high $|\Delta y|$, the figures are plotted in a limited range of $|\Delta y|<0.7$.
The correlations of like-sign mesons (see \Cref{fig-a,fig-b,fig-c,fig-d,fig-e,fig-f}) show a pronounced enhancement in the near-side region ($\Delta y, \Delta\varphi) \sim (0,0)$, primarily due to mini jets and Bose--Einstein quantum-statistics. 
Mini jets~\cite{Ali:jets,Landshoff:minijets}, arise from parton fragmentation at low momenta, emerge with minimal difference in azimuthal and polar angles, thereby contributing significantly to the observed signal. 
Additionally, Bose--Einstein quantum statistics amplify the near-side correlation for like-sign pions and kaons; their bosonic nature allows them to occupy the same quantum state -- unlike fermions that are subject to Pauli’s exclusion principle -- leading to similar values of $\varphi$ and $y$. The opposite behavior is observed for the baryon--baryon and antibaryon--antibaryon correlation functions of identical proton pairs, as shown in \Cref{fig-g,fig-h,fig-i}. Instead of a near-side peak, the correlation functions exhibit a wide anticorrelation in this region, accompanied by a narrow peak at (\(\Delta y, \Delta\varphi\)) = (0,0), which arises due to final-state interactions (FSI)~\cite{unfolding}. Additionally, a pronounced away-side correlation is observed. This behavior has previously been reported in pp collisions at \(\sqrt{s} = 7\) TeV~\cite{ALICEInsight} and in Au--Au collisions across energies \(\sqrt{s_{\rm NN}} = 7.7 - 200\) GeV~\cite{STARbeamenergy}.

In contrast, correlations between particle--antiparticle pairs for mesons (~\Cref{fig2-a,fig2-b,fig2-c,fig2-d,fig2-e,fig2-f}) and baryons (~\Cref{fig2-g,fig2-h,fig2-i}) exhibit a mini-jet structure, as well as contributions from Coulomb and strong final-state interactions, which further enhance the near-side peak. While Bose--Einstein and Fermi--Dirac quantum-statistic effects do not manifest in pairs of non-identical particles, charge conservation and resonance decays can significantly influence the correlation structure. 
In particular, the highly pronounced and narrow peak observed for unlike-sign kaons (~\Cref{fig2-d,fig2-e,fig2-f}) is largely driven by the decay of the \(\phi\) meson and strangeness conservation.
When a resonance decay releases a large amount of kinetic energy, it can produce a ridge-like structure at \(\Delta y \approx 0\) that extends across the entire \(\Delta\varphi\) range, as observed in \Cref{fig2-a,fig2-b,fig2-c} for unlike-sign pion correlations, particularly in the lower multiplicity classes. This effect is primarily attributed to the decay of the \(\rho^0\) meson into \(\pi^+\pi^-\), which contributes significantly to the away-side enhancement. Additional contributions may arise from \(\eta\) and \(\omega\) mesons, which decay into three-pion final states (\(\pi^+\pi^-\pi^0\)), as discussed in Ref.~\cite{Eggert:1974ek}.

\begin{figure}[!h]
    \centering
    \includegraphics[width=7cm]{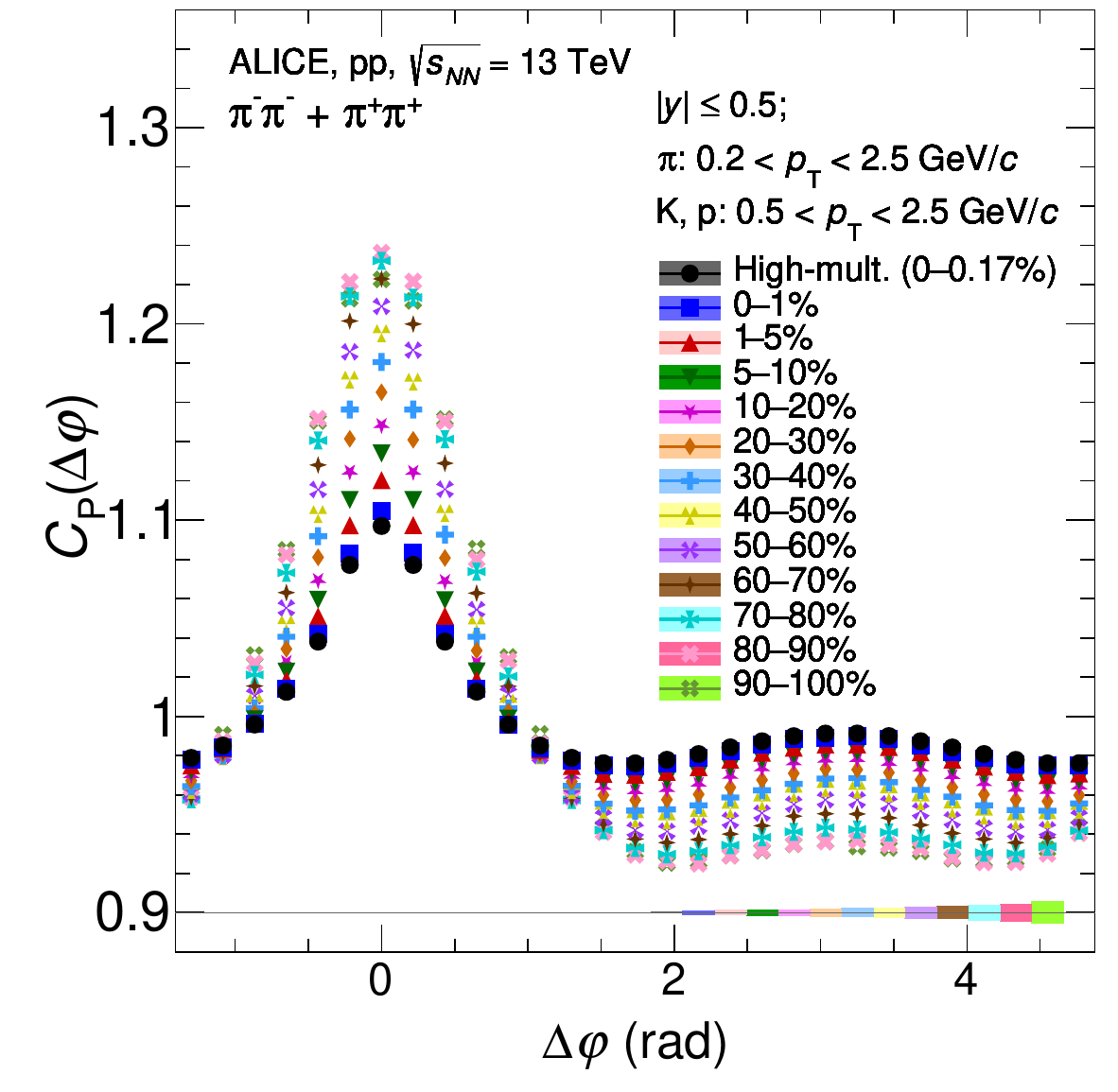}
    \hspace{1cm}
    \includegraphics[width=7cm]{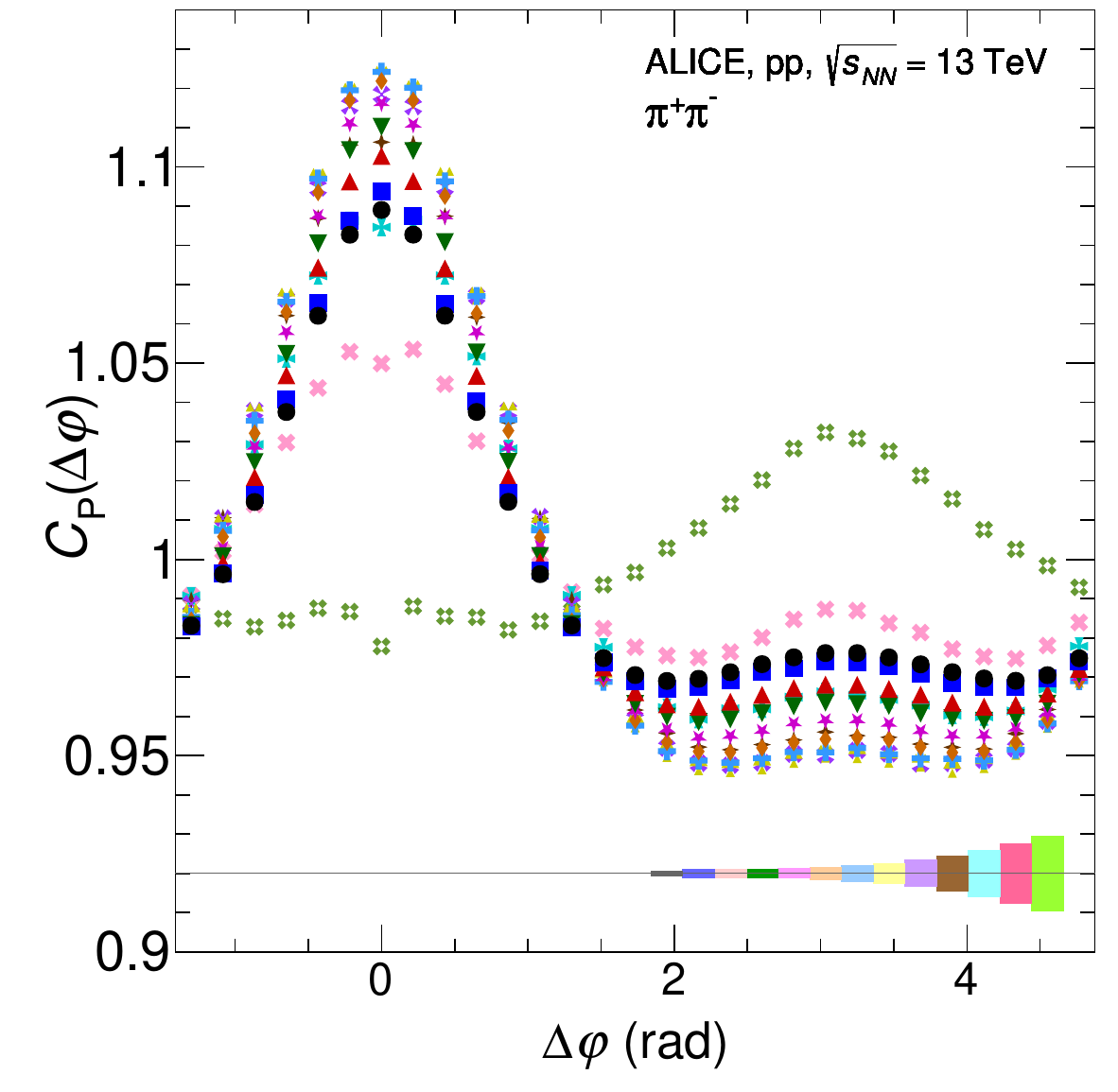}
    \includegraphics[width=7cm]{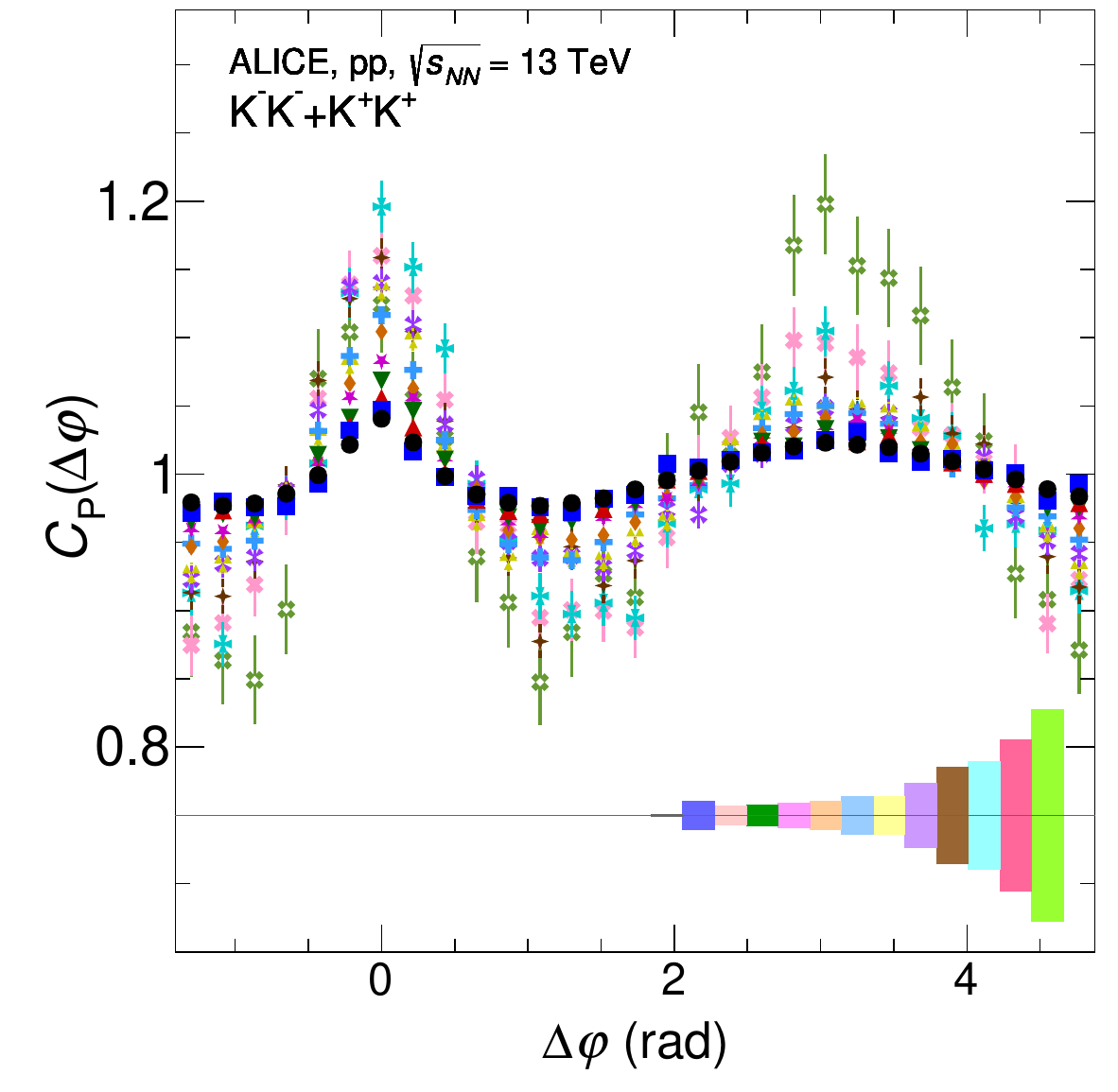}
    \hspace{1cm}
    \includegraphics[width=7cm]{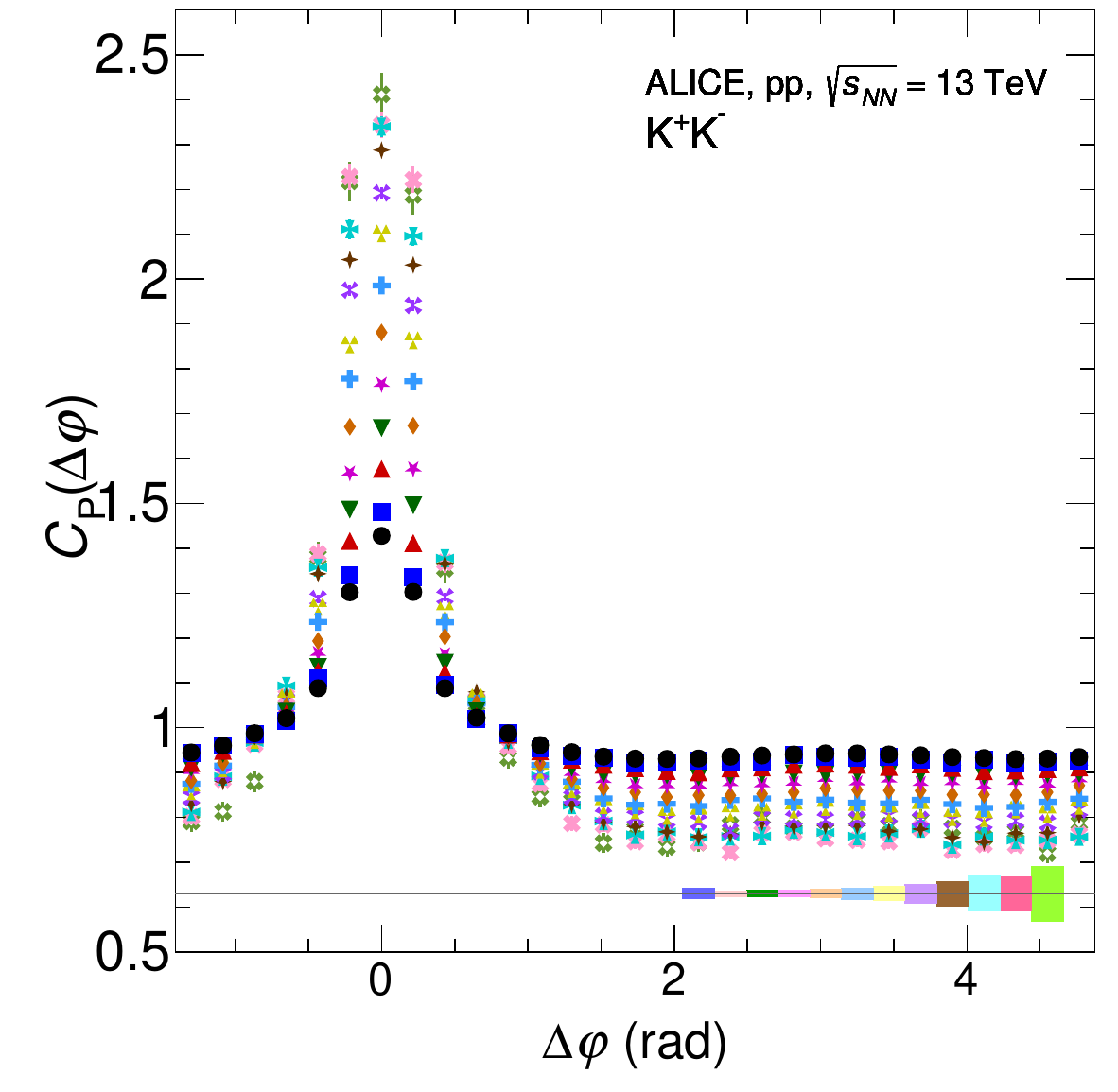}
    \includegraphics[width=7cm]{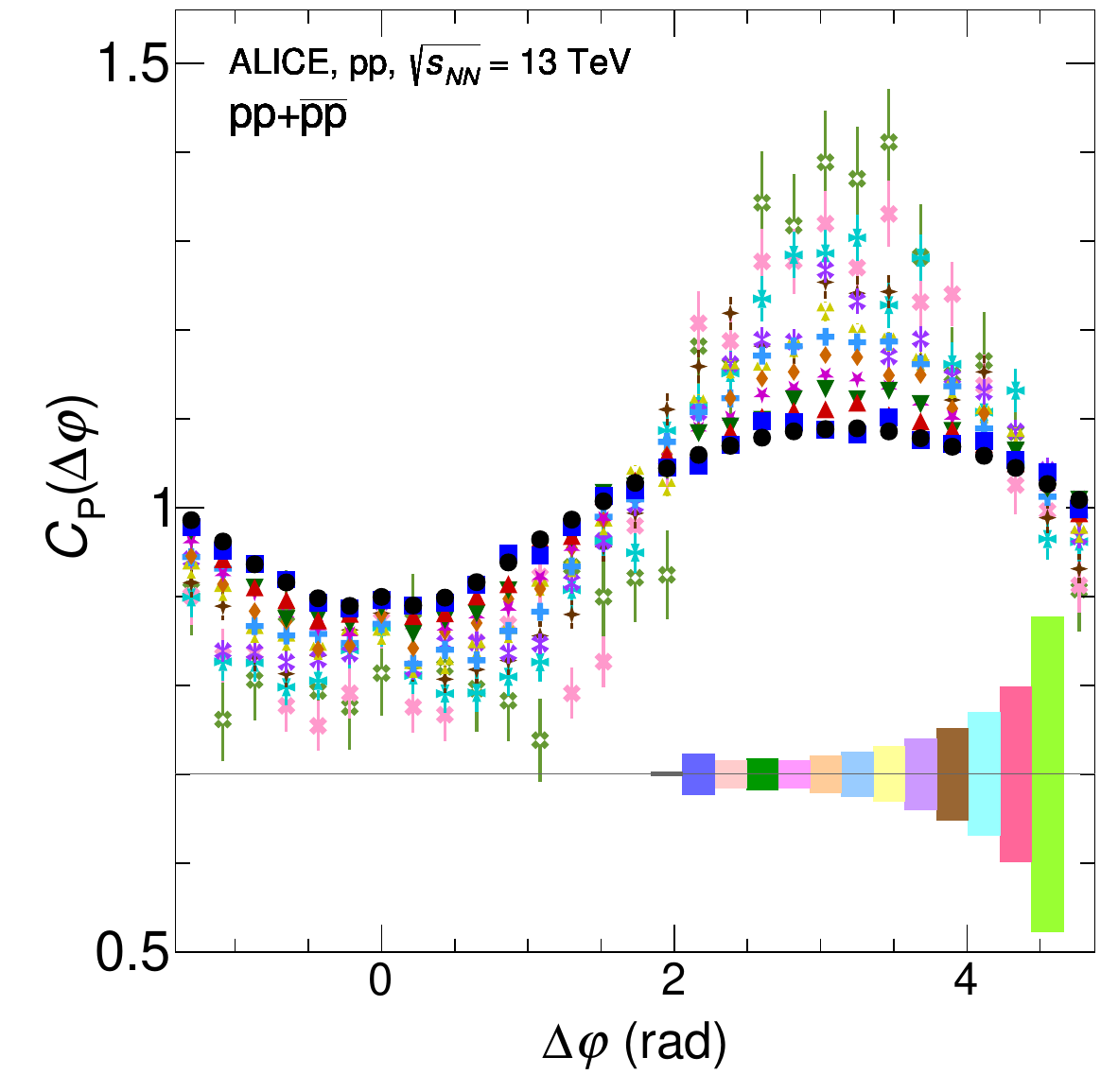}
    \hspace{1cm}
    \includegraphics[width=7cm]{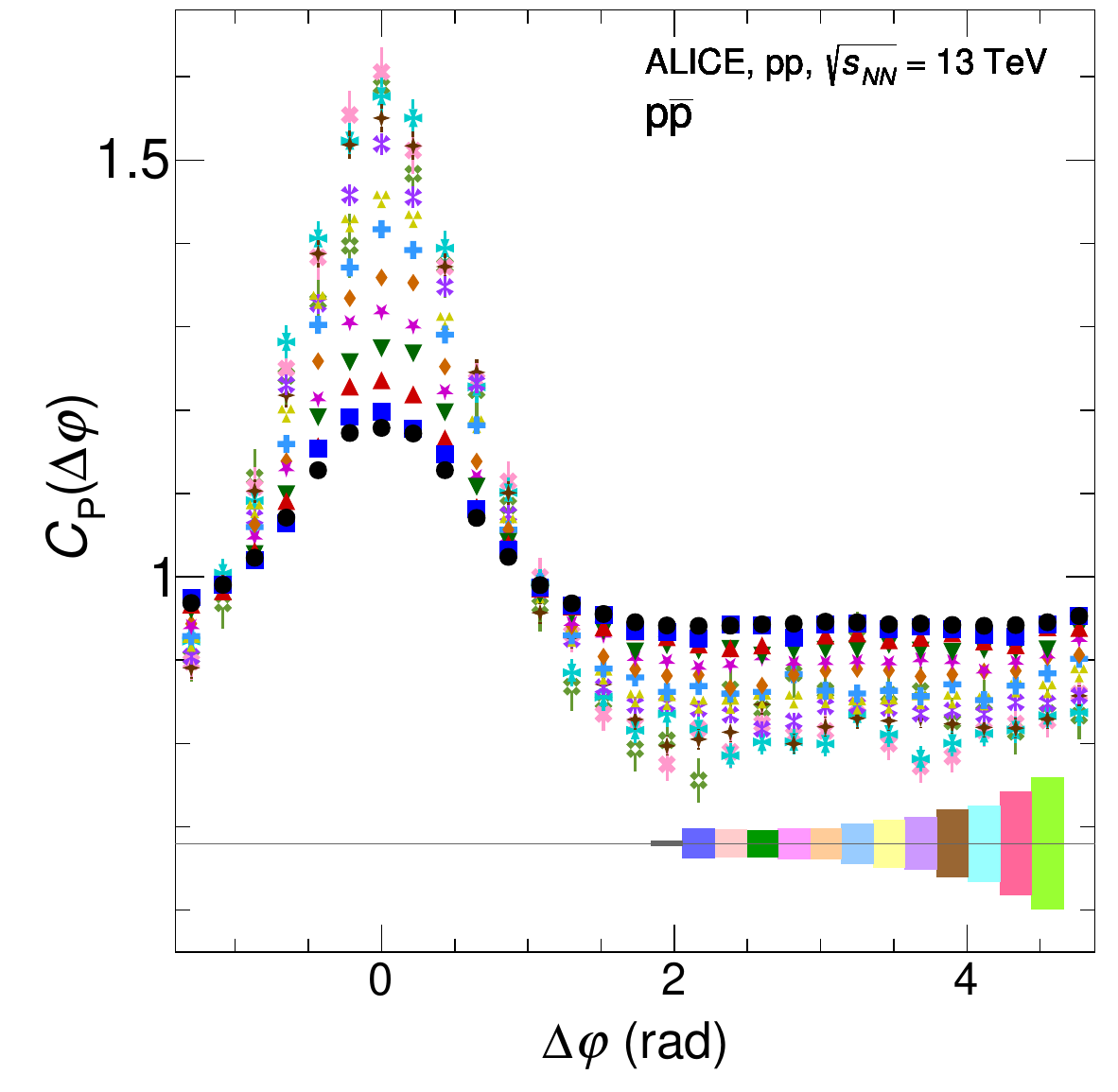}
    \caption{$\Delta\varphi$ projections of correlation functions within $\abs{\Delta y} < 1$ for like- and unlike-sign pions, kaons, and protons in multiplicity classes, using the $C_{\rm{P}}$ definition. Statistical uncertainties (bars) and systematic uncertainties (boxes) are both drawn with the color associated with the corresponding multiplicity class.}
    \label{fig:projection_probabilityratio}
\end{figure}

\begin{figure}[!h]
    \centering
    \includegraphics[width=7cm]{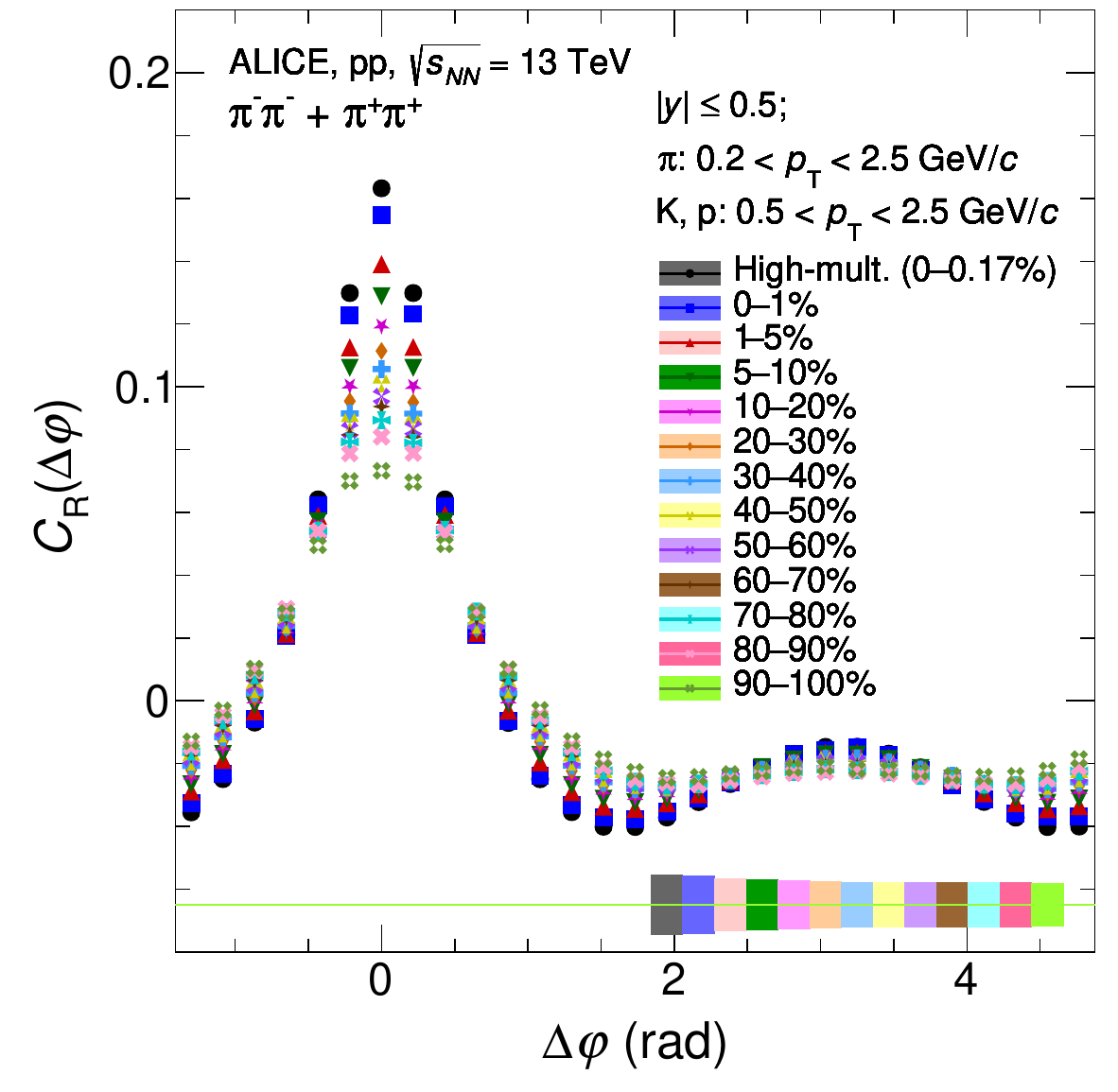}
    \hspace{1cm}
    \includegraphics[width=7cm]{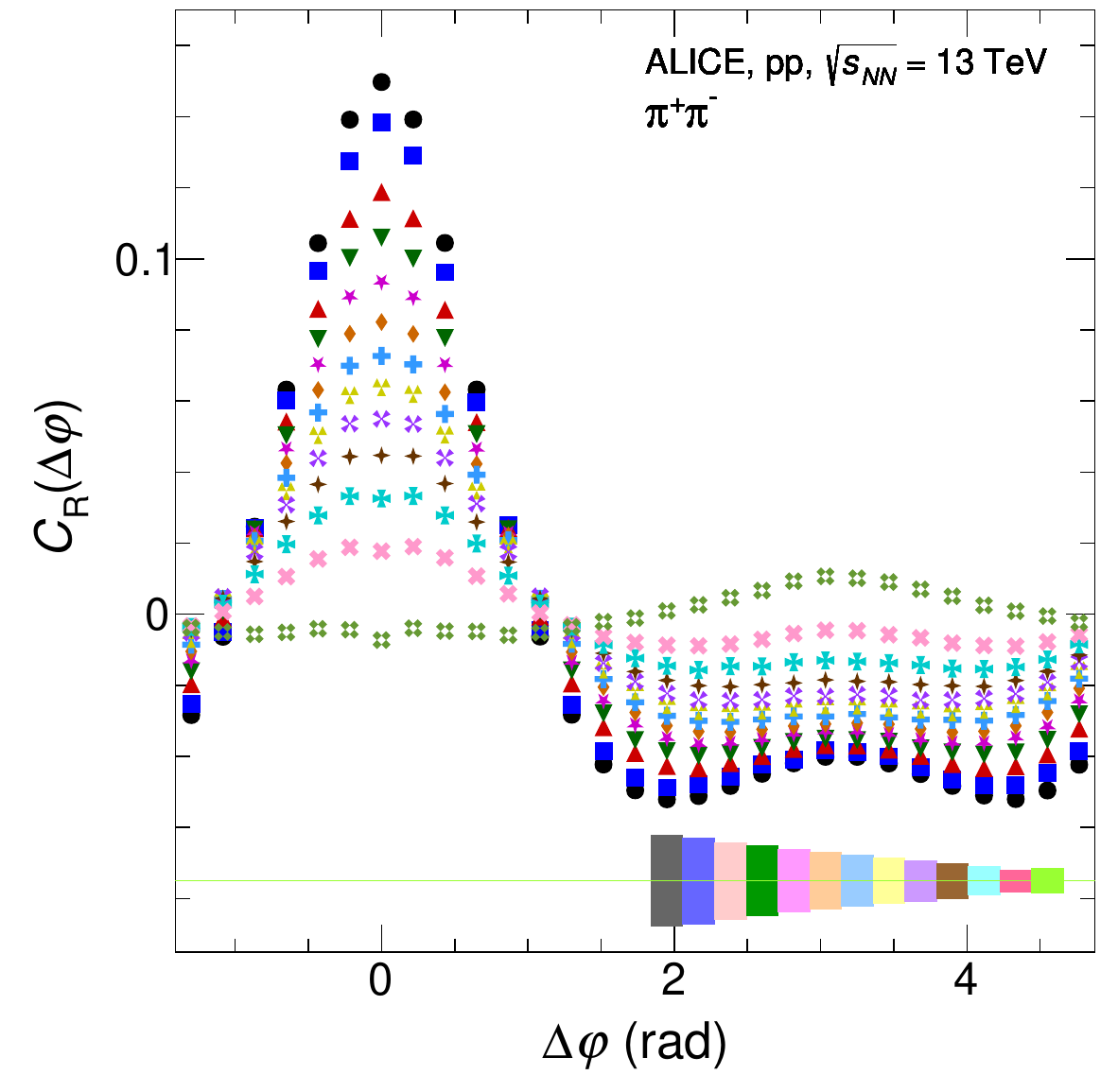}
    \includegraphics[width=7cm]{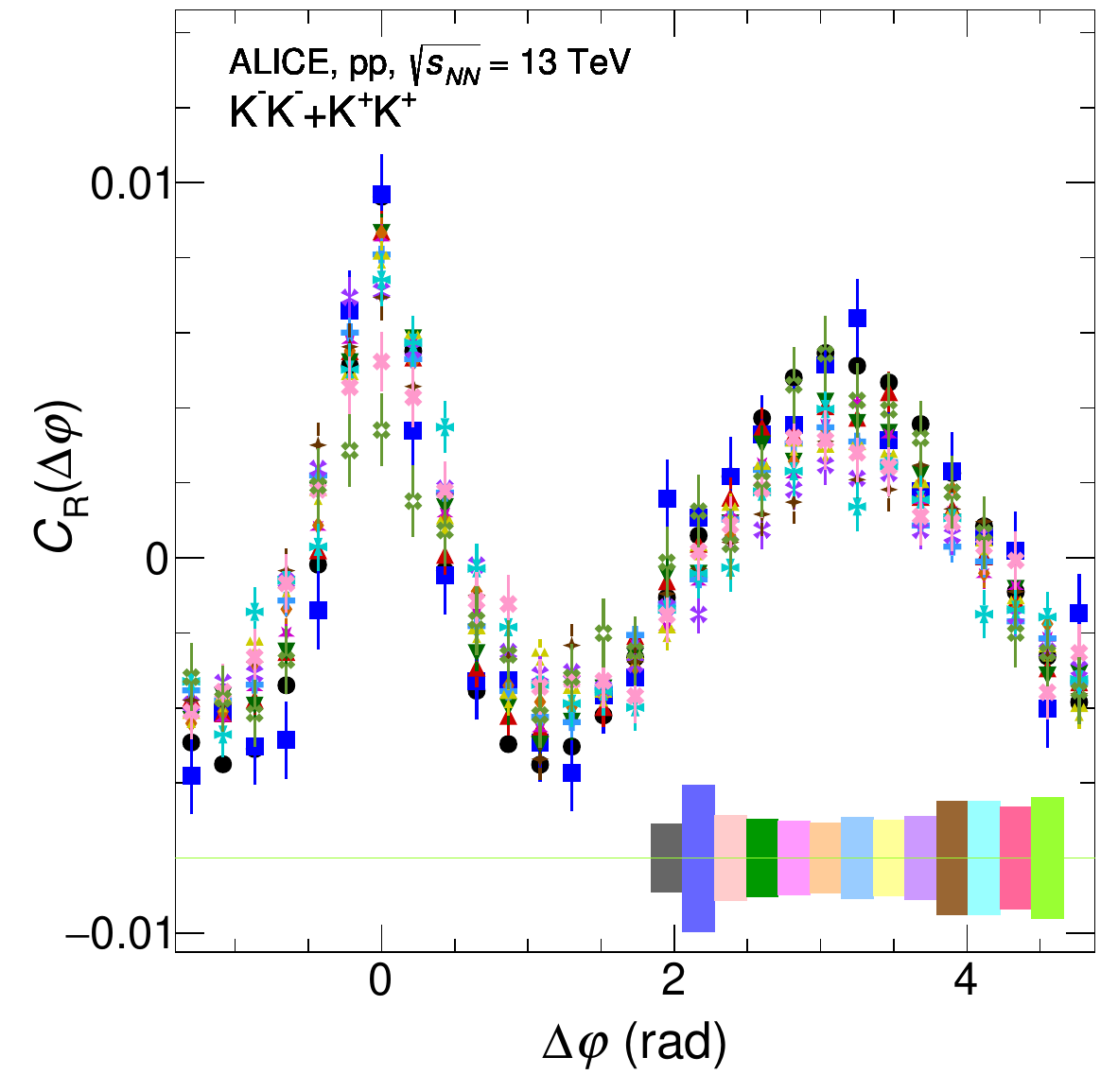}
    \hspace{1cm}
    \includegraphics[width=7cm]{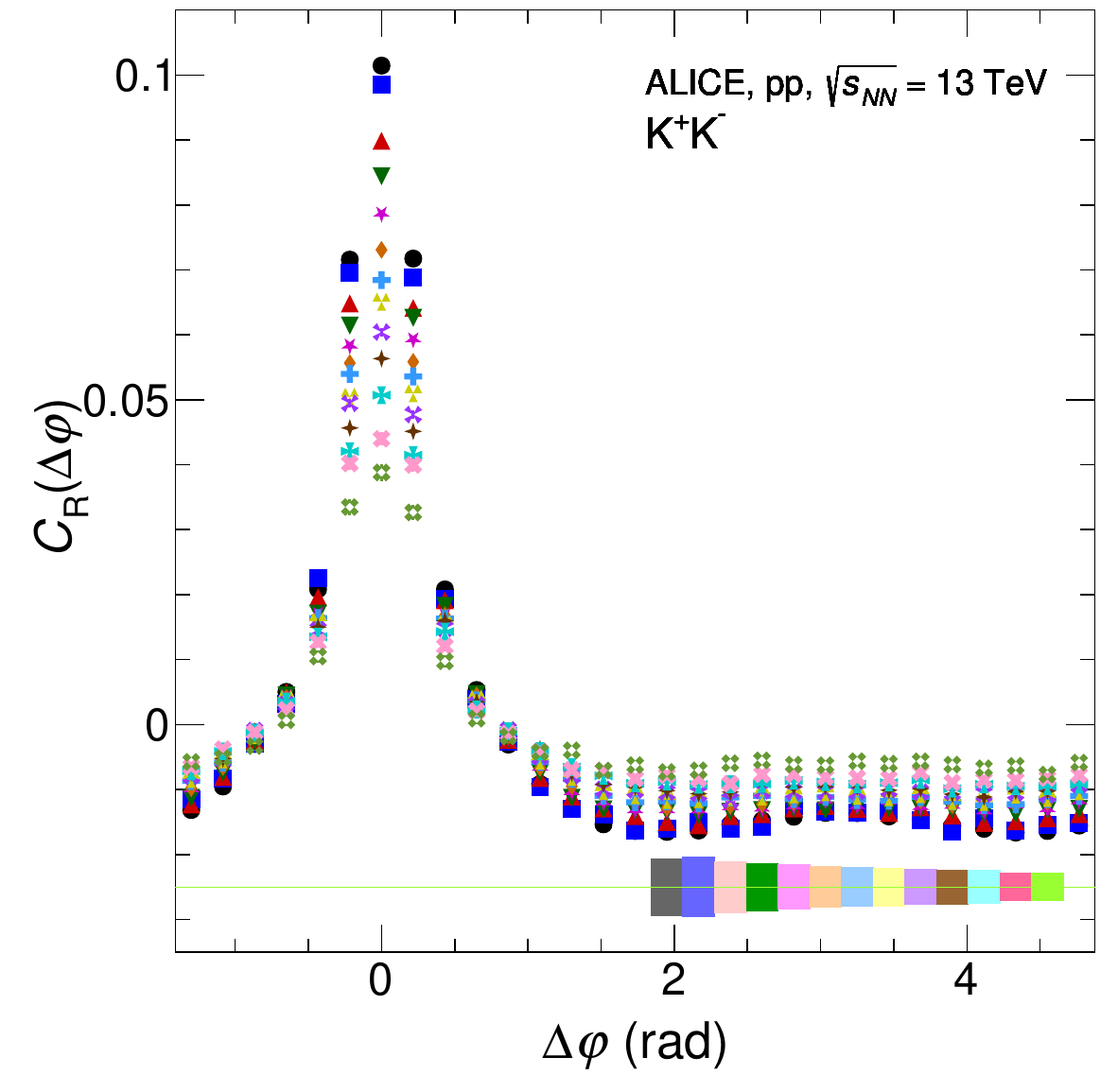}
    \includegraphics[width=7cm]{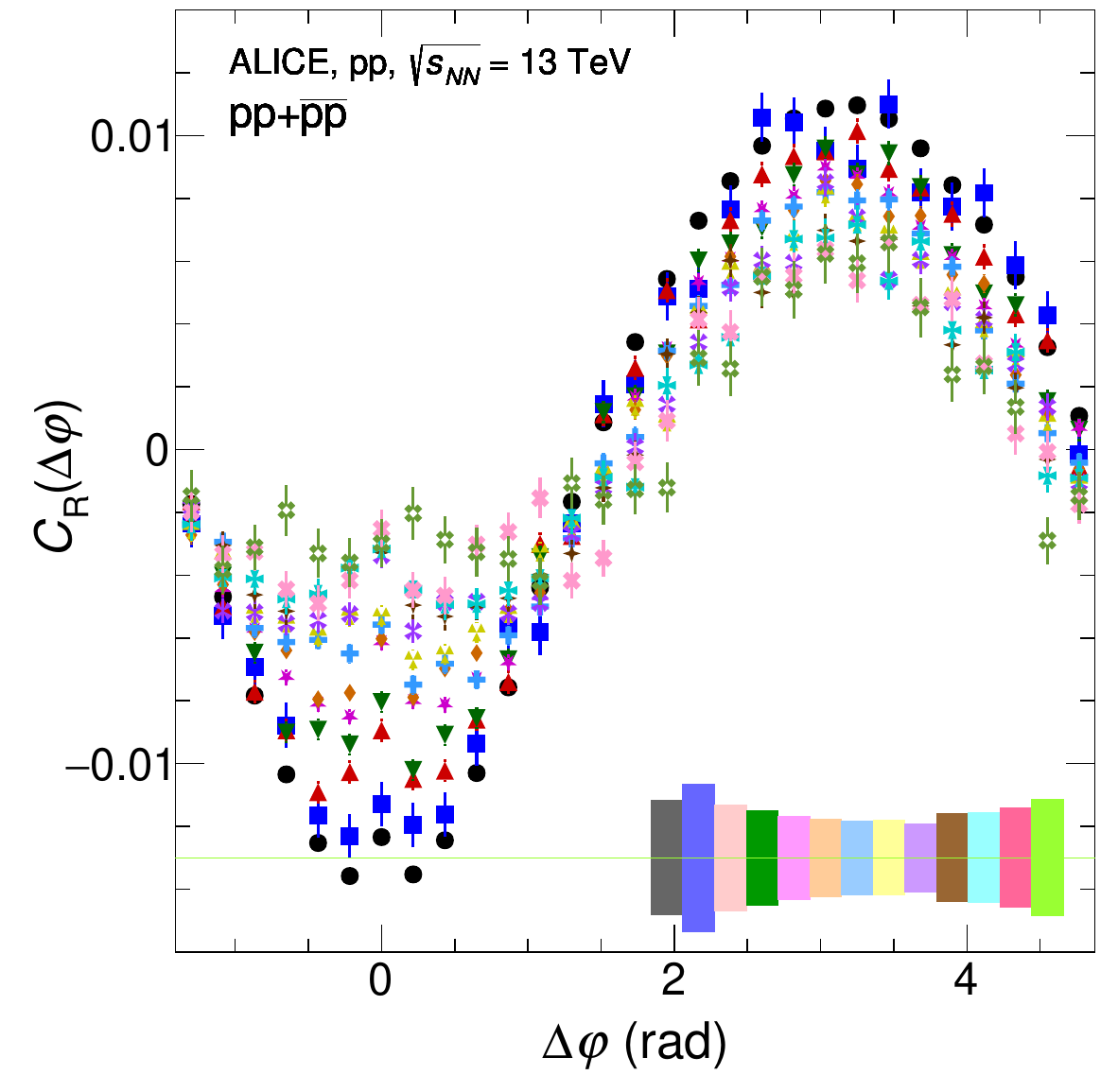}
    \hspace{1cm}
    \includegraphics[width=7cm]{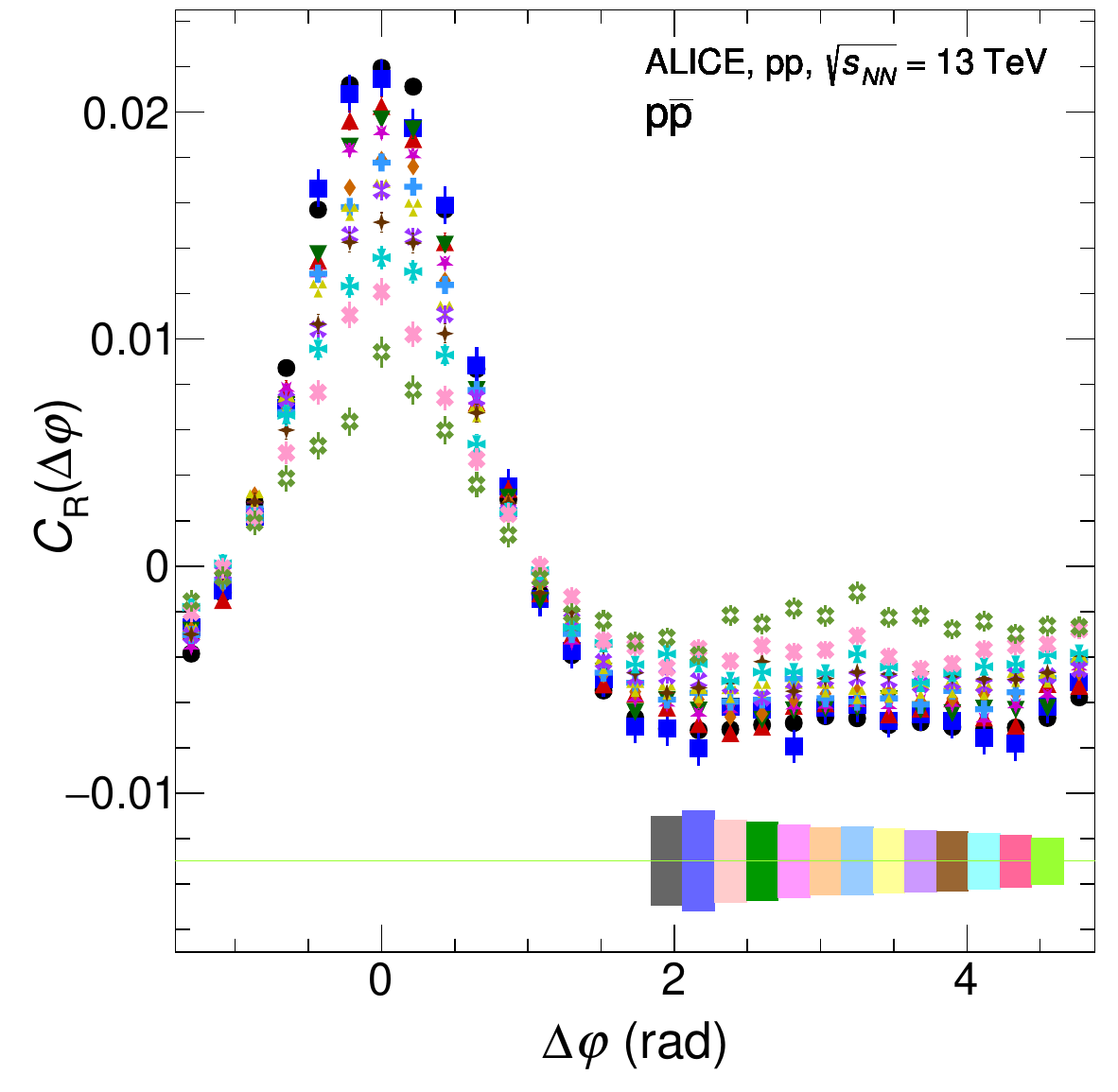}
    \caption{$\Delta\varphi$ projections of correlation functions within $\abs{\Delta y} < 1$ for like- and unlike-sign pions, kaons, and protons in multiplicity classes, using the $C_{\rm{R}}$ definition. Statistical uncertainties (bars) and systematic uncertainties (boxes) are both drawn with the colour associated with the corresponding multiplicity class.}
    \label{fig:projection_rescaledtwopartcumulant}
\end{figure}

In \Cref{fig:projection_probabilityratio,fig:projection_rescaledtwopartcumulant}, $\Delta\varphi$ projections integrated over the full $\abs{\Delta y} < 1$ acceptance range of like- and unlike-sign correlation functions for the $C_{\rm{P}}$ and the $C_{\rm{R}}$ definitions, respectively, are shown. 
The plots include both statistical and systematic uncertainties. Statistical uncertainties are represented by error bars directly on the correlation functions, while systematic uncertainties are depicted as colored boxes at the bottom of each plot. Each color corresponds to a specific correlation function for a given multiplicity class, with the same color scheme consistently applied across all classes. To improve readability, the systematic uncertainties are displayed as the absolute value of the total systematic uncertainty averaged over all $\Delta\varphi$ bins. By using the $C_{\rm{P}}$, the systematic uncertainties increase for lower multiplicity classes, while in the case of $C_{\rm{R}}$, similar values of the systematics are obtained across all multiplicity classes.

This analysis, initially focused on anticorrelations, has also revealed other distinct correlation patterns across different particle species and multiplicity classes. As shown in~\Cref{fig:projection_probabilityratio,fig:projection_rescaledtwopartcumulant}, the unlike-sign pairs show a small correlation in the away-side region as in the case of pions, while almost absent in the case of unlike-sign kaons and protons. In contrast, like-sign correlations vary significantly: kaons display a pronounced back-to-back jet structure that is comparable in magnitude to the near-side peak, whereas protons show no near-side mini jet signatures but feature a dominant away-side jet structure.

Another important observation concerns how the correlation functions evolve with multiplicity. As shown in Fig.~\ref{fig:projection_probabilityratio}, by using the $C_{\rm{P}}$ definition, the correlation strength increases at lower multiplicities. 
In the case of unlike-sign pion pairs, a non-monotonic behavior in the correlation function is observed, primarily attributed to the decay of $\rho^0$ resonances. Additionally, a significant contribution is expected from diffractive processes in low-multiplicity events, where a flat near-side correlation is seen. These events typically fragment in the forward direction, and their decay products can acquire sufficient transverse momentum to populate opposite azimuthal regions. Consequently, the near-side peak in the correlation function is suppressed, while the away-side structure becomes significantly enhanced.
This effect is particularly pronounced for pions due to their low mass and the possibility of events containing only one or two pions, where the two particles are more likely to appear in opposite hemispheres. In contrast, the production of heavier particles such as kaons and protons follows different mechanisms. Unlike-sign kaons, for instance, are predominantly produced in pairs due to strangeness conservation, often through intermediate resonances like the $\phi$ meson, which enhances the near-side peak in the correlation function. For unlike-sign protons, baryon-number conservation plays a crucial role, as they typically originate from baryon--antibaryon pair production. Additionally, the higher energy threshold required for kaon and proton production means that events containing these particles are more likely to produce additional hadrons, leading to a broader phase-space distribution and reducing the suppression of near-side correlations observed for pions.

After applying a rescaling factor to remove the trivial multiplicity dependence, as observed in Fig.~\ref{fig:projection_rescaledtwopartcumulant}, the $C_{\rm{R}}$ increases with multiplicity. Notably, once the 1/N scaling is removed, unlike-sign pions exhibit a monotonic behavior with multiplicity. An interesting observation is that unlike-sign kaons show no multiplicity dependence after rescaling, suggesting a distinct behavior compared to other species.
\section{Comparison to MC models}
\begin{figure}[!h]
   \centering
 \subfigure[]{
   \includegraphics[width=5cm]{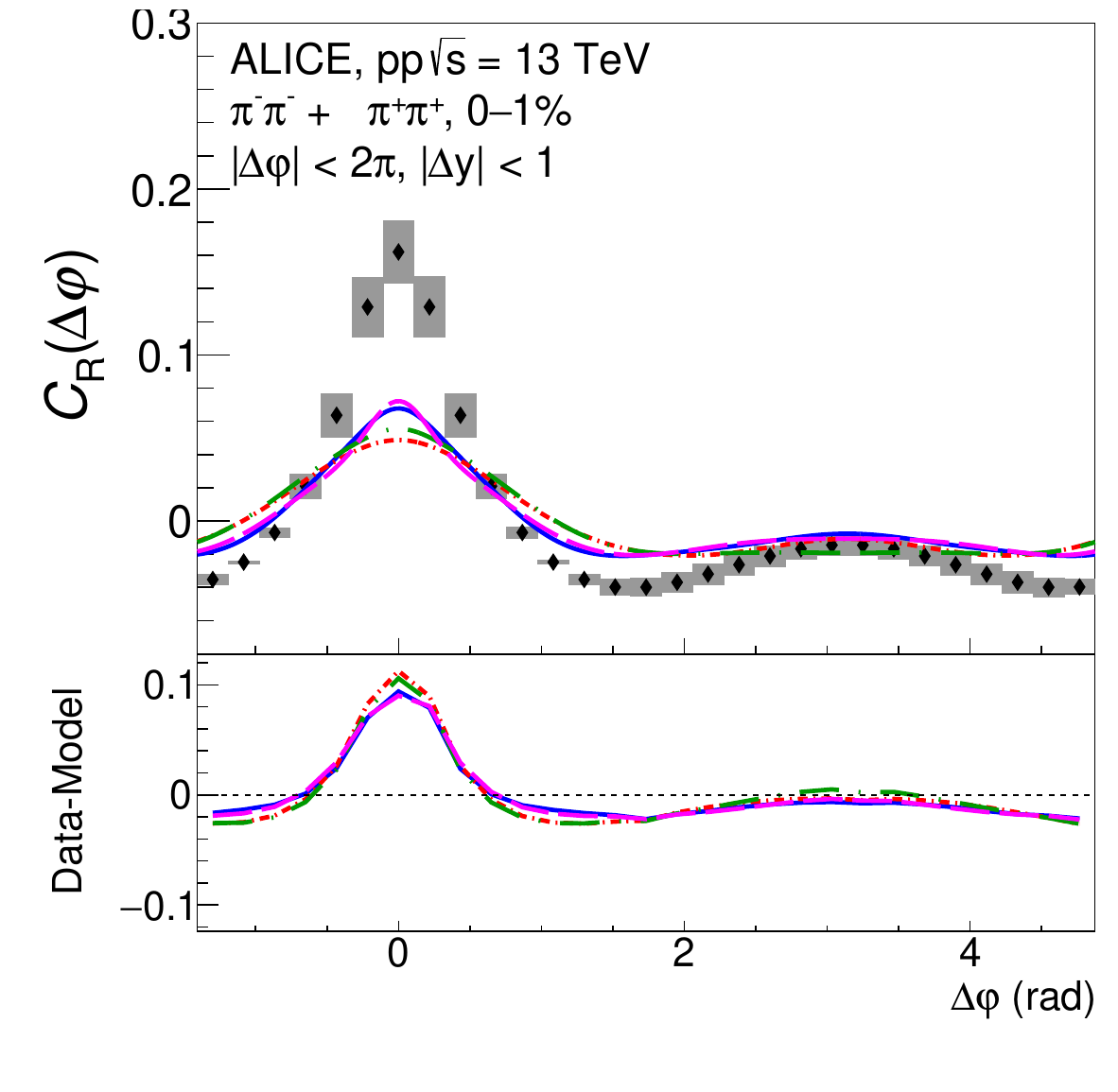} 
   \label{fig3-a}}
    \subfigure[]{
   \includegraphics[width=5cm]{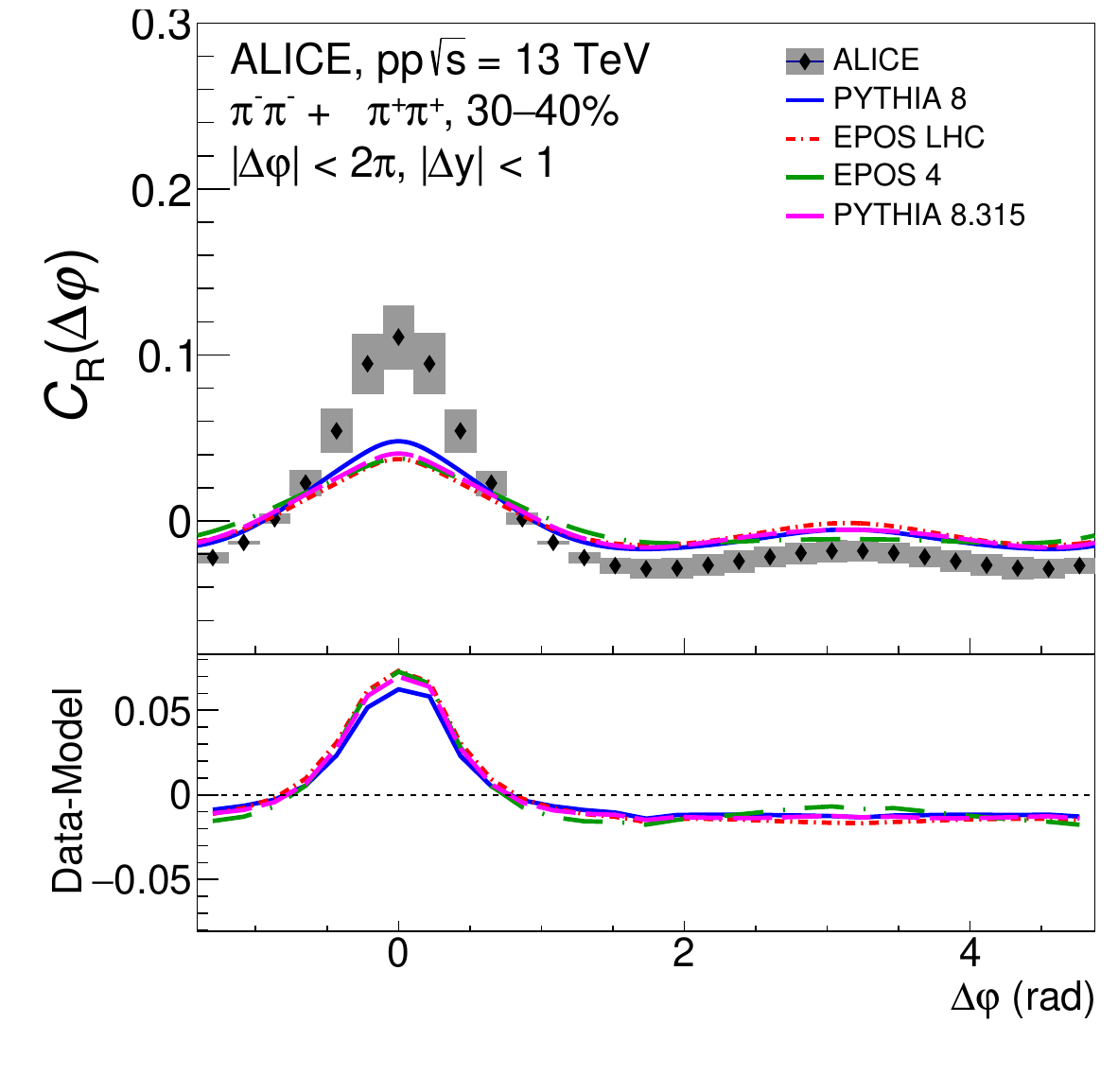}
   \label{fig3-b}}
 \subfigure[]{
   \includegraphics[width=5cm]{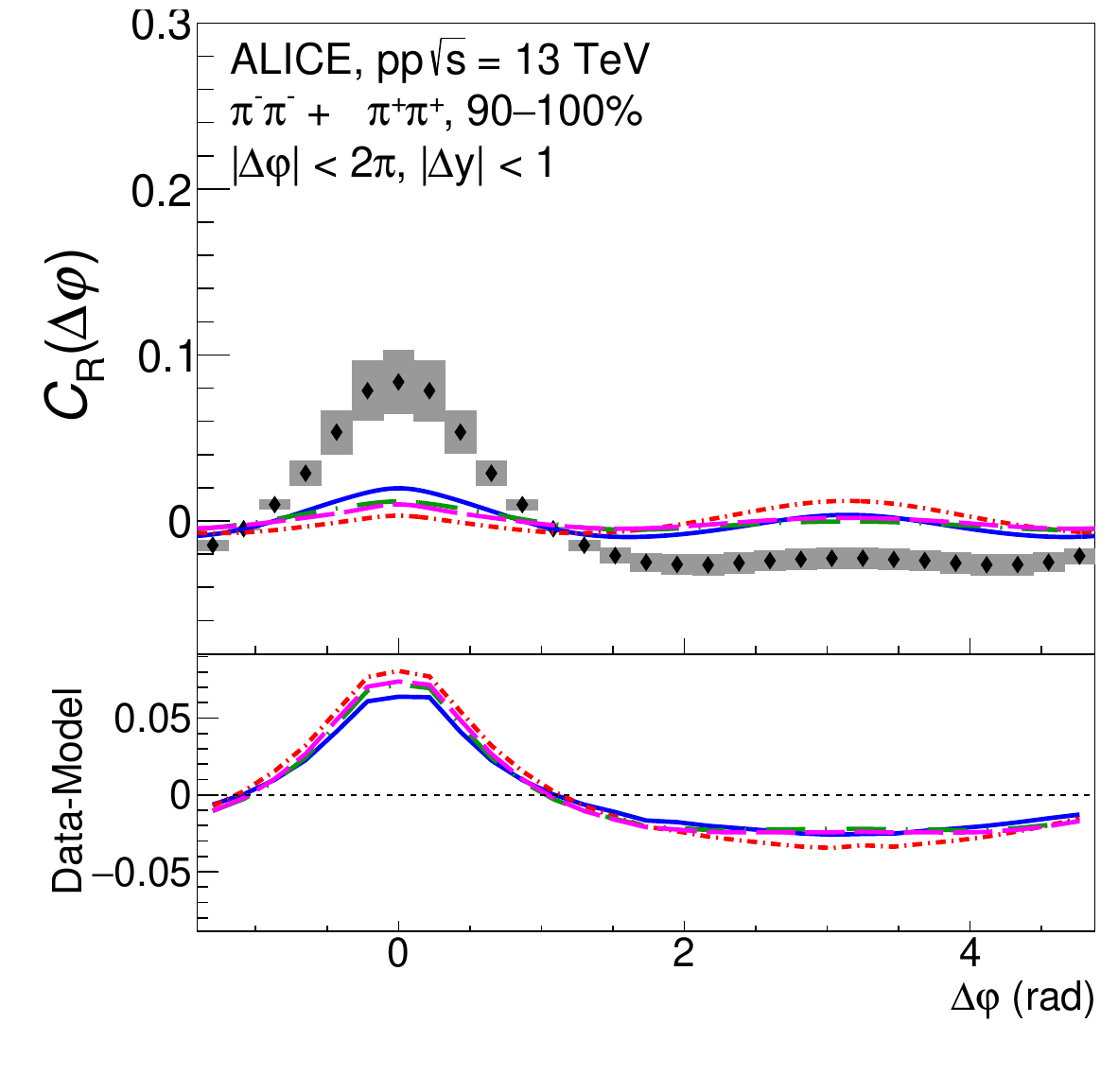}
   \label{fig3-c}}
\subfigure[]{
   \includegraphics[width=5cm]{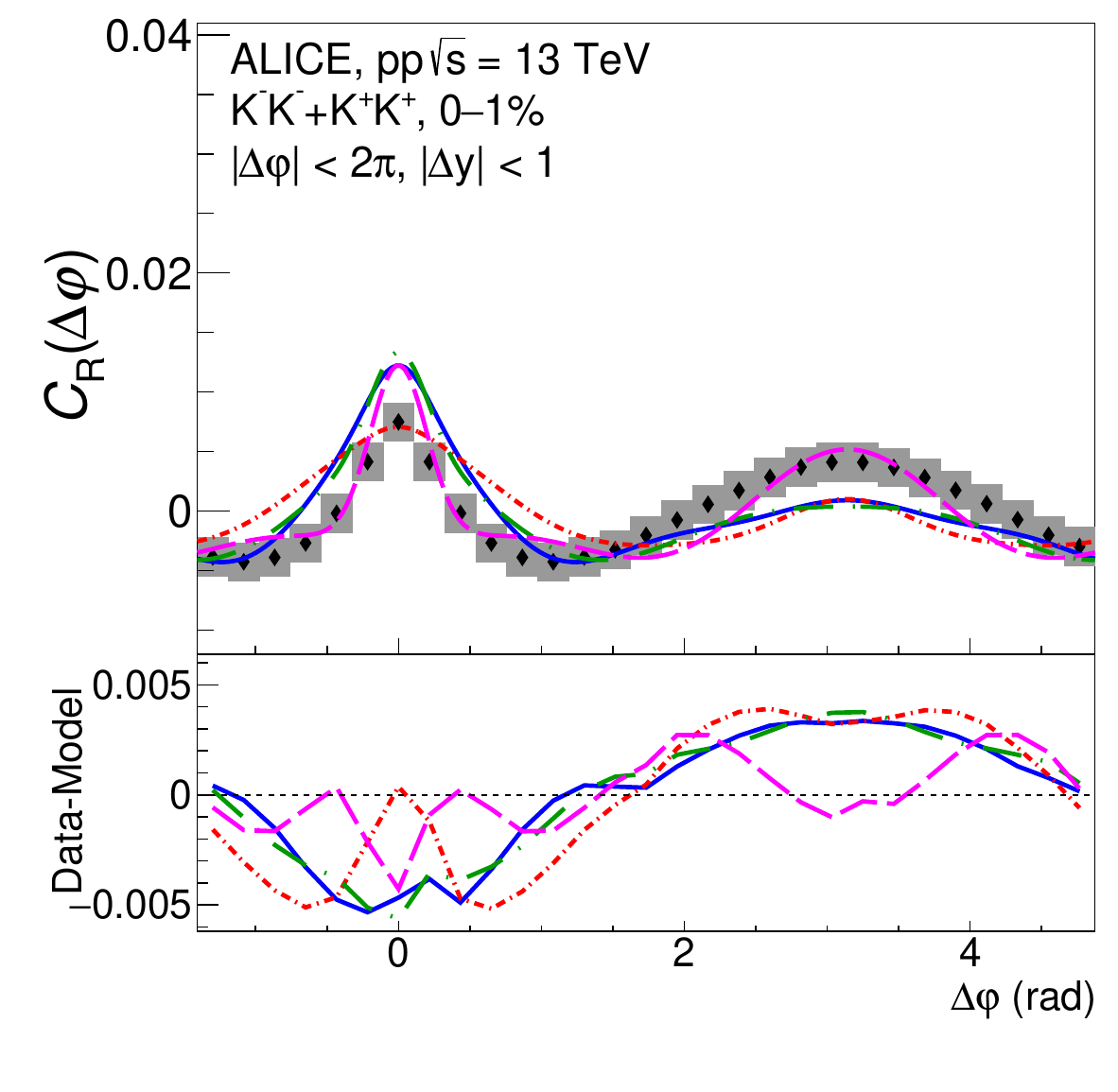}
   \label{fig3-d}}
   \subfigure[]{
   \includegraphics[width=5cm]{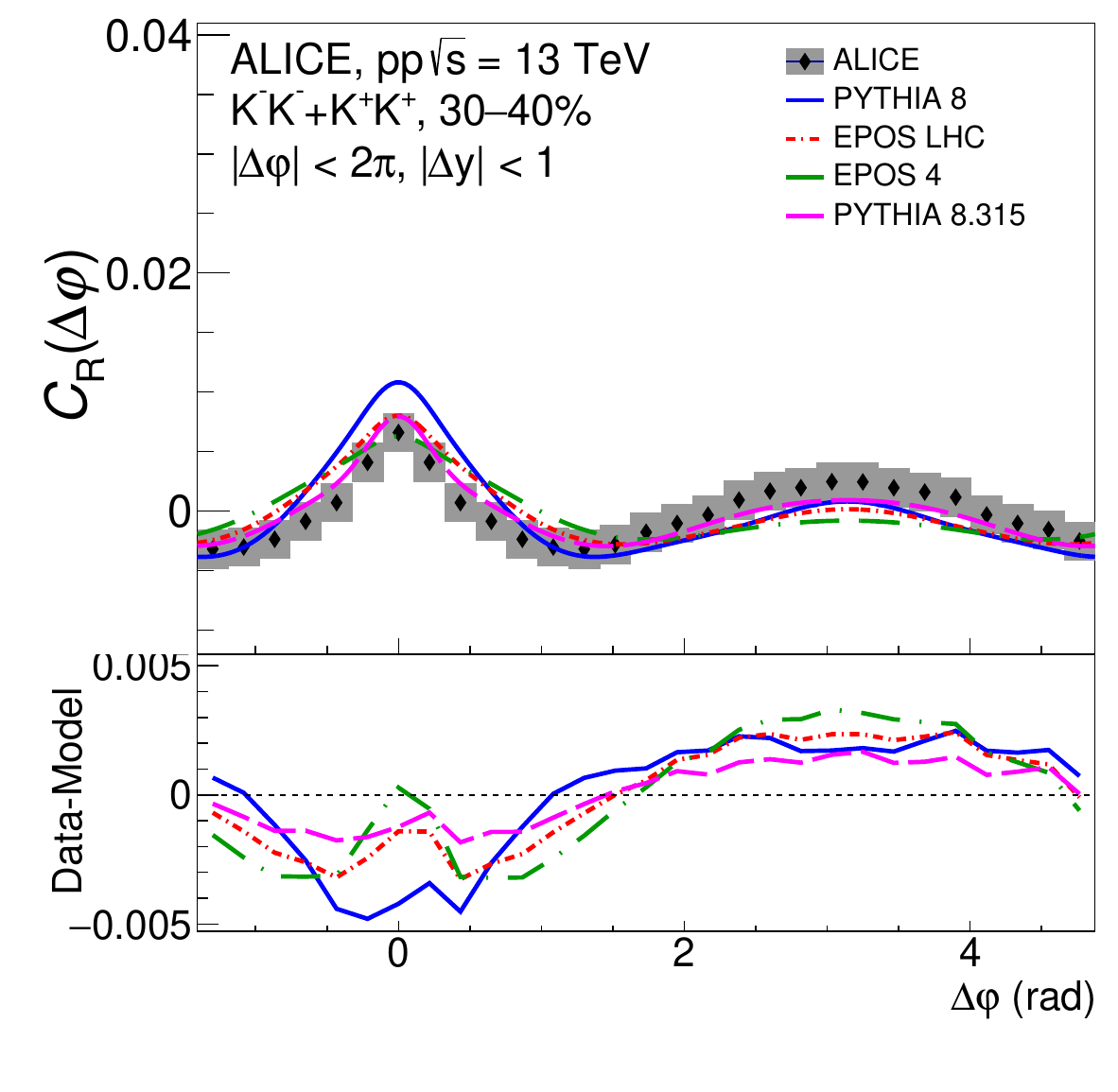}
   \label{fig3-e}}
 \subfigure[]{
   \includegraphics[width=5cm]{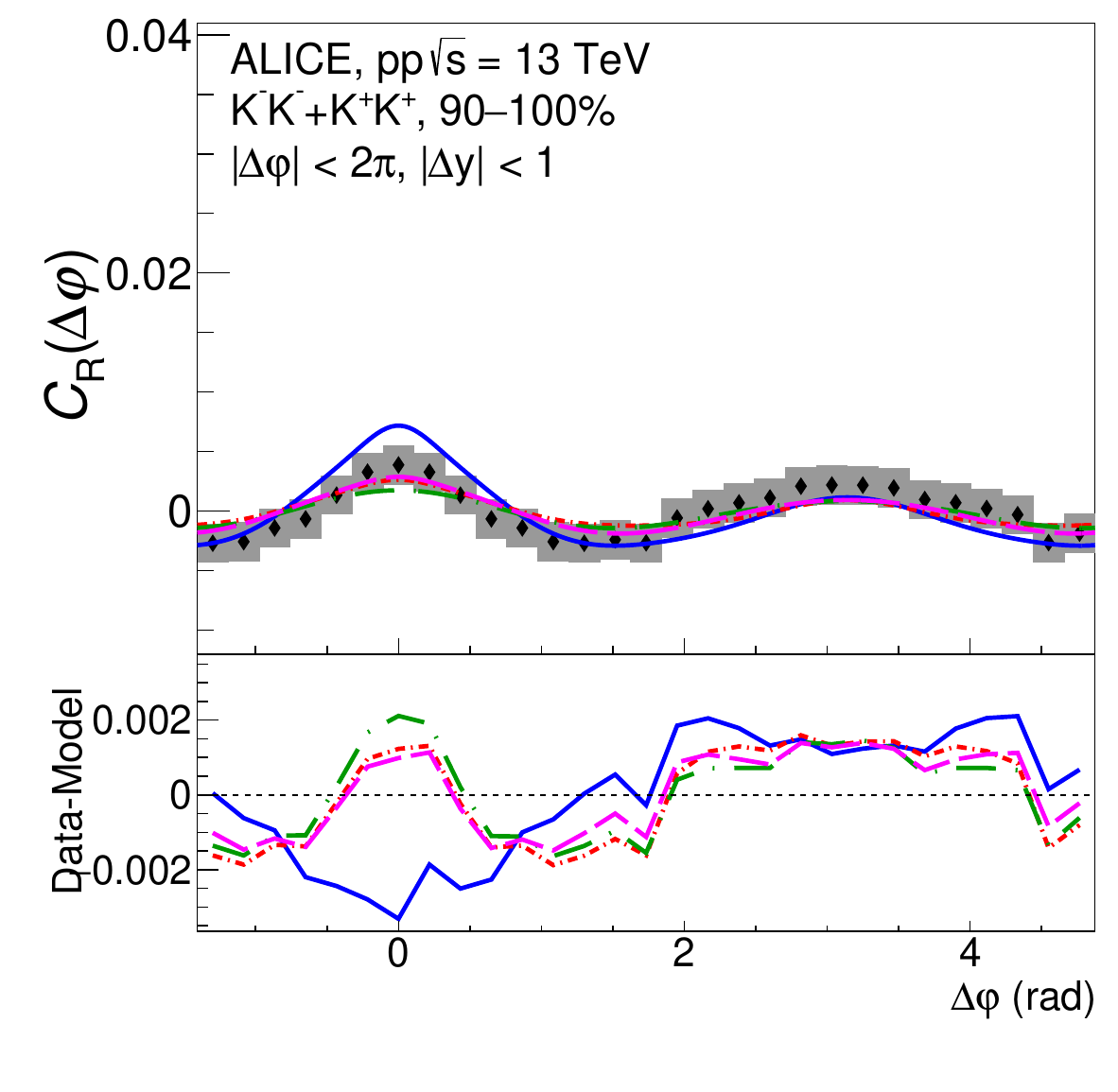}
   \label{fig3-f}}
\subfigure[]{
   \includegraphics[width=5cm]{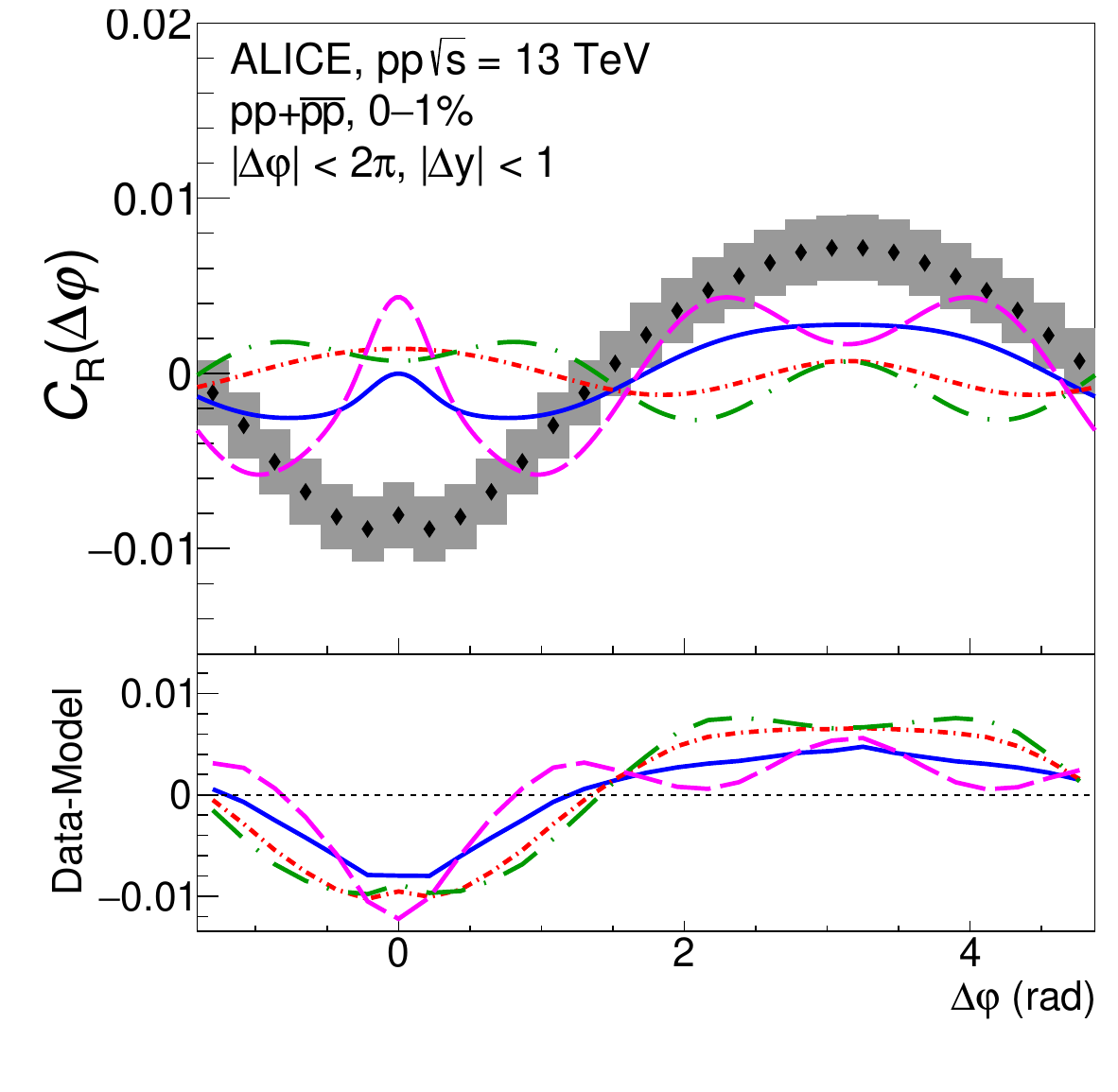}
   \label{fig3-g}}
    \subfigure[]{
   \includegraphics[width=5cm]{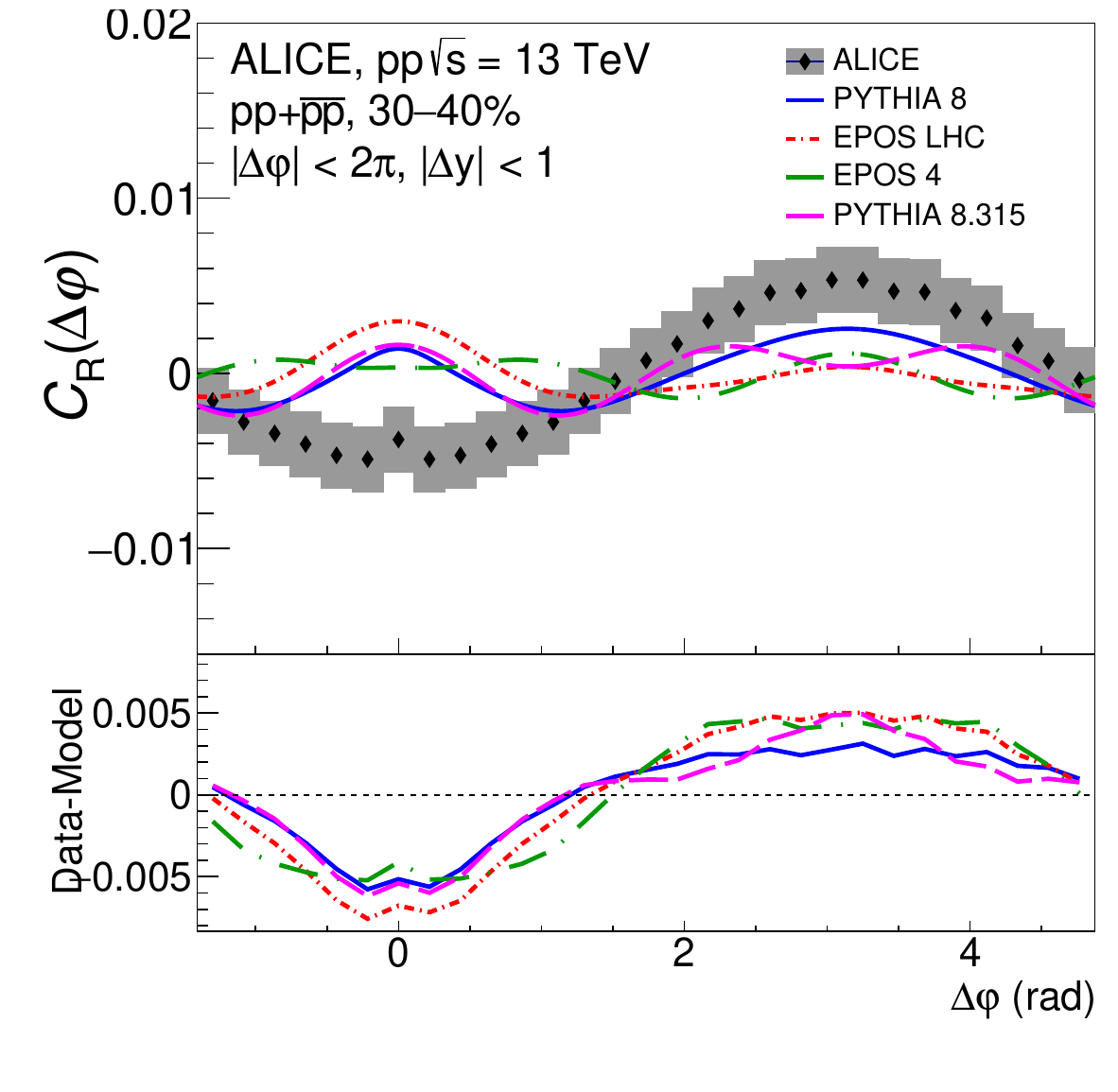}
   \label{fig3-h}}
 \subfigure[]{
   \includegraphics[width=5cm]{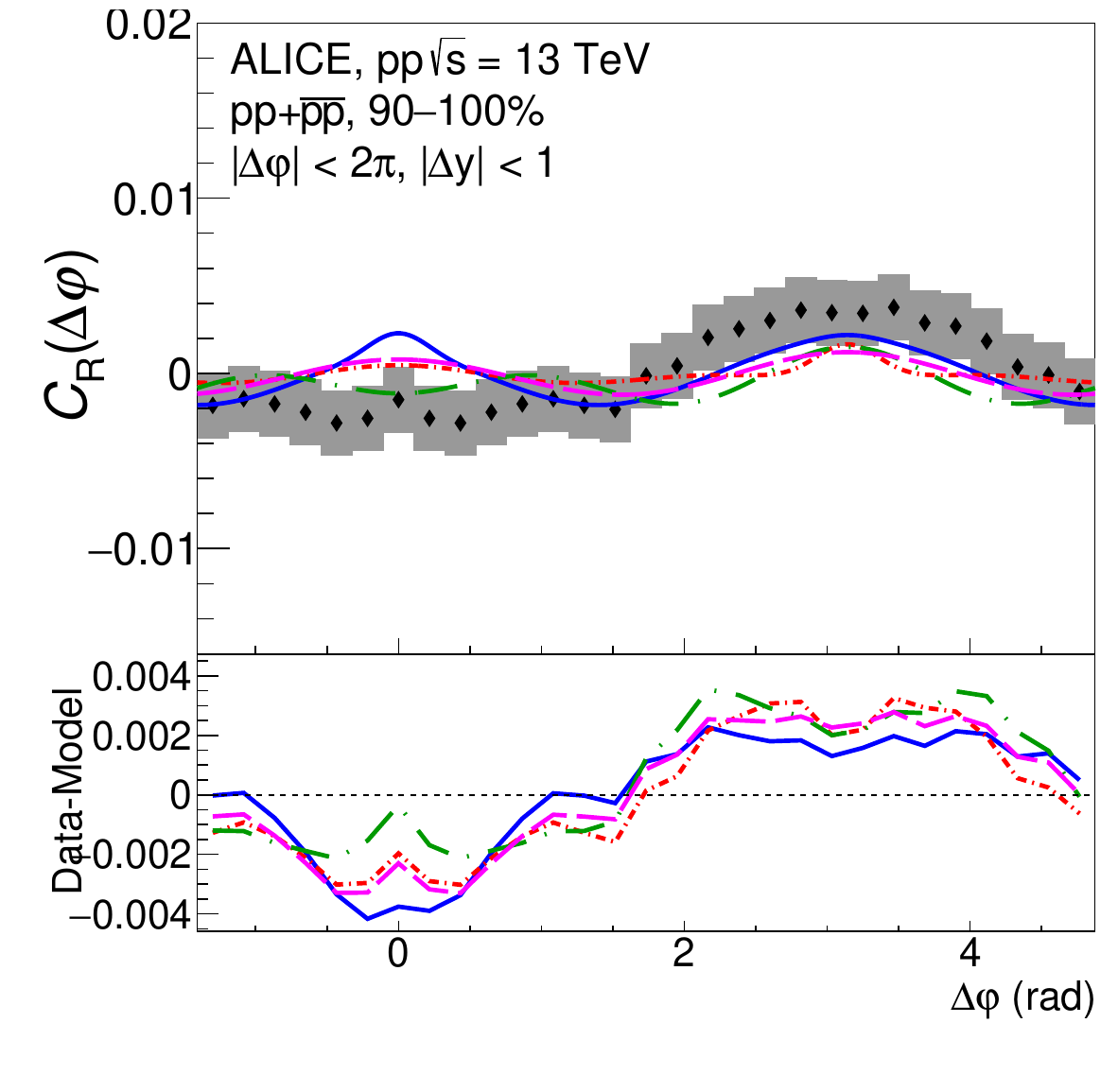}
   \label{fig3-i}}
      \caption{Comparison between $\Delta\varphi$-correlation functions and MC models (PYTHIA8 Monash 2013 in blue, EPOS LHC in red, EPOS4 in green, and PYTHIA 8.315 in magenta) in pp collision at $\sqrt{s} = 13$ TeV by using the $C_{\rm{R}}$ definition for like-sign pions, kaons, and protons pairs (top-to-bottom) for the highest (0--1\%), middle (30--40\%), and lowest (90--100\%) multiplicity classes (left-to-right).}
\label{comparisonMCmodel_like-sign}
\end{figure}

\begin{figure}[!h]
   \centering
 \subfigure[]{
   \includegraphics[width=5cm]{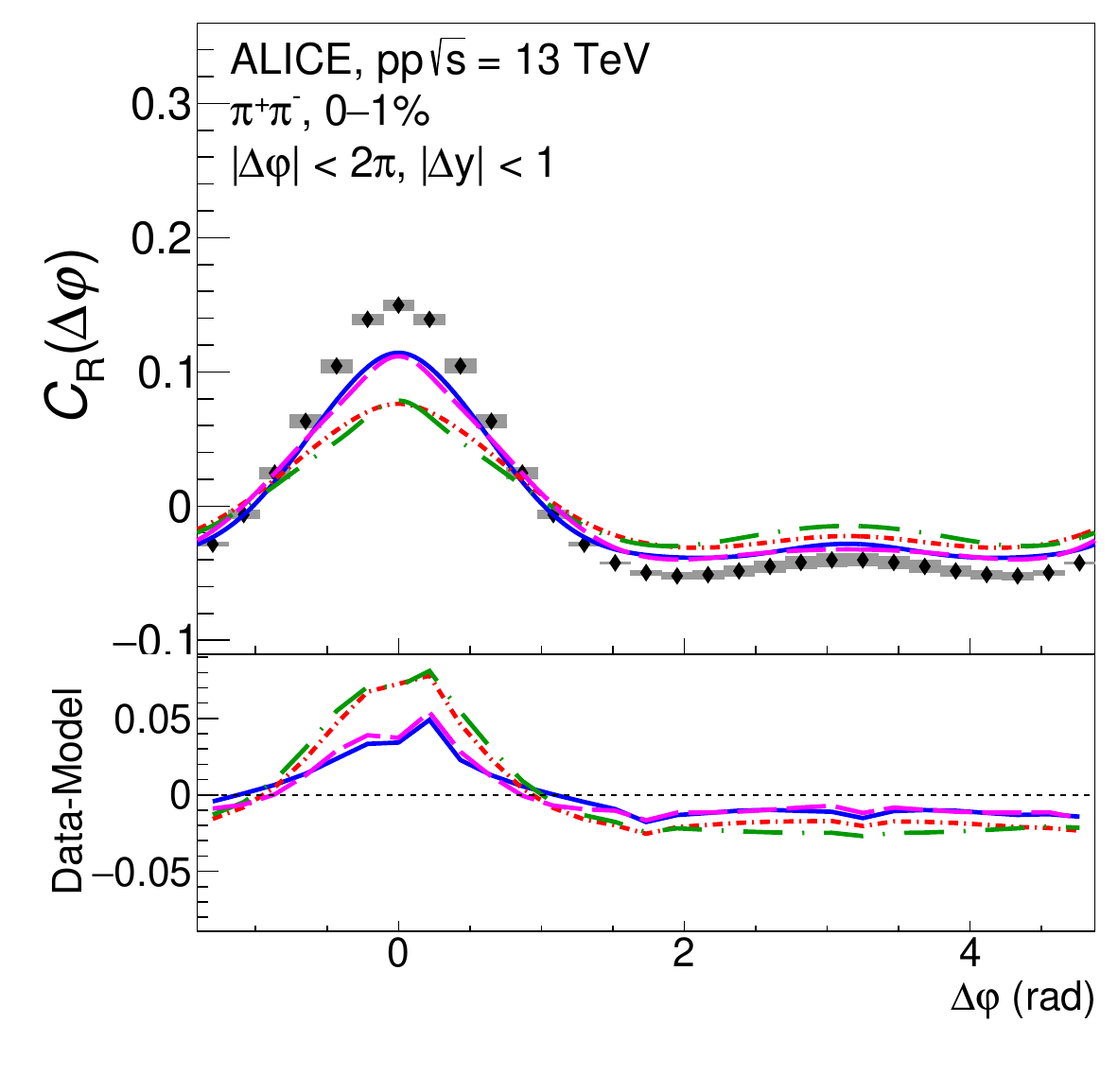} 
   \label{fig4-a}}
 \subfigure[]{
   \includegraphics[width=5cm]{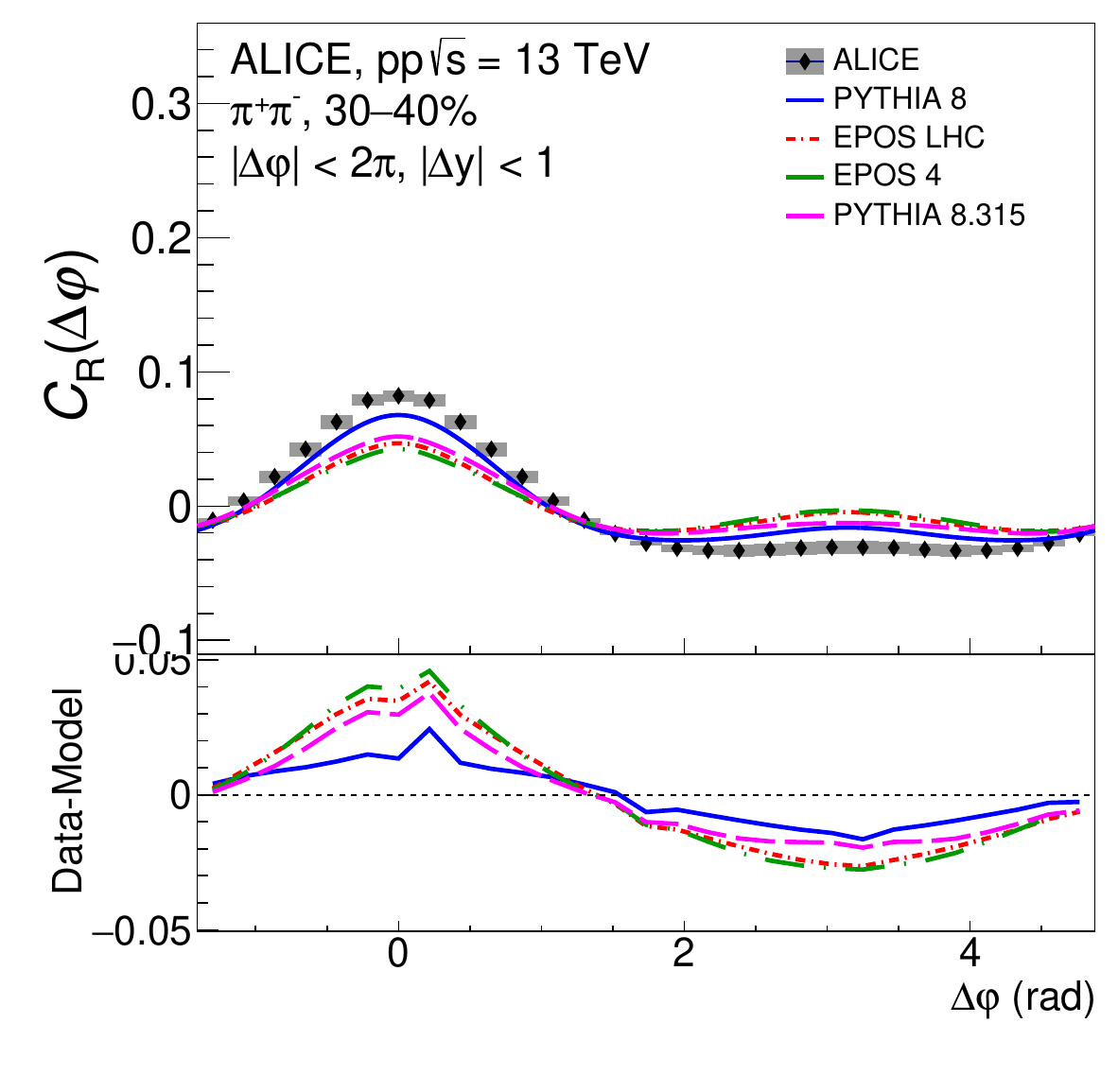} 
   \label{fig4-b}}
 \subfigure[]{
   \includegraphics[width=5cm]{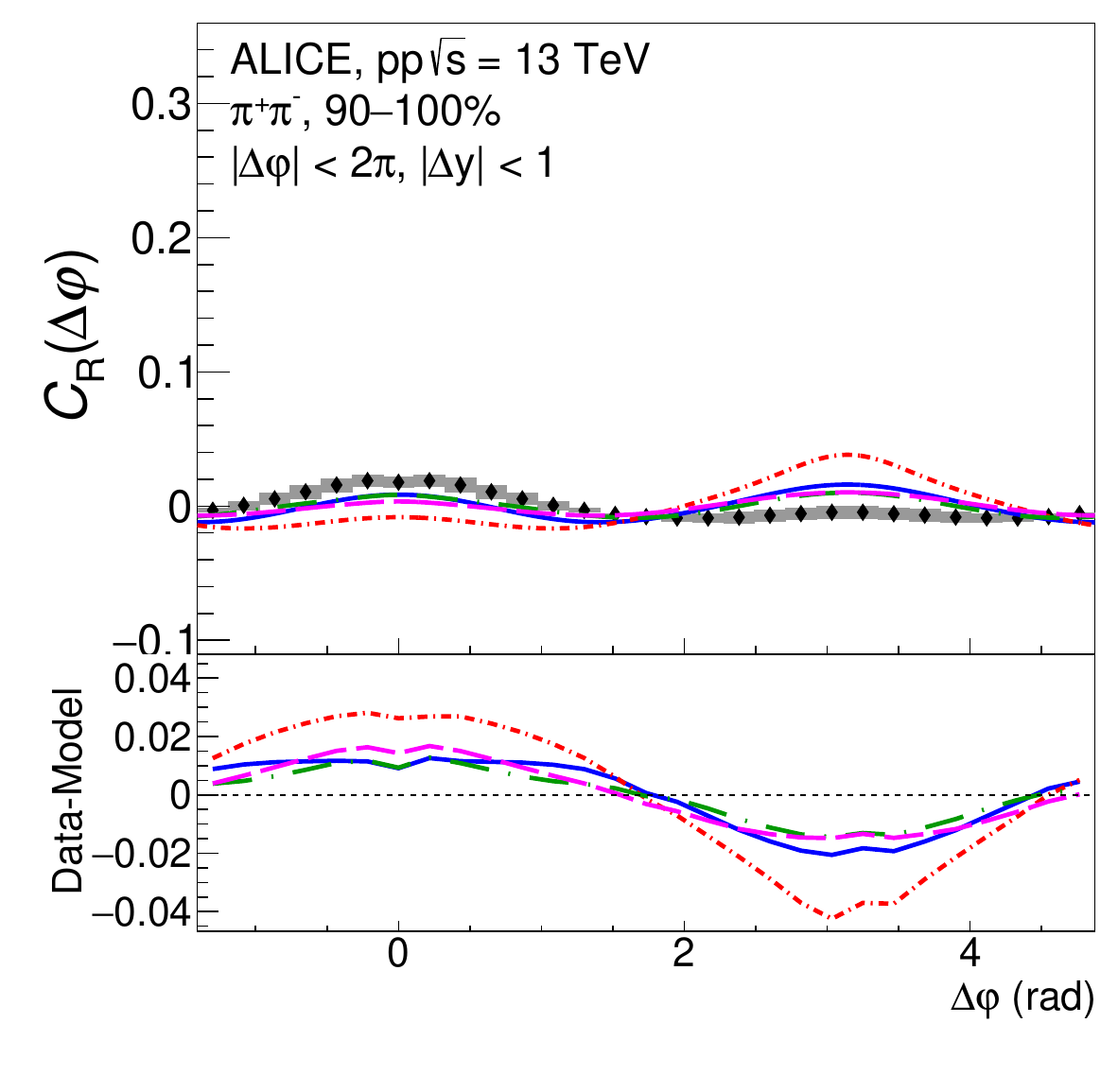}
   \label{fig4-c}}
\subfigure[]{
   \includegraphics[width=5cm]{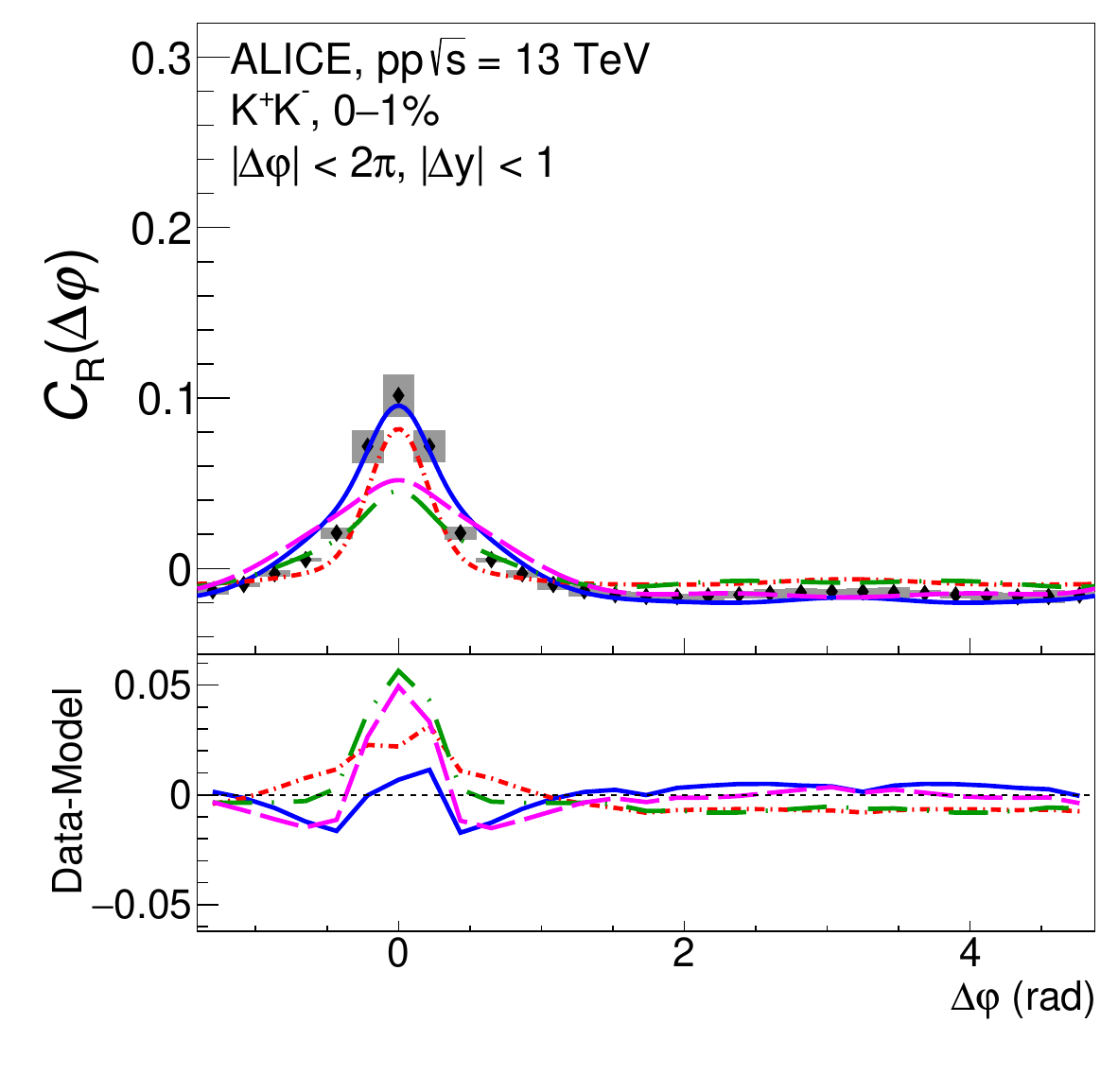}
   \label{fig4-d}}
   \subfigure[]{
   \includegraphics[width=5cm]{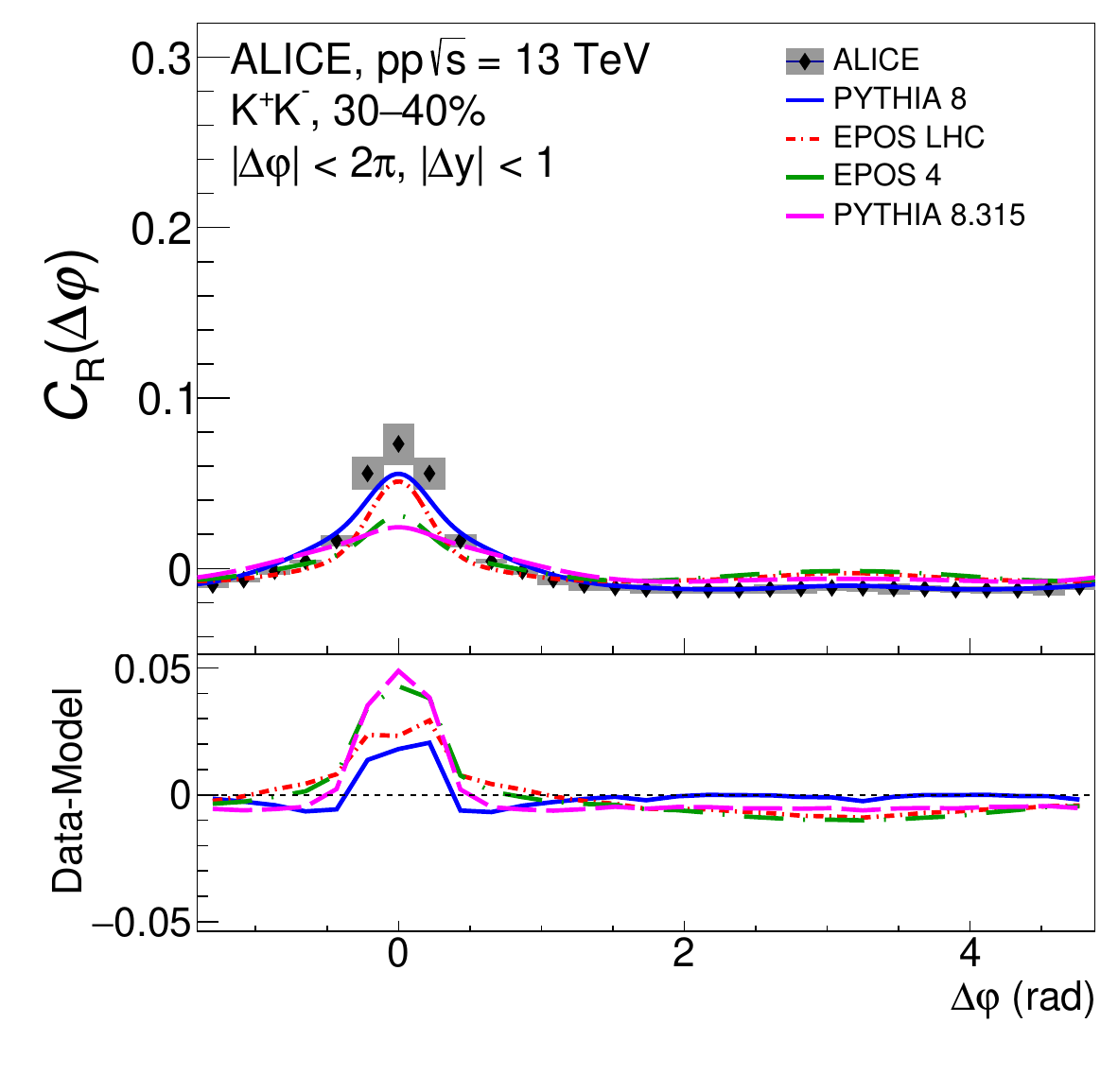}
   \label{fig4-e}}
 \subfigure[]{
   \includegraphics[width=5cm]{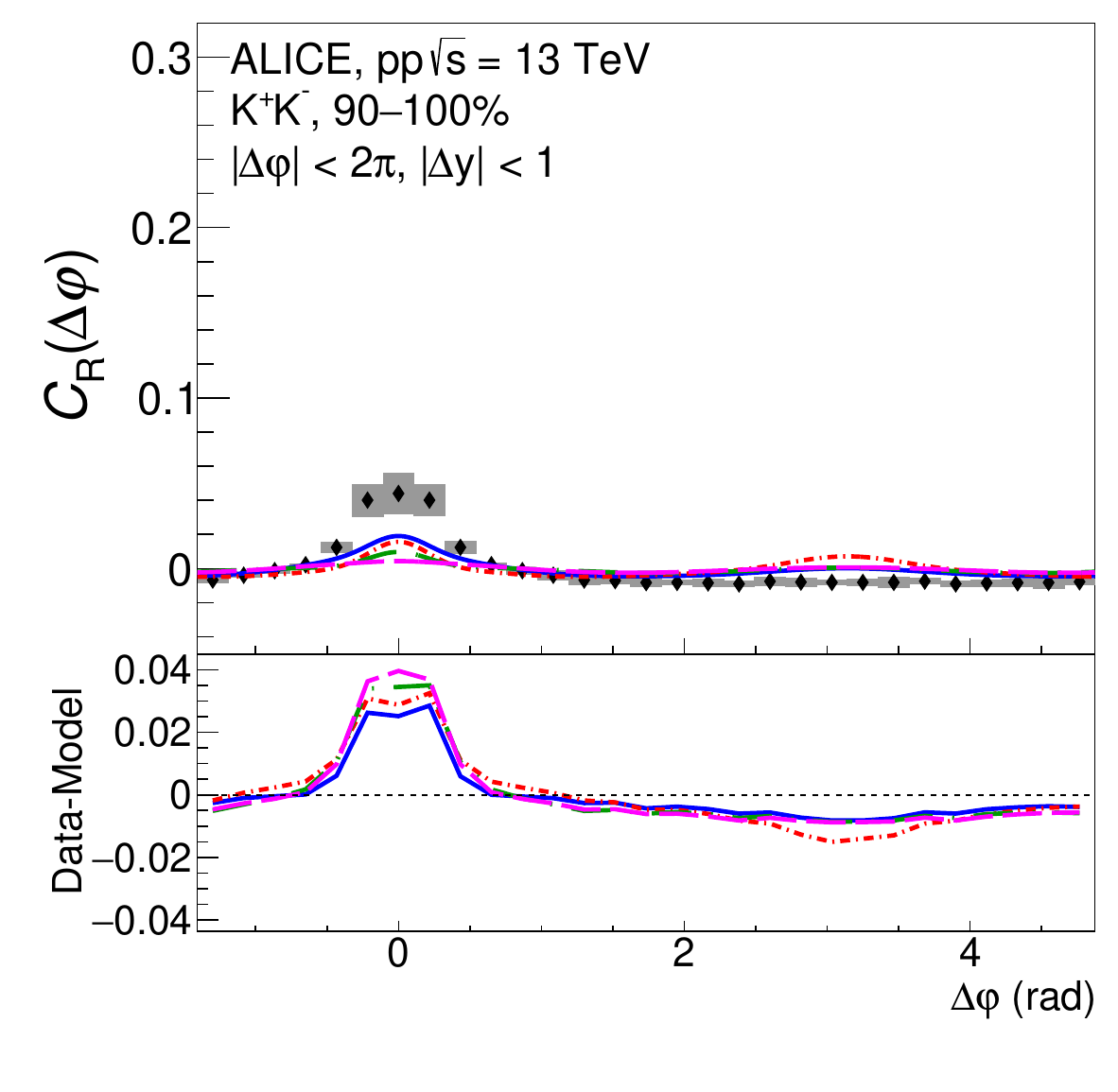}
   \label{fig4-f}}
\subfigure[]{
   \includegraphics[width=5cm]{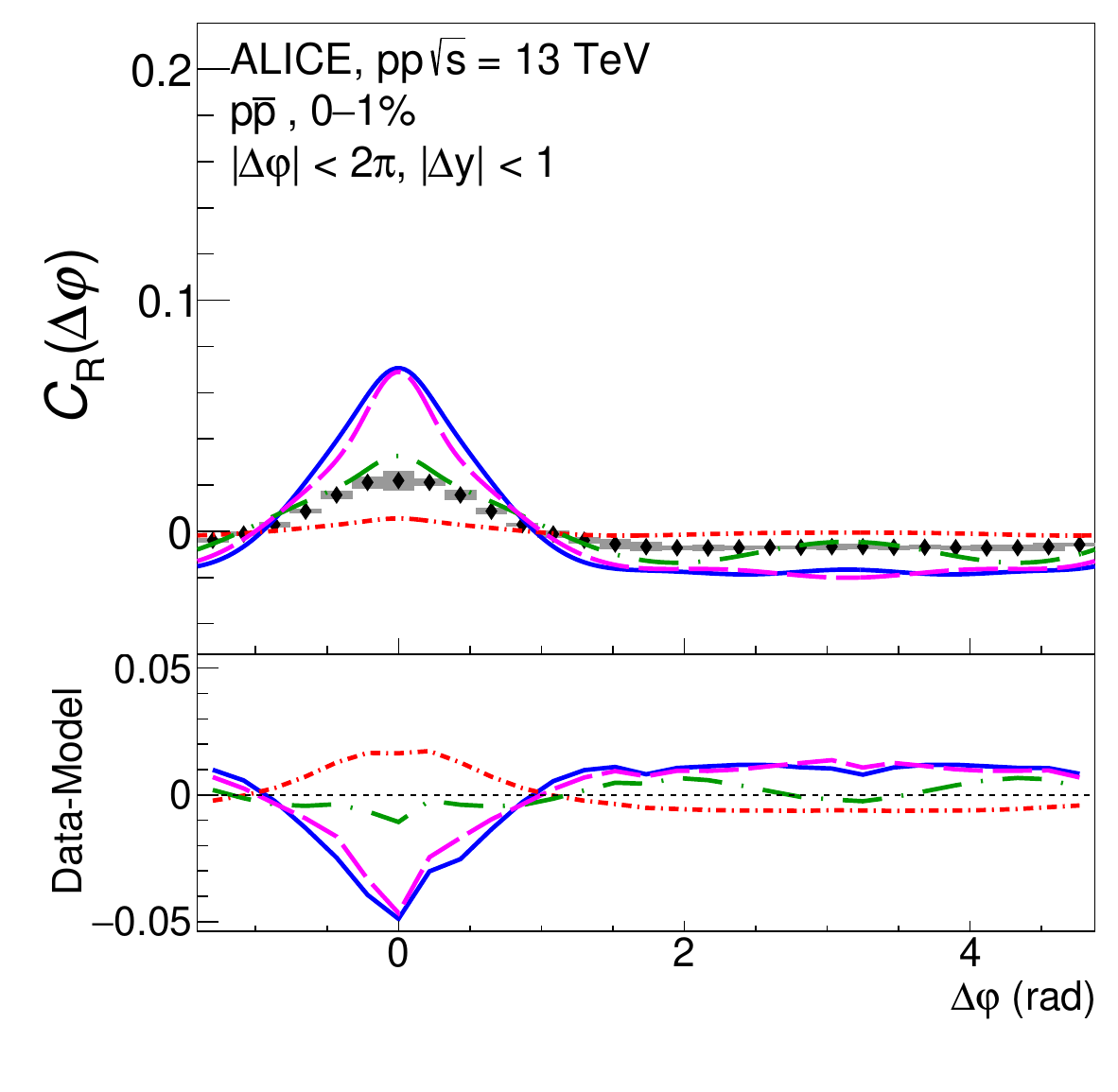}
   \label{fig4-g}}
   \subfigure[]{
   \includegraphics[width=5cm]{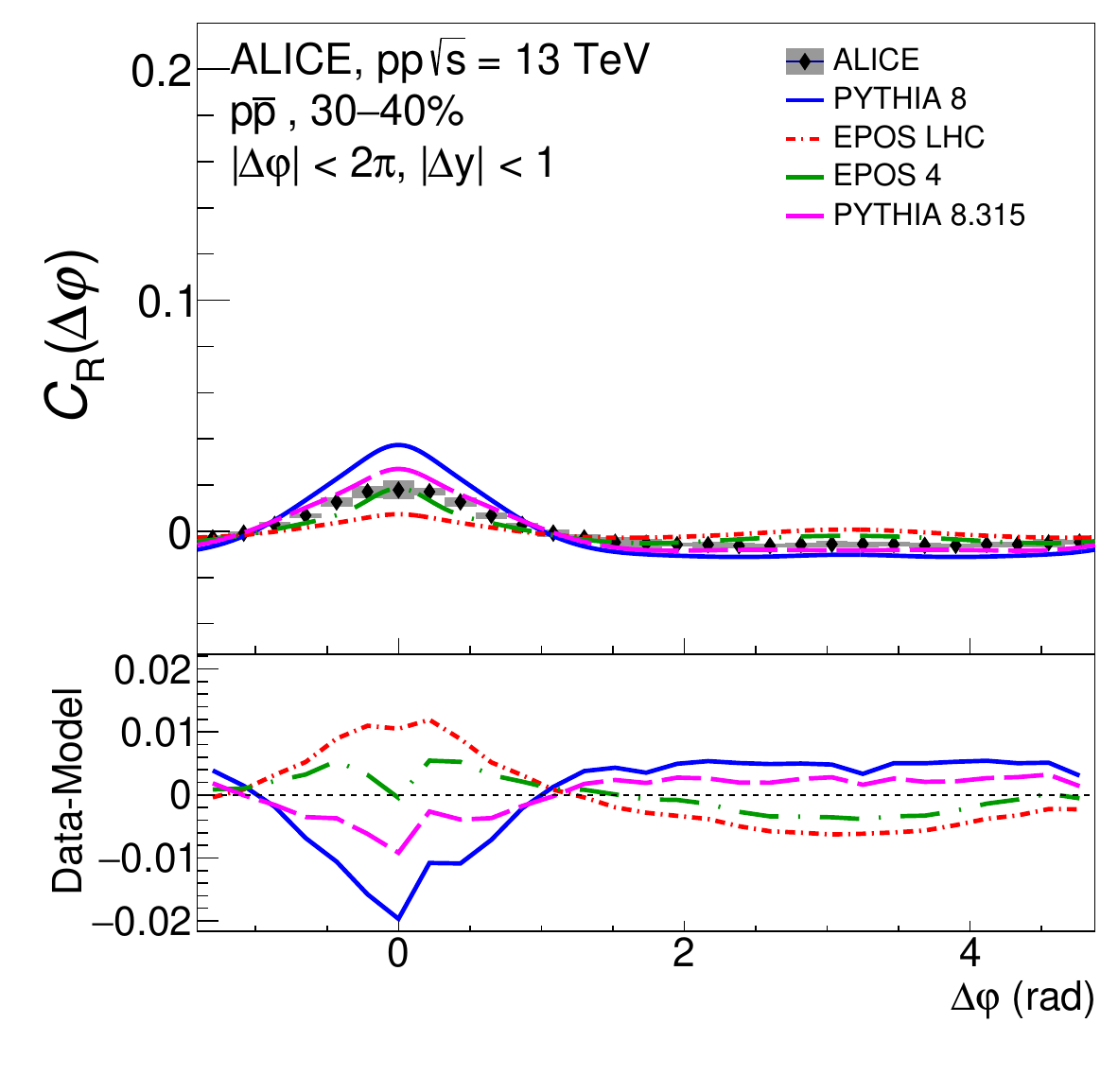}
   \label{fig4-h}}
 \subfigure[]{
   \includegraphics[width=5cm]{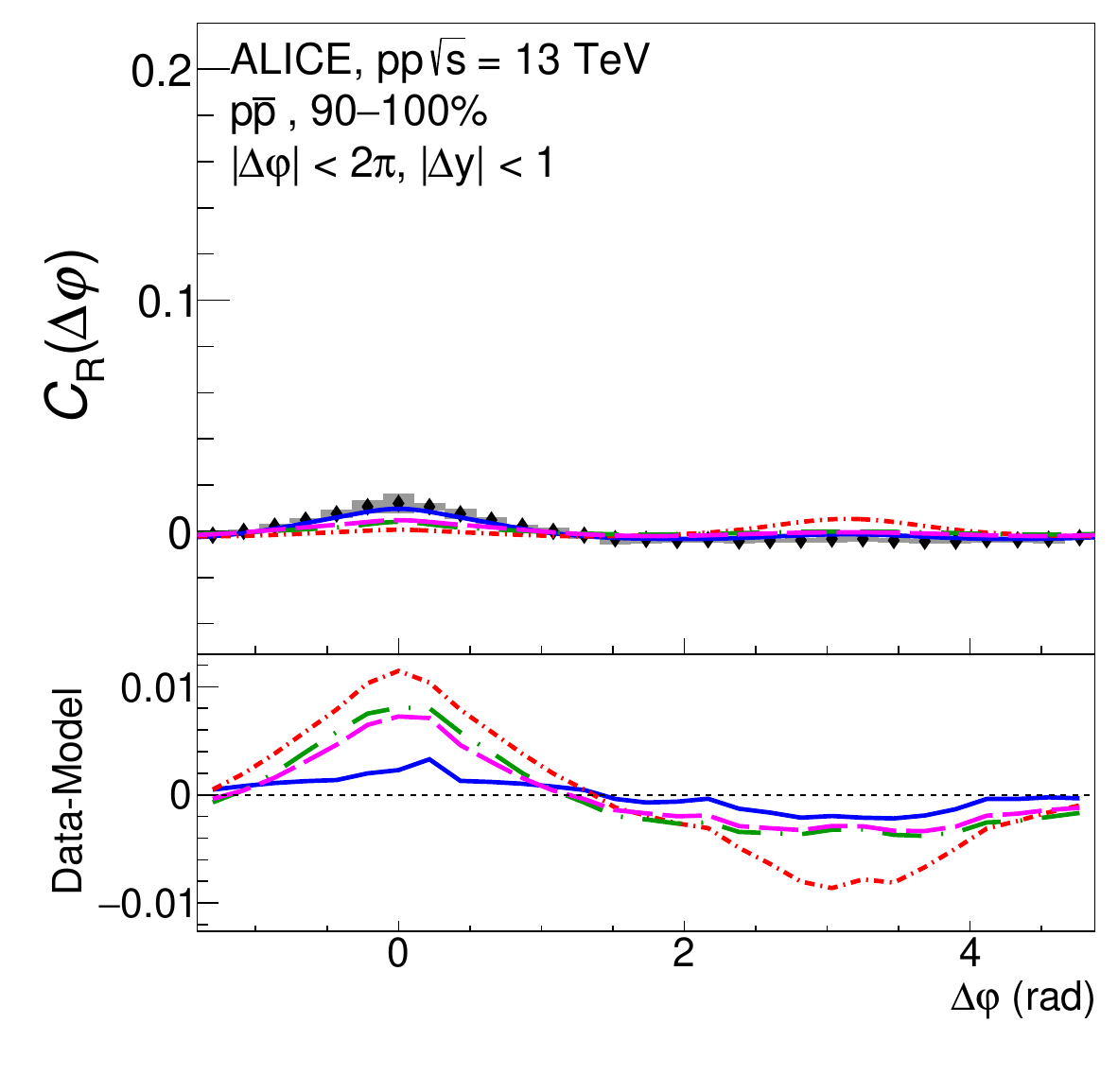}
   \label{fig4-i}}
\caption{Comparison between $\Delta\varphi$-correlation functions and MC models (PYTHIA8 Monash 2013 in blue, EPOS LHC in red, EPOS4 in green, and PYTHIA 8.315 in magenta) in pp collision at $\sqrt{s} = 13$ TeV by using the $C_{\rm{R}}$ definition for unlike-sign pairs of pions, kaons, and protons (top-to-bottom) for the highest (0--1\%), middle (30--40\%), and lowest (90--100\%) multiplicity classes (left-to-right).}
\label{comparisonMCmodel_unlike-sign}
\end{figure}

The results obtained using the $C_{\rm{R}}$ definition were compared with MC models: PYTHIA8 Monash 2013~\cite{Skands_2014}, PYTHIA 8.315~\cite{newpythia}, EPOS LHC~\cite{Pierog:eposlhc}, and EPOS 4~\cite{epos4}, all generated for pp collisions at $\sqrt{s} = 13$ TeV. Both PYTHIA versions employ the Lund string-fragmentation model to describe hadronization, ensuring a consistent microscopic picture of final-state particle production. PYTHIA8 Monash 2013 corresponds to a well-established tune optimized to reproduce a broad range of LHC observables, whereas PYTHIA 8.315 includes a crucial bug fix in the parton shower but has not been extensively retuned, which generally leads to a poorer description of the particle yields. For EPOS, both versions are based on the core–corona framework, where the dense medium formed in high-multiplicity events (the core) is treated separately from the surrounding dilute region (the corona). EPOS LHC introduces collective effects via parton-ladder splitting but systematically underestimates the away-side correlations, particularly in low-multiplicity events. EPOS 4 retains the same core–corona picture but implements a more consistent treatment of multiple scatterings and saturation, resulting in a somewhat improved description of the away side compared to EPOS LHC.
\Cref{comparisonMCmodel_like-sign,comparisonMCmodel_unlike-sign} show the two-particle correlations on the azimuthal distribution for like- and unlike-sign pions, kaons, and protons in the highest, middle, and lowest multiplicity classes. 

All models, when using the $C_{\rm{R}}$ definition, show an increase in the correlation with growing multiplicity, which is in agreement with the trend observed in the data. In the utilized MC models, quantum-statistics effects are not accounted for. This limitation becomes evident when comparing correlations of like-sign pions, as observed in \Cref{fig3-a,fig3-b,fig3-c}, where the models underestimate correlations on the near side compared to experimental data.
For like-sign kaons, shown in \Cref{fig3-d,fig3-e,fig3-f}, the model values exceed the data on the near side, although better agreement is observed on the away side. The models fail to reproduce the anticorrelation observed in like-sign protons.

For unlike-sign pions in the lowest multiplicity class (Fig.~\ref{fig4-c}), the models fail to accurately describe the correlation due to the lack of pairs of pions close in the $\Delta y,\Delta\varphi$ space.
More generally, PYTHIA simulations, benefitting from their effective description of mini-jet structures, qualitatively capture both the near- and away-side peaks for unlike-sign pions and kaons. EPOS LHC, in contrast, fails to describe the away side, especially at low multiplicity, while EPOS 4 shows a noticeable improvement in this region.
Interestingly, EPOS 4 reproduces the unlike-sign proton correlations both qualitatively and quantitatively, showing good agreement in the highest and middle multiplicity classes. PYTHIA 8, on the other hand, exhibits larger discrepancies at high-multiplicity but approaches good agreement within uncertainties in the lower multiplicity classes, as shown in~\Cref{fig4-g,fig4-h,fig4-i}.

\begin{figure}[!h]
   \centering
 \subfigure[]{
   \includegraphics[width=5cm]{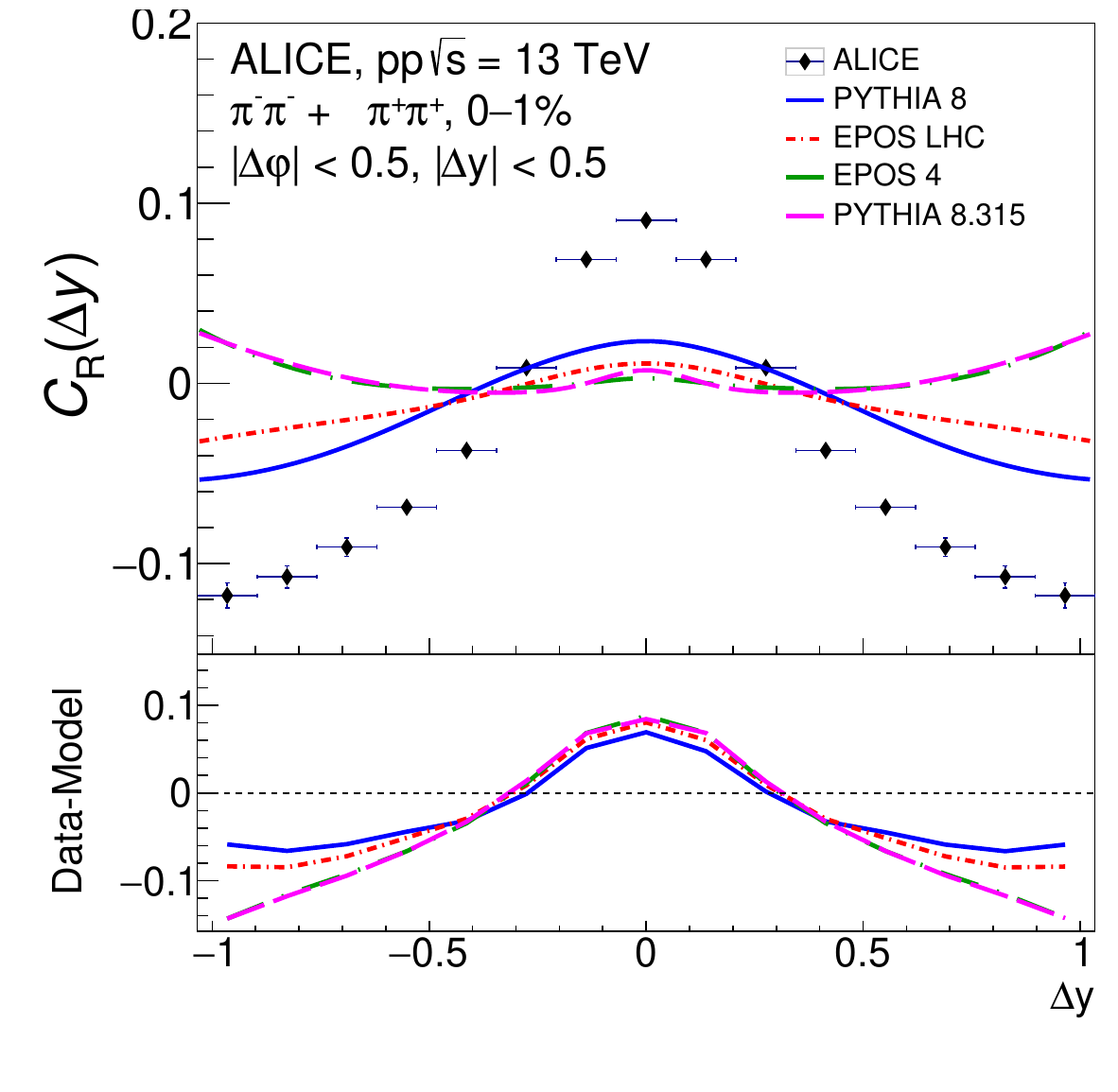} 
   \label{fig5-a}}
    \subfigure[]{
   \includegraphics[width=5cm]{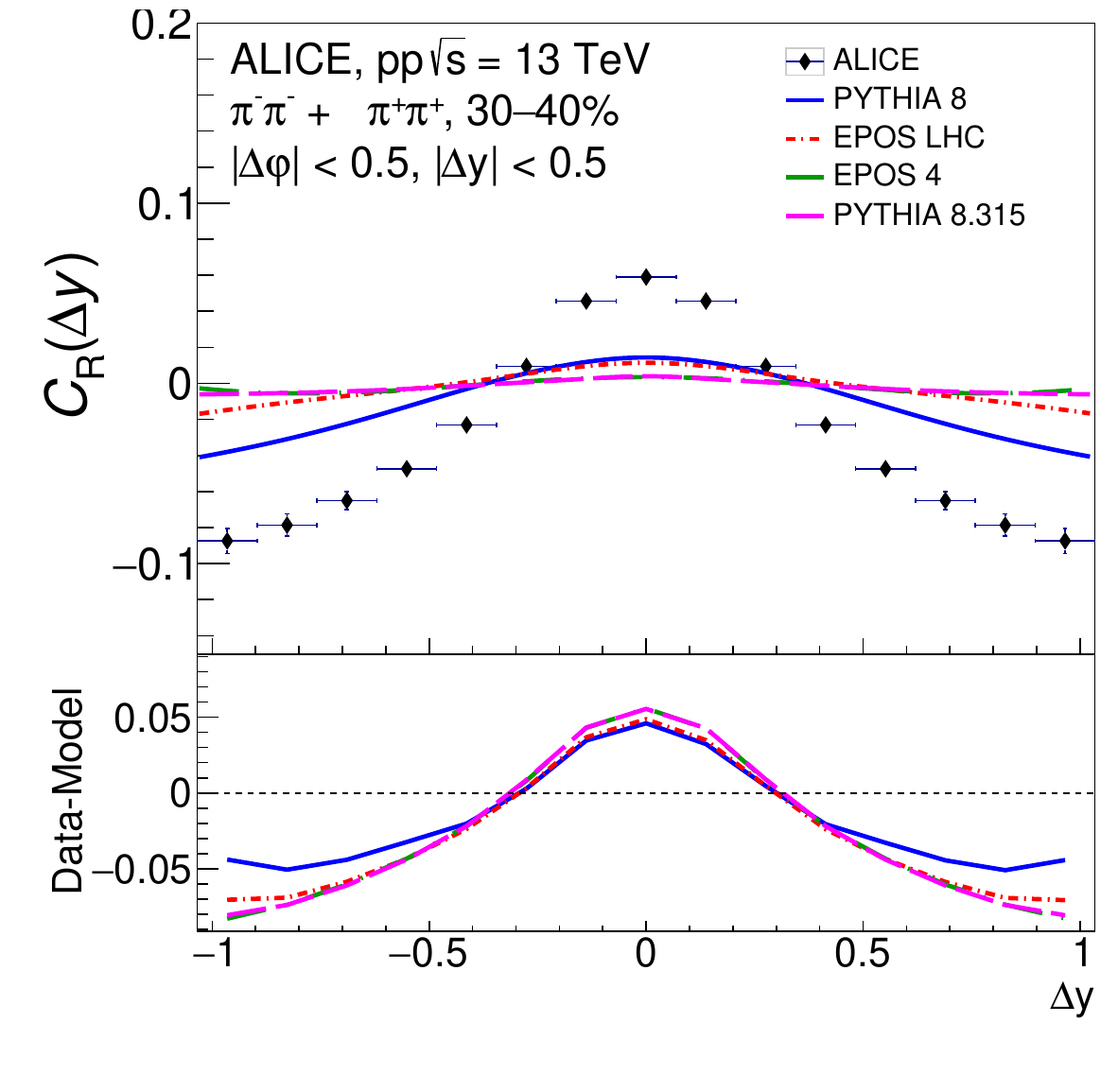} 
   \label{fig5-b}}
 \subfigure[]{
   \includegraphics[width=5cm]{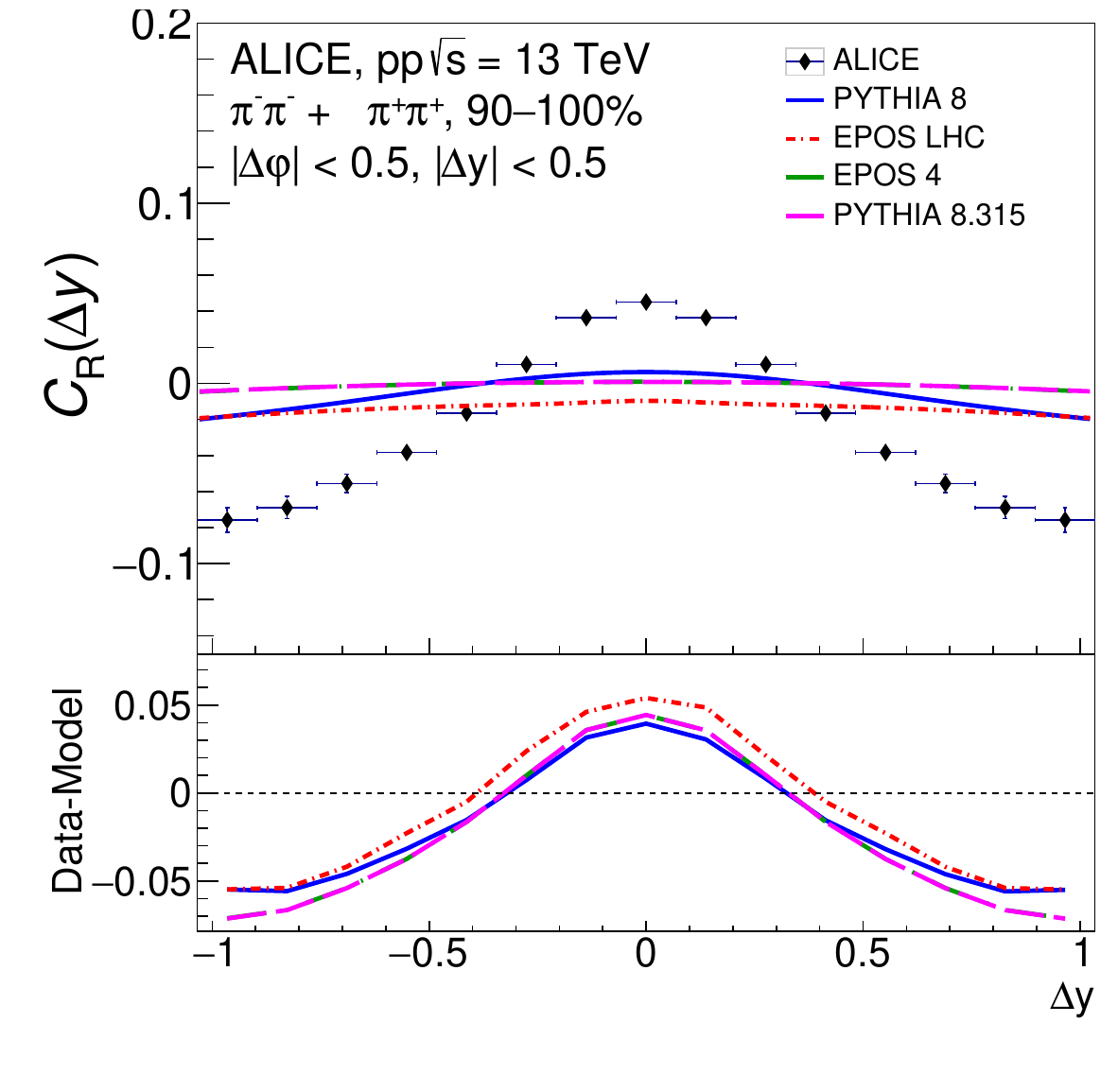}
   \label{fig5-c}}
\subfigure[]{
   \includegraphics[width=5cm]{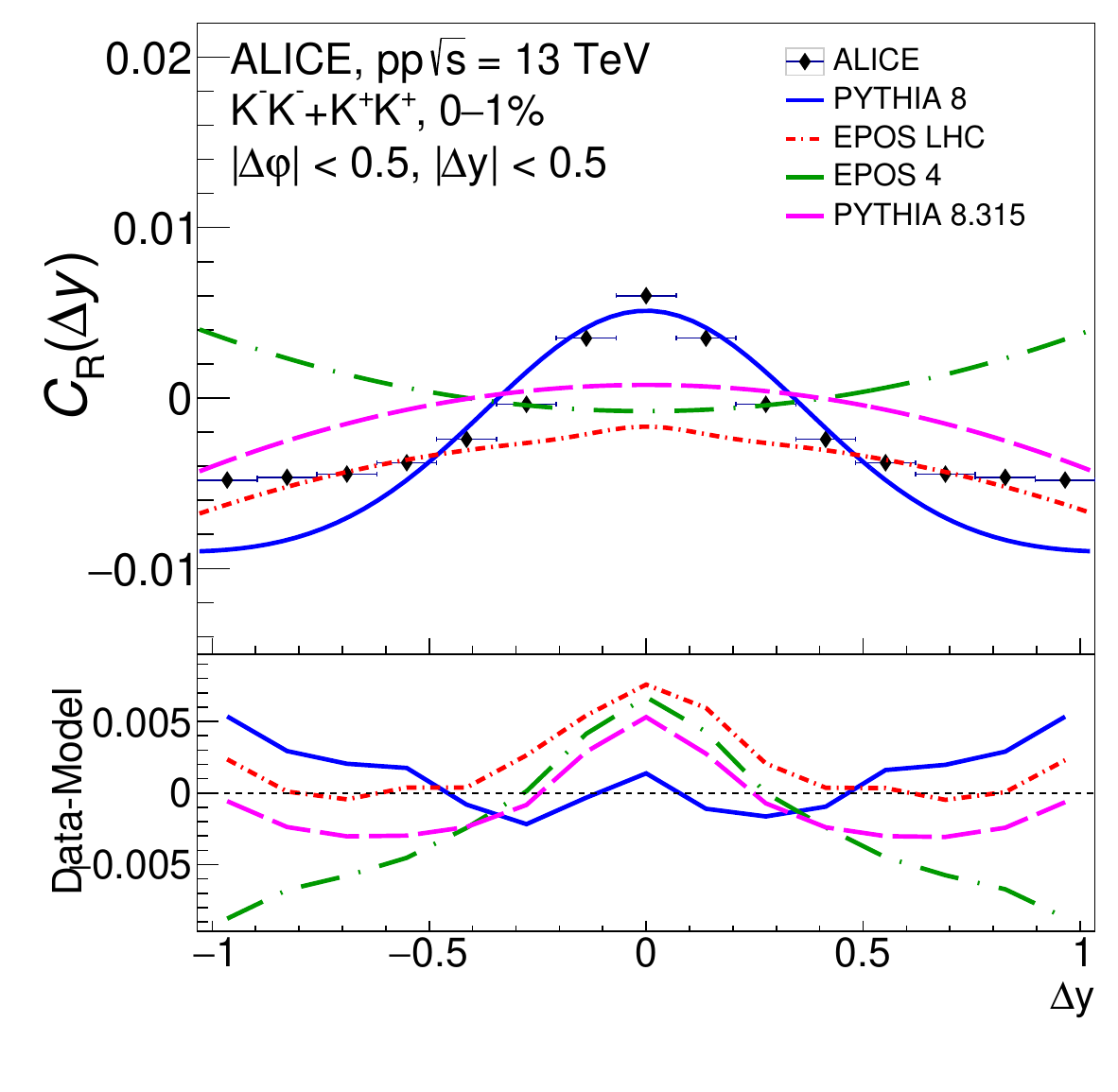}
   \label{fig5-d}}
   \subfigure[]{
   \includegraphics[width=5cm]{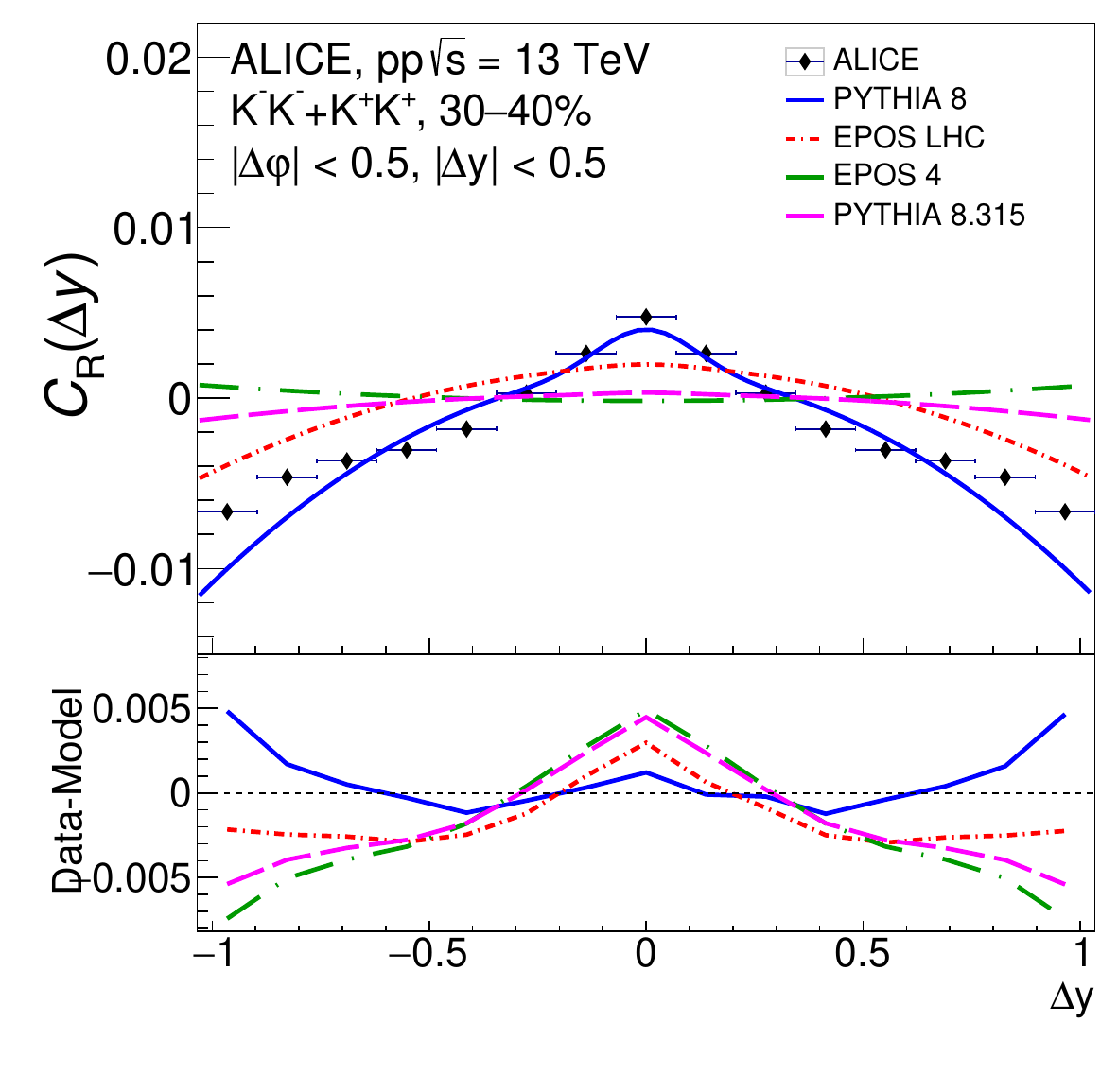}
   \label{fig5-e}}
 \subfigure[]{
   \includegraphics[width=5cm]{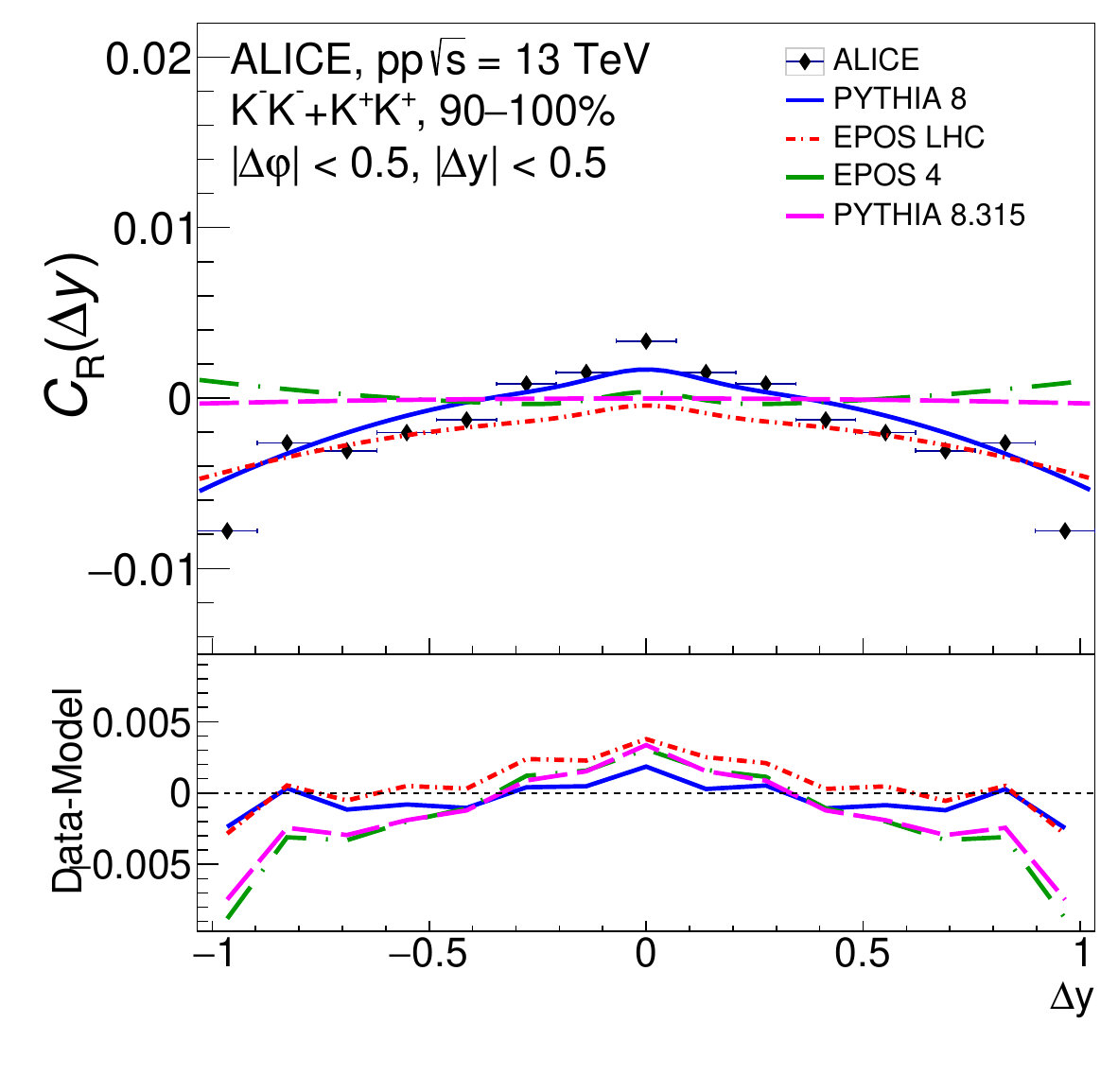}
   \label{fig5-f}}
\subfigure[]{
   \includegraphics[width=5cm]{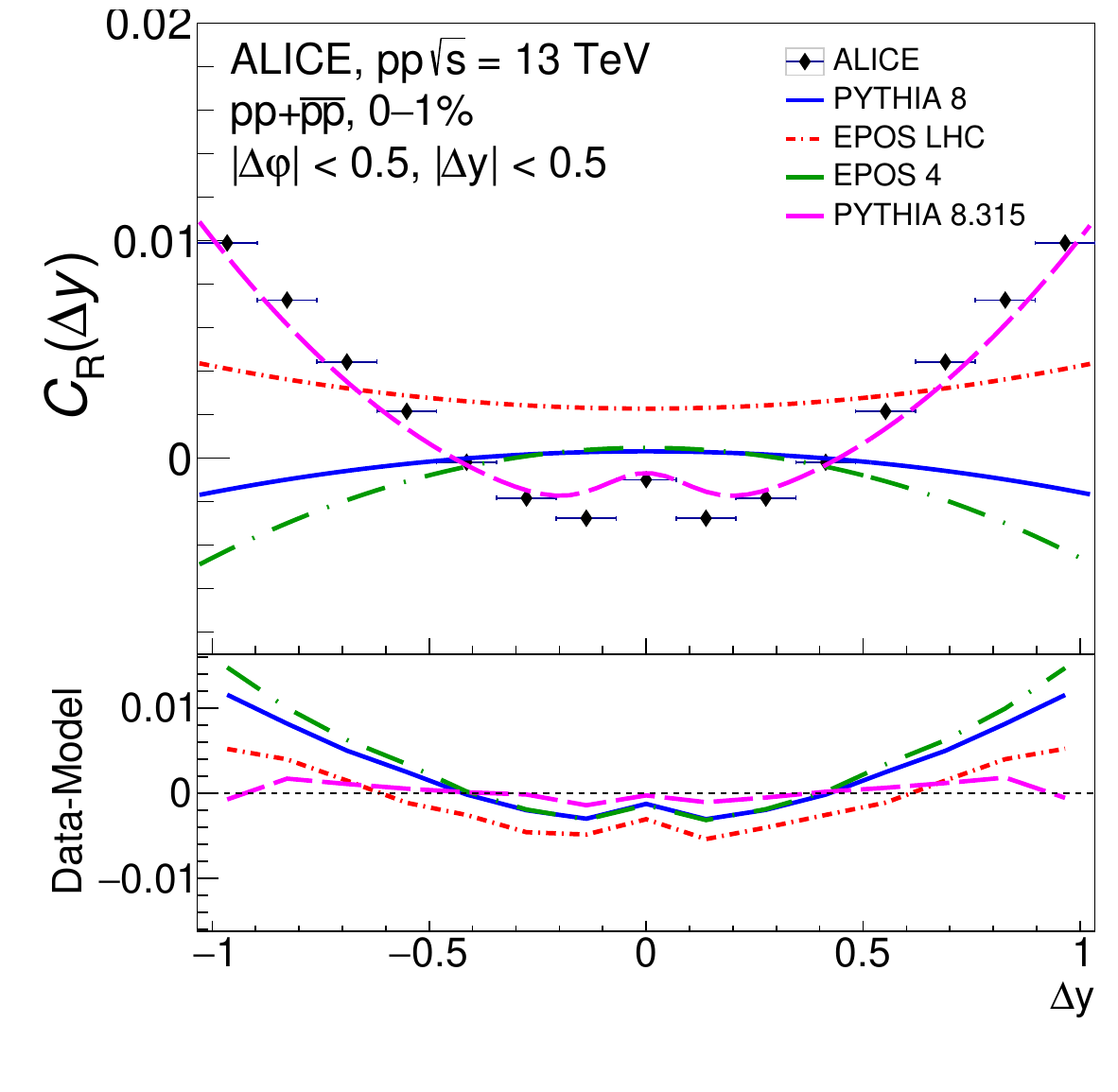}
   \label{fig5-g}}
   \subfigure[]{
   \includegraphics[width=5cm]{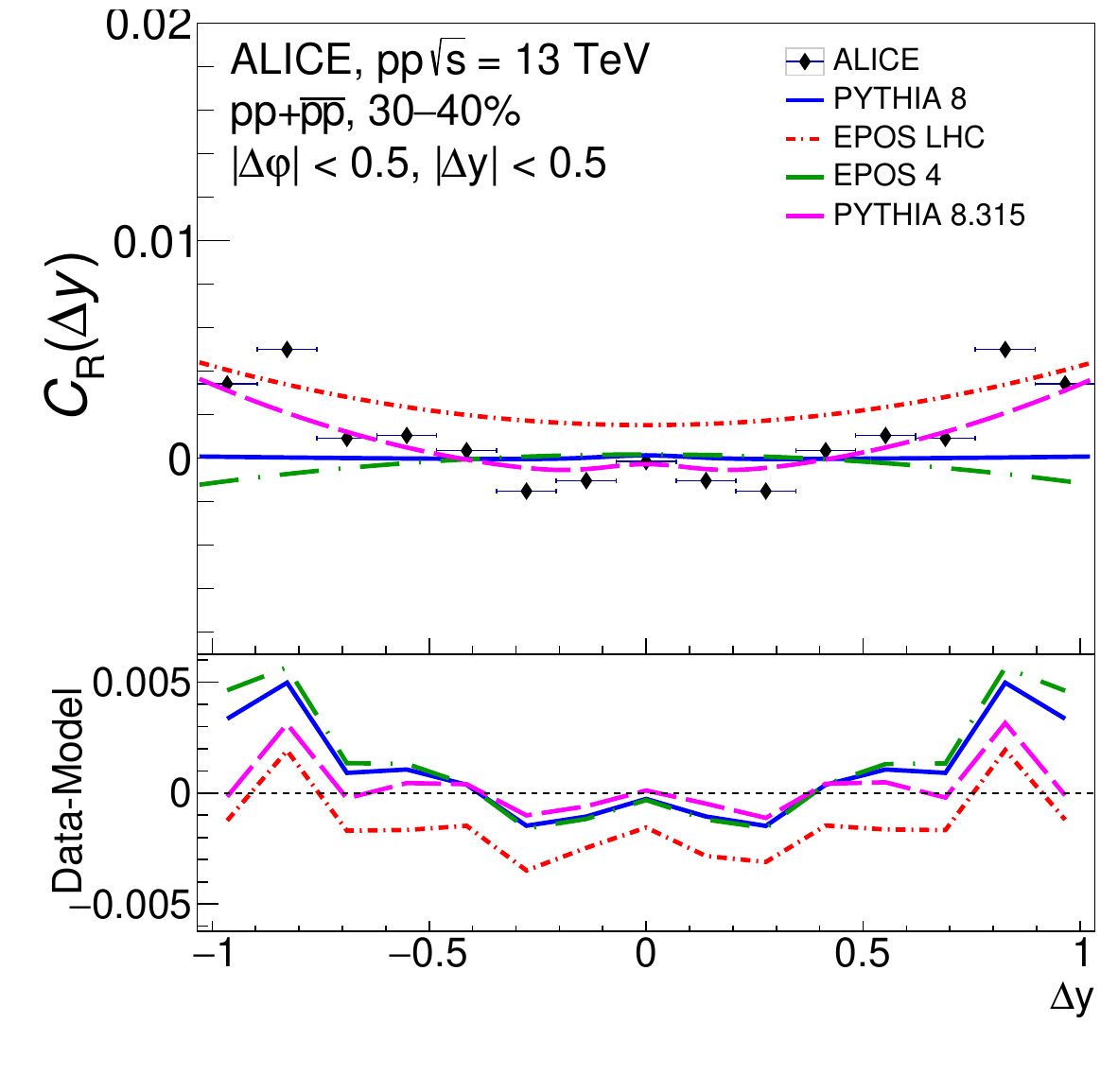}
   \label{fig5-h}}
 \subfigure[]{
   \includegraphics[width=5cm]{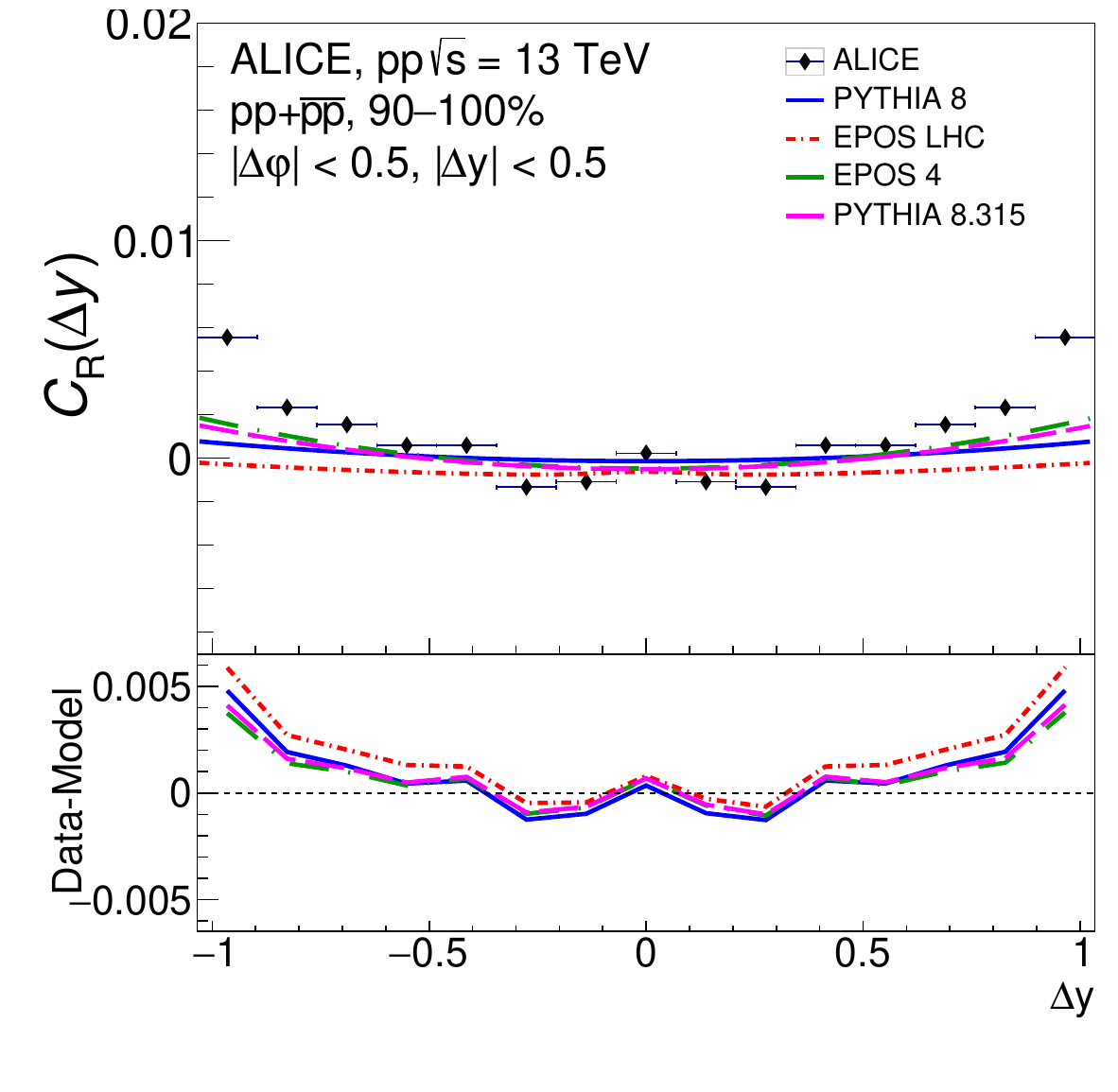}
   \label{fig5-i}}
   
      \caption{$\Delta y$-correlation function distribution compared with MC models (PYTHIA8 Monash 2013 in blue, EPOS LHC in red, EPOS4 in green, and PYTHIA 8.315 in magenta) in pp collision at $\sqrt{s} = 13$ TeV by using the $C_{\rm{R}}$ definition for like-sign pions, kaons, and protons (top-to-bottom) and for the highest (0--1\%), middle (30--40\%), and lowest (90--100\%) multiplcity classes (left-to-right).}
\label{comparisonMCmodel_rapidity_like-sign}
\end{figure}

\begin{figure}[!h]
   \centering
 \subfigure[]{
   \includegraphics[width=5cm]{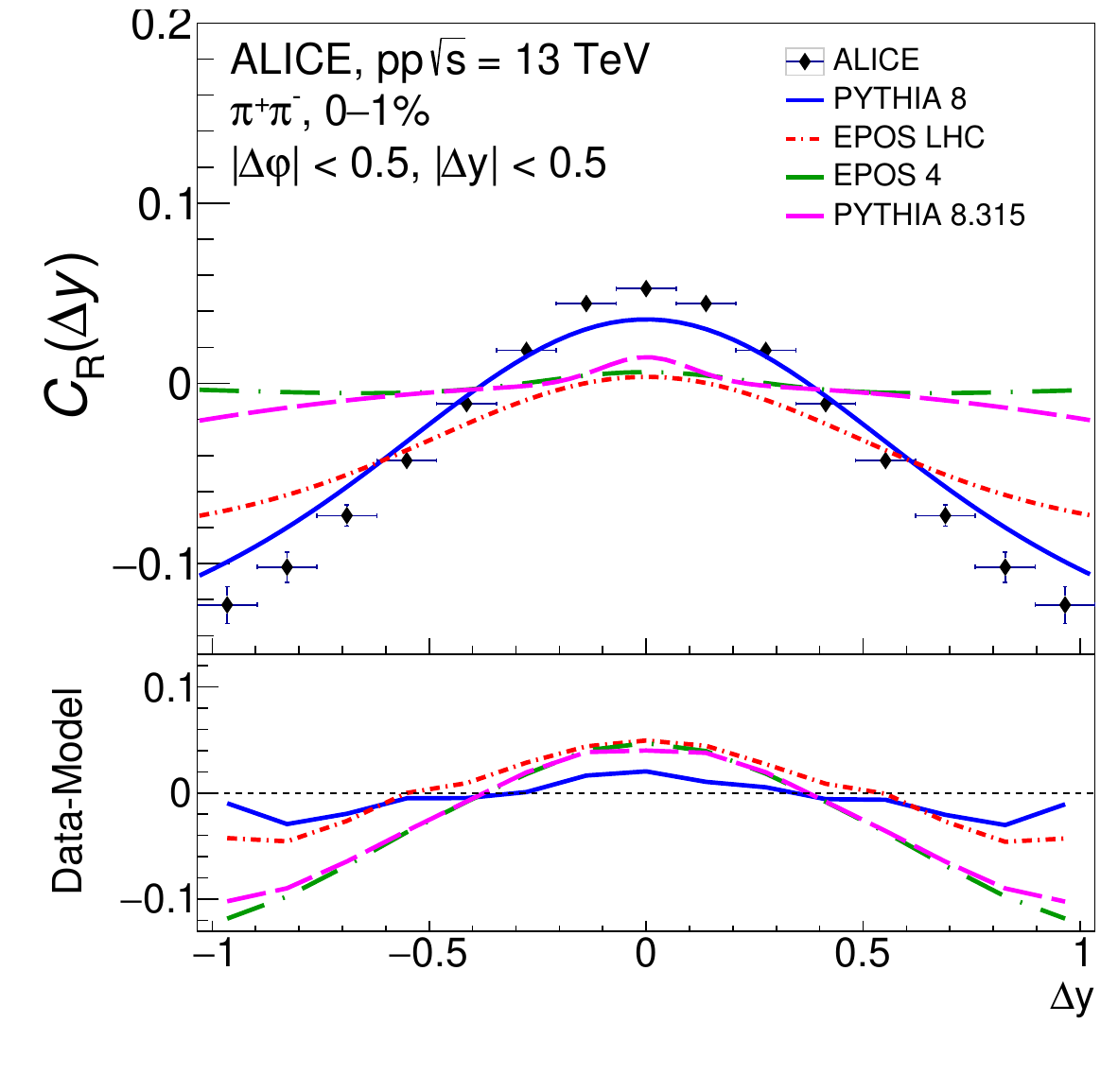} 
   \label{fig6-a}}
 \subfigure[]{
   \includegraphics[width=5cm]{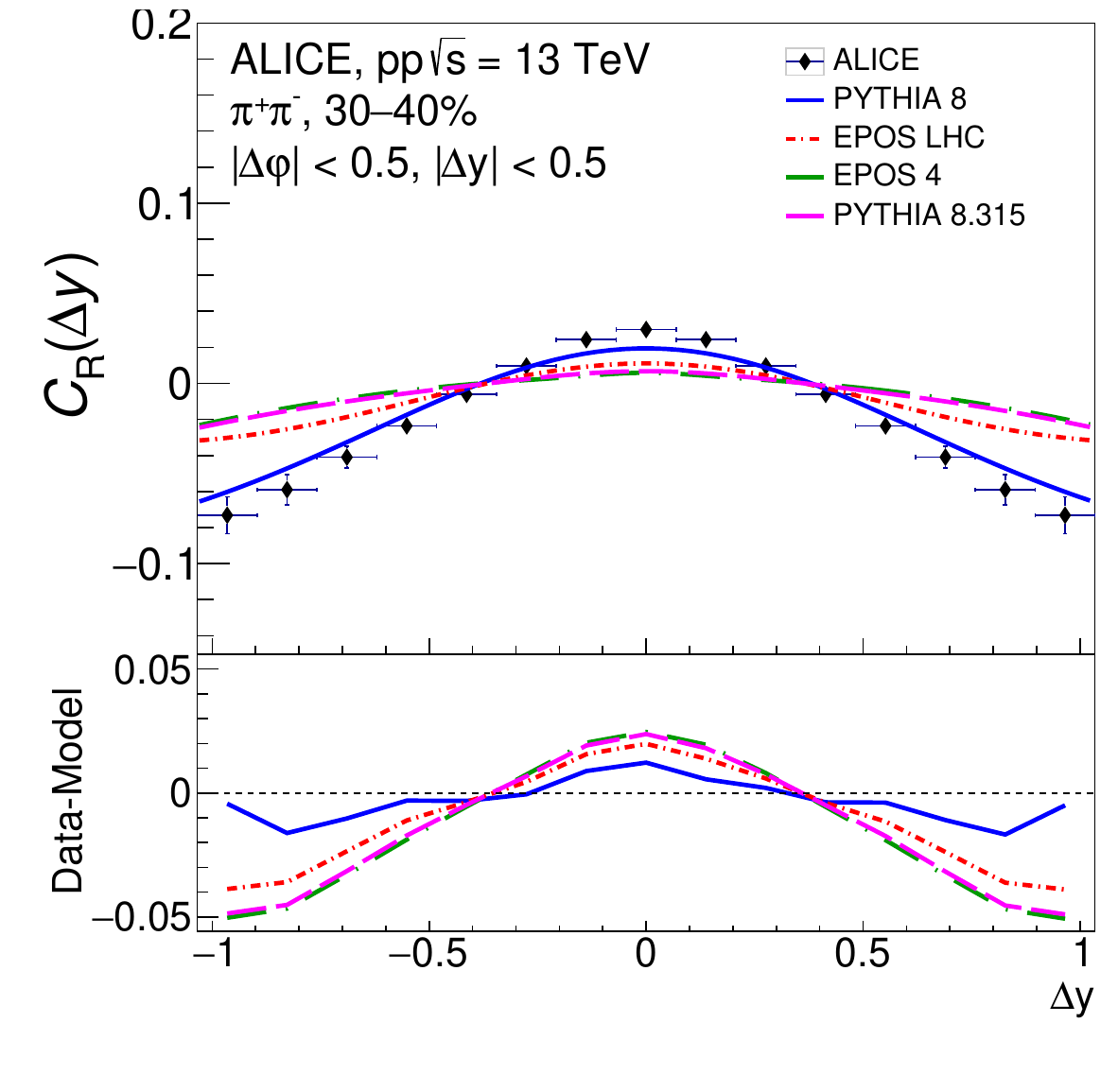} 
   \label{fig6-b}}
 \subfigure[]{
   \includegraphics[width=5cm]{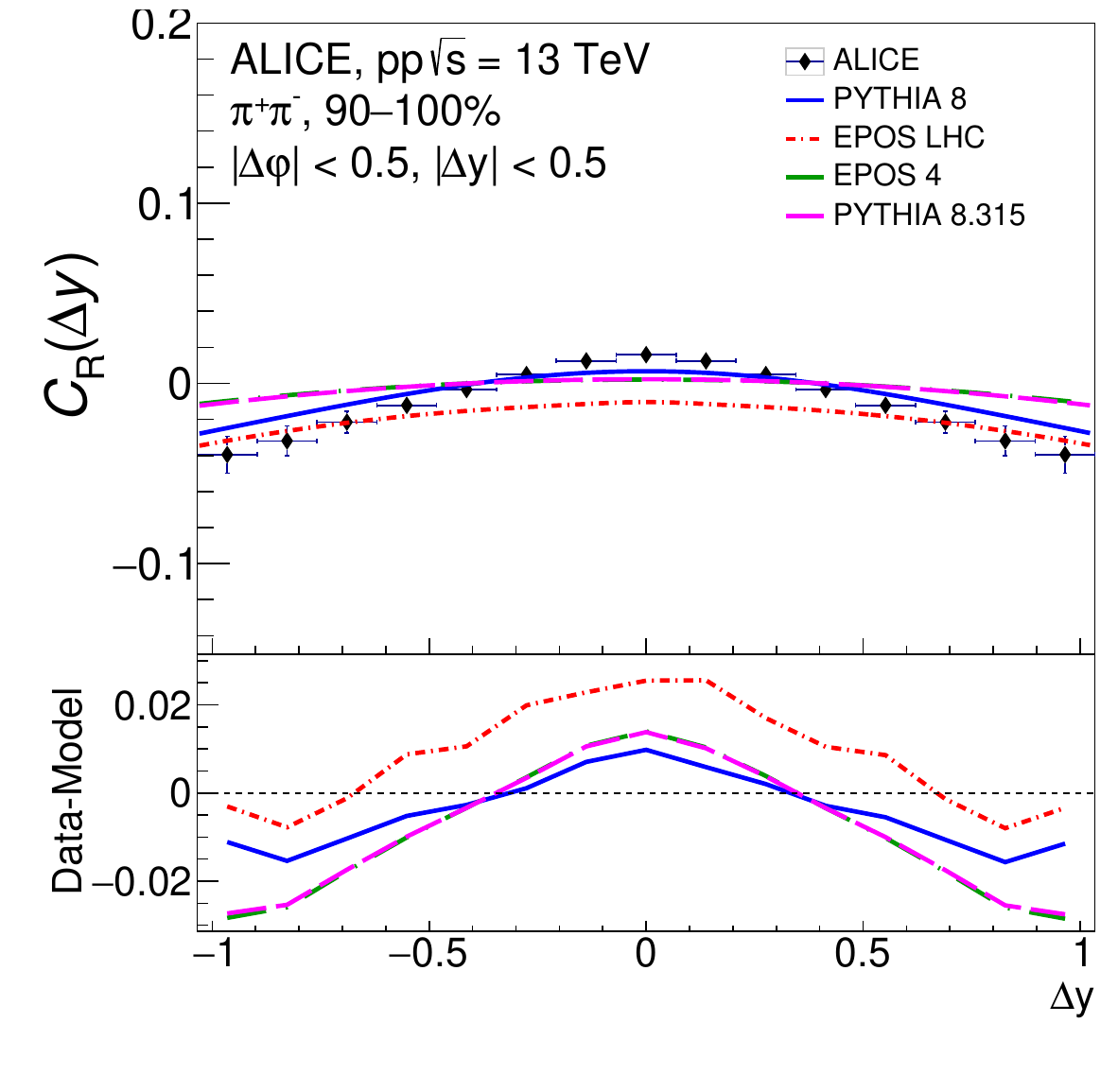}
   \label{fig6-c}}
\subfigure[]{
   \includegraphics[width=5cm]{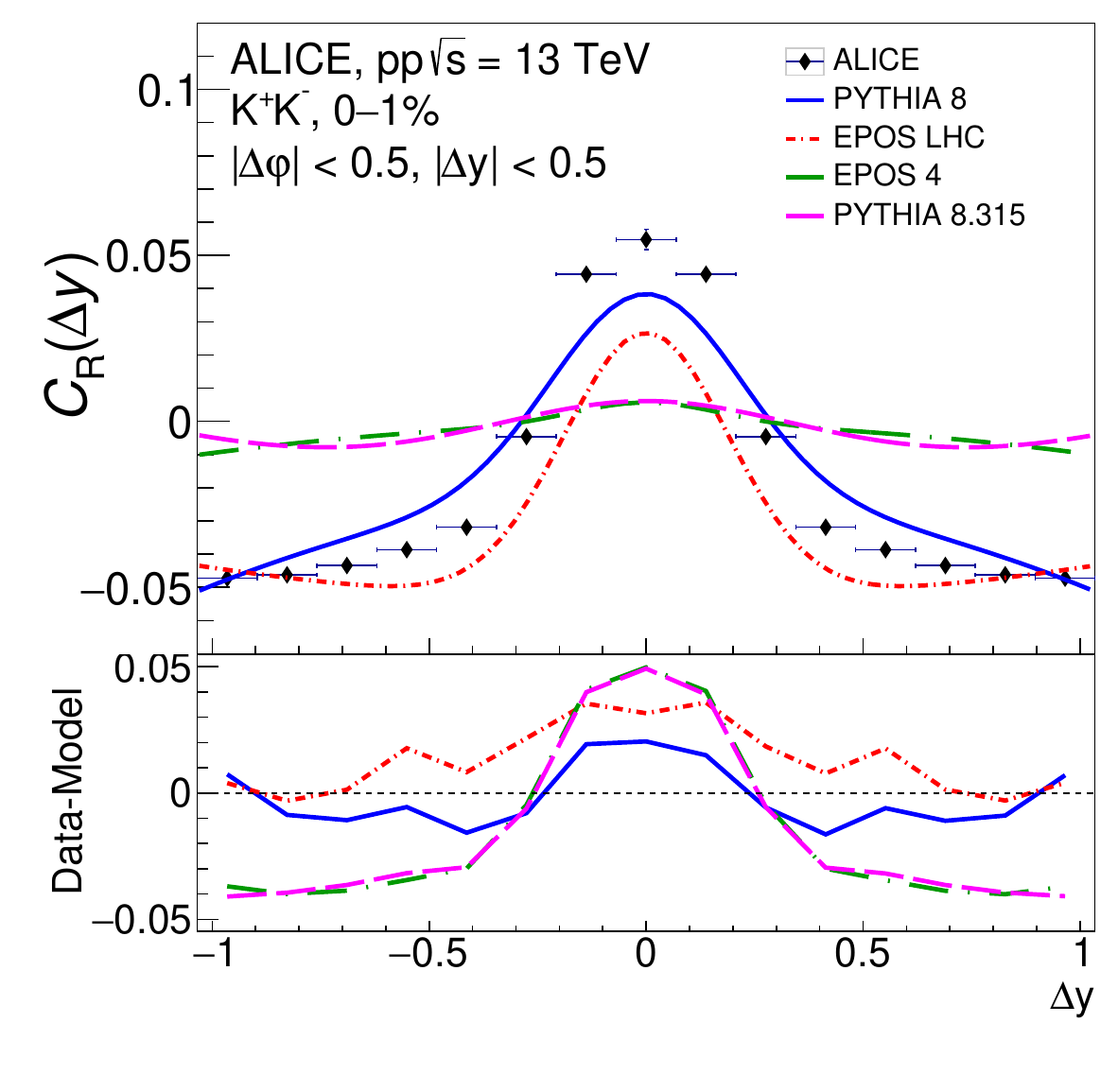}
   \label{fig6-d}}
   \subfigure[]{
   \includegraphics[width=5cm]{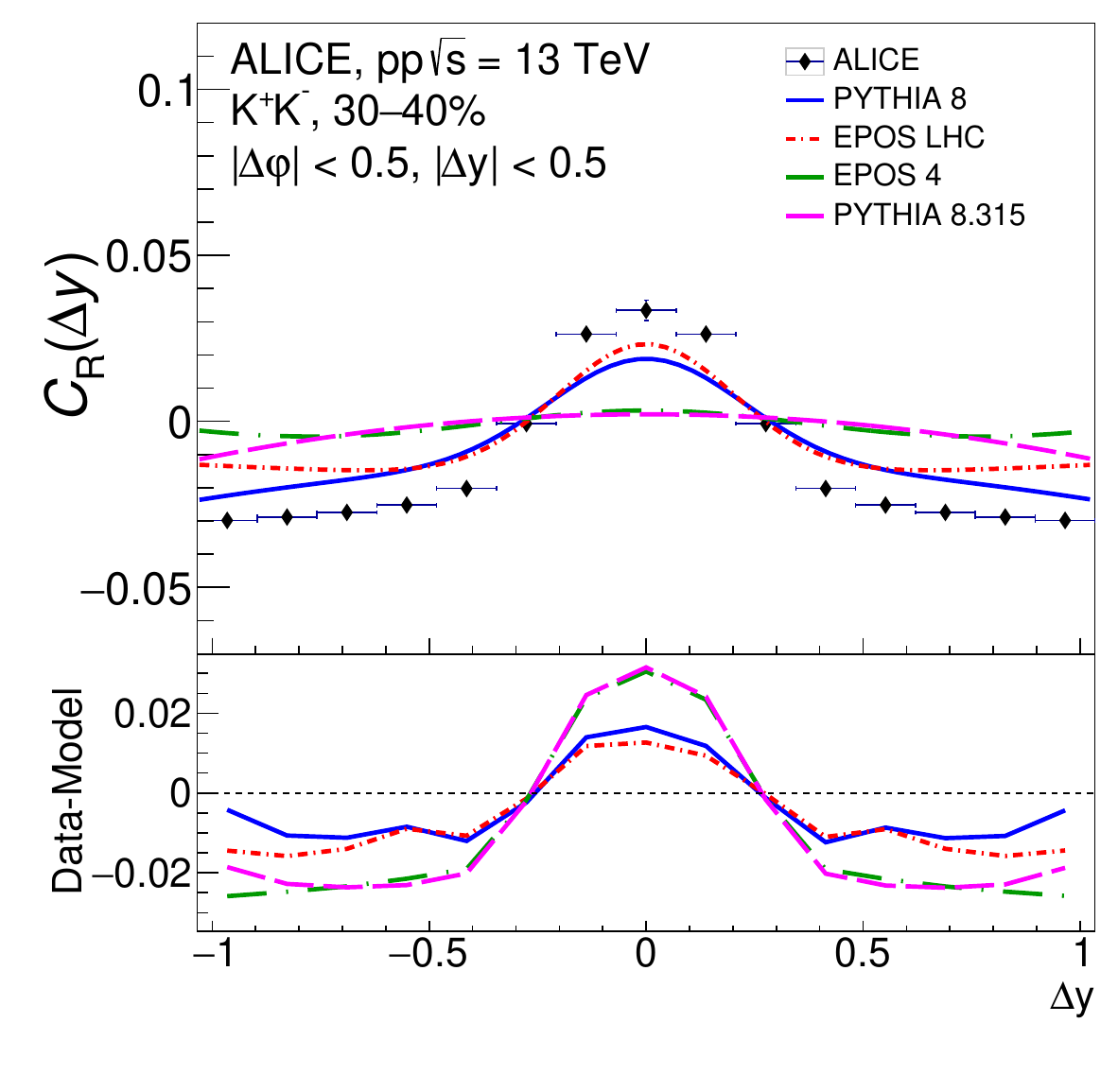}
   \label{fig6-e}}
 \subfigure[]{
   \includegraphics[width=5cm]{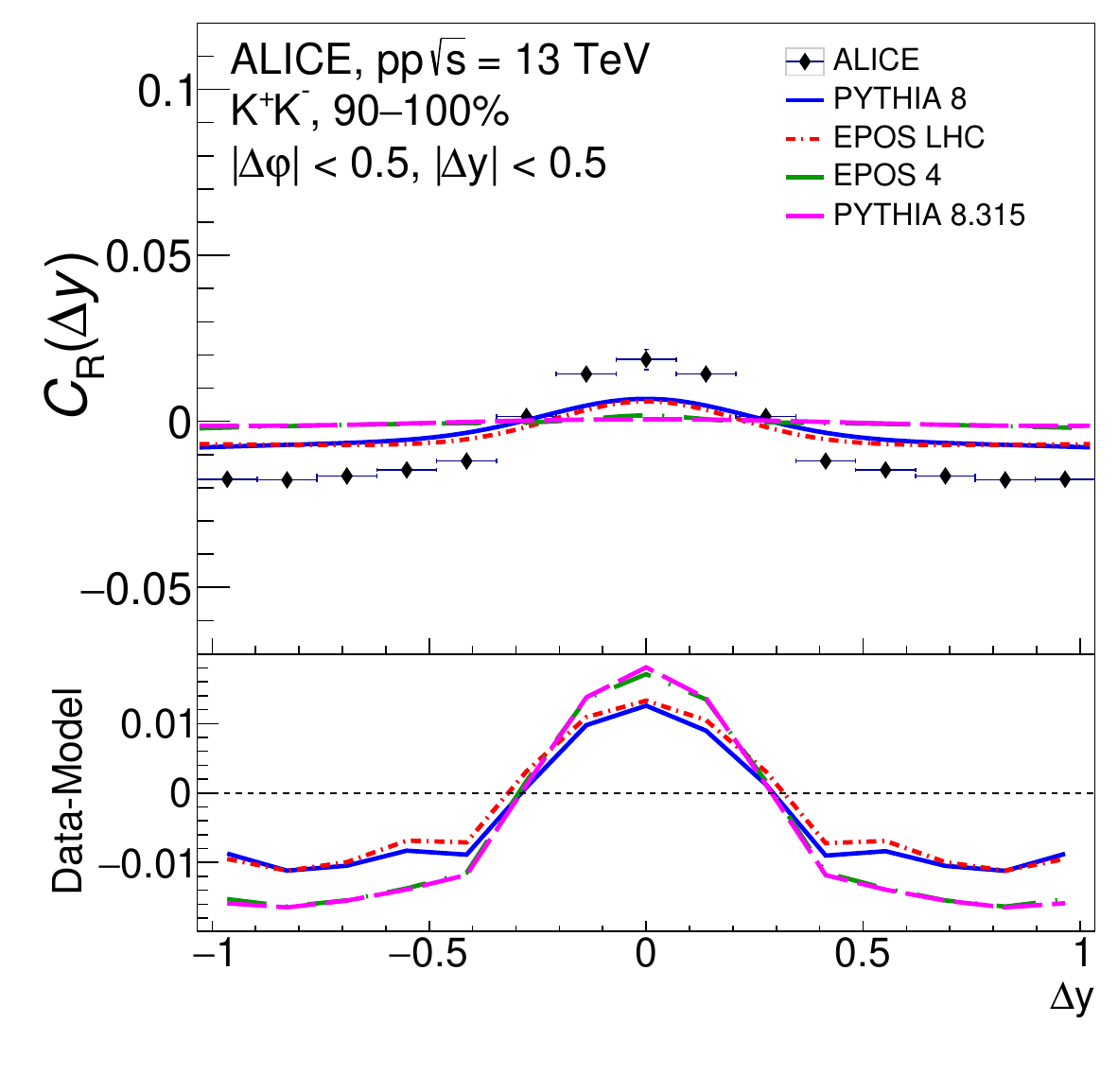}
   \label{fig6-f}}

   \subfigure[]{
   \includegraphics[width=5cm]{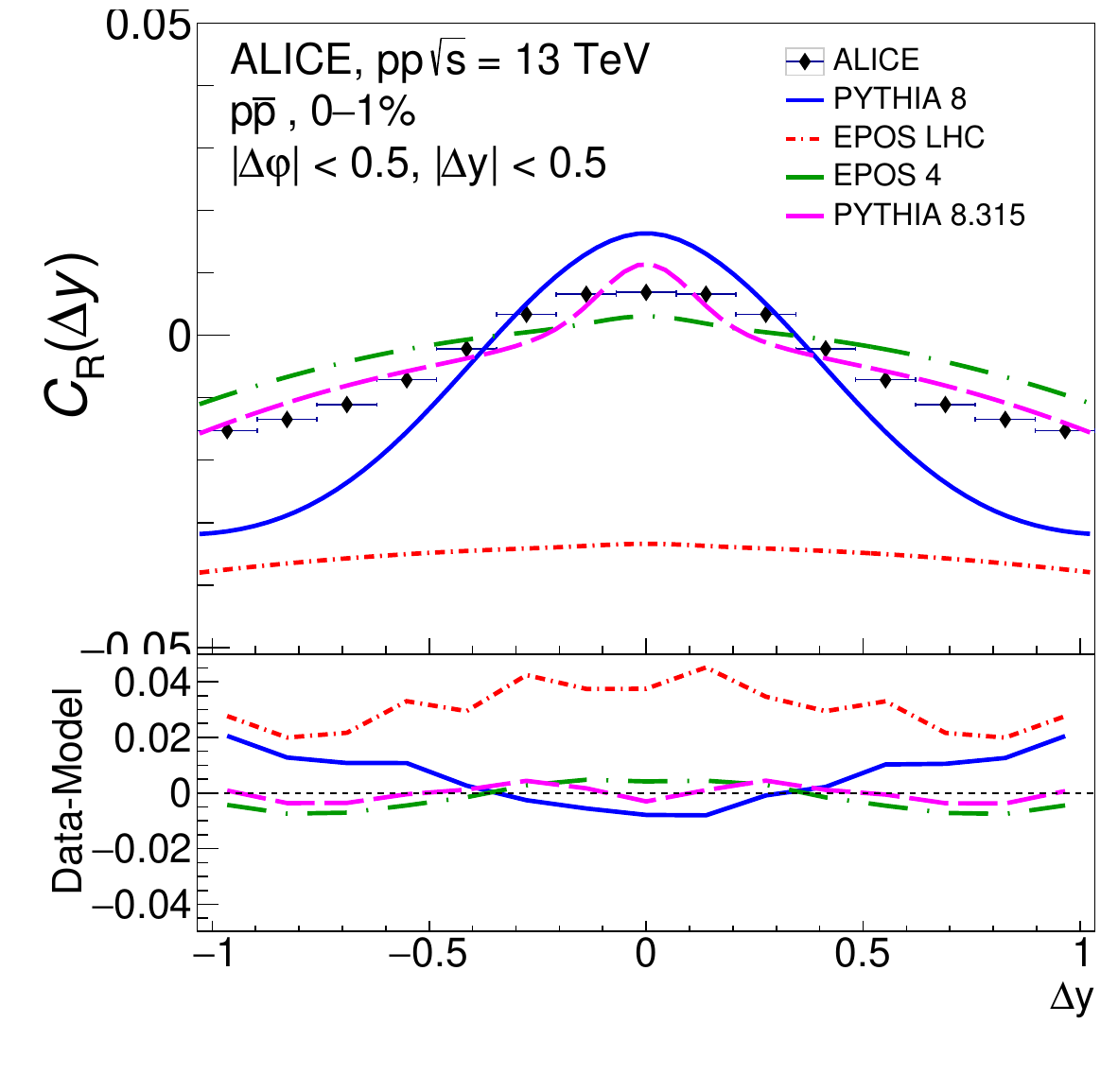}
   \label{fig6-g}}
   \subfigure[]{
   \includegraphics[width=5cm]{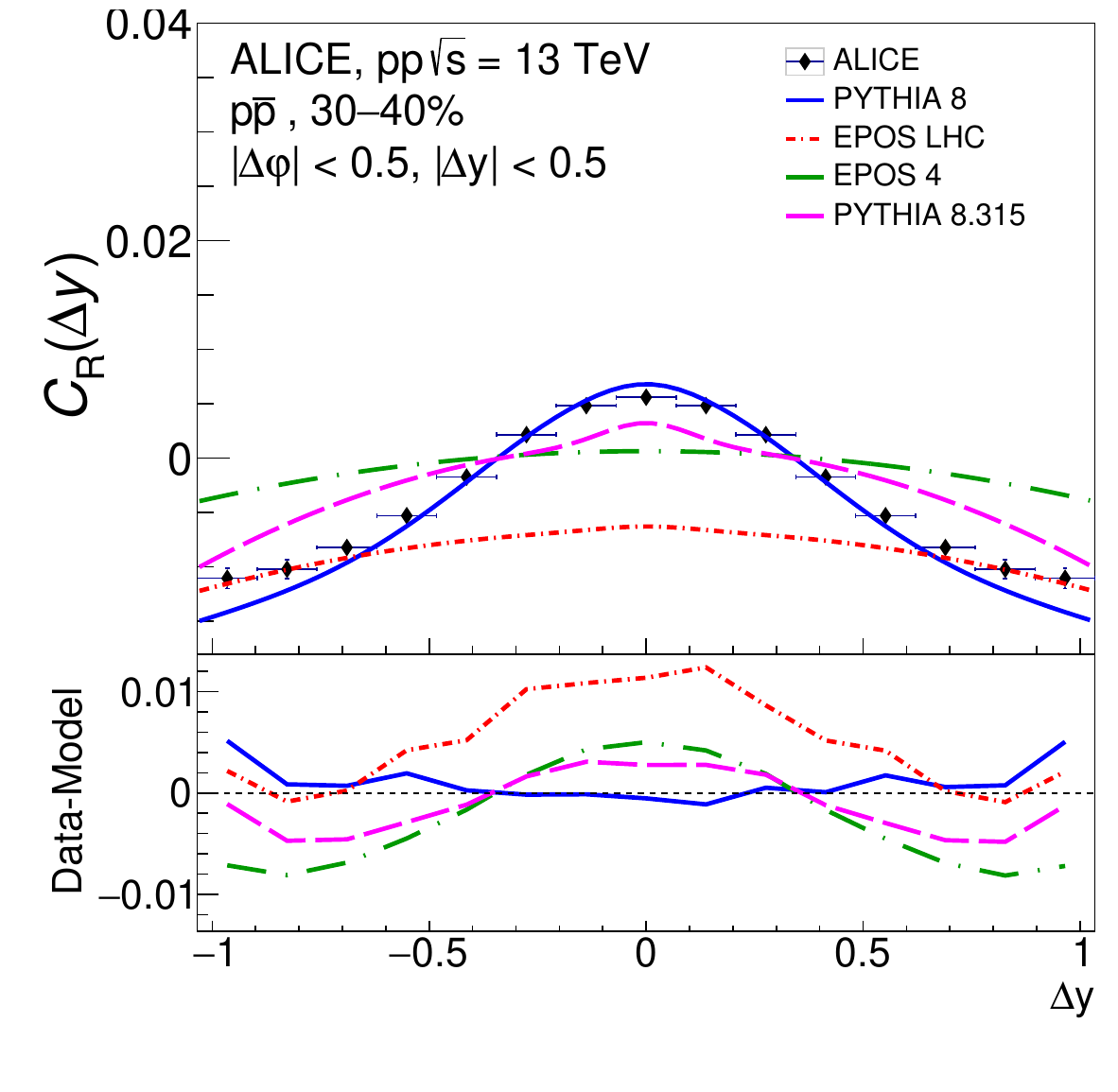}
   \label{fig6-h}}
 \subfigure[]{
   \includegraphics[width=5cm]{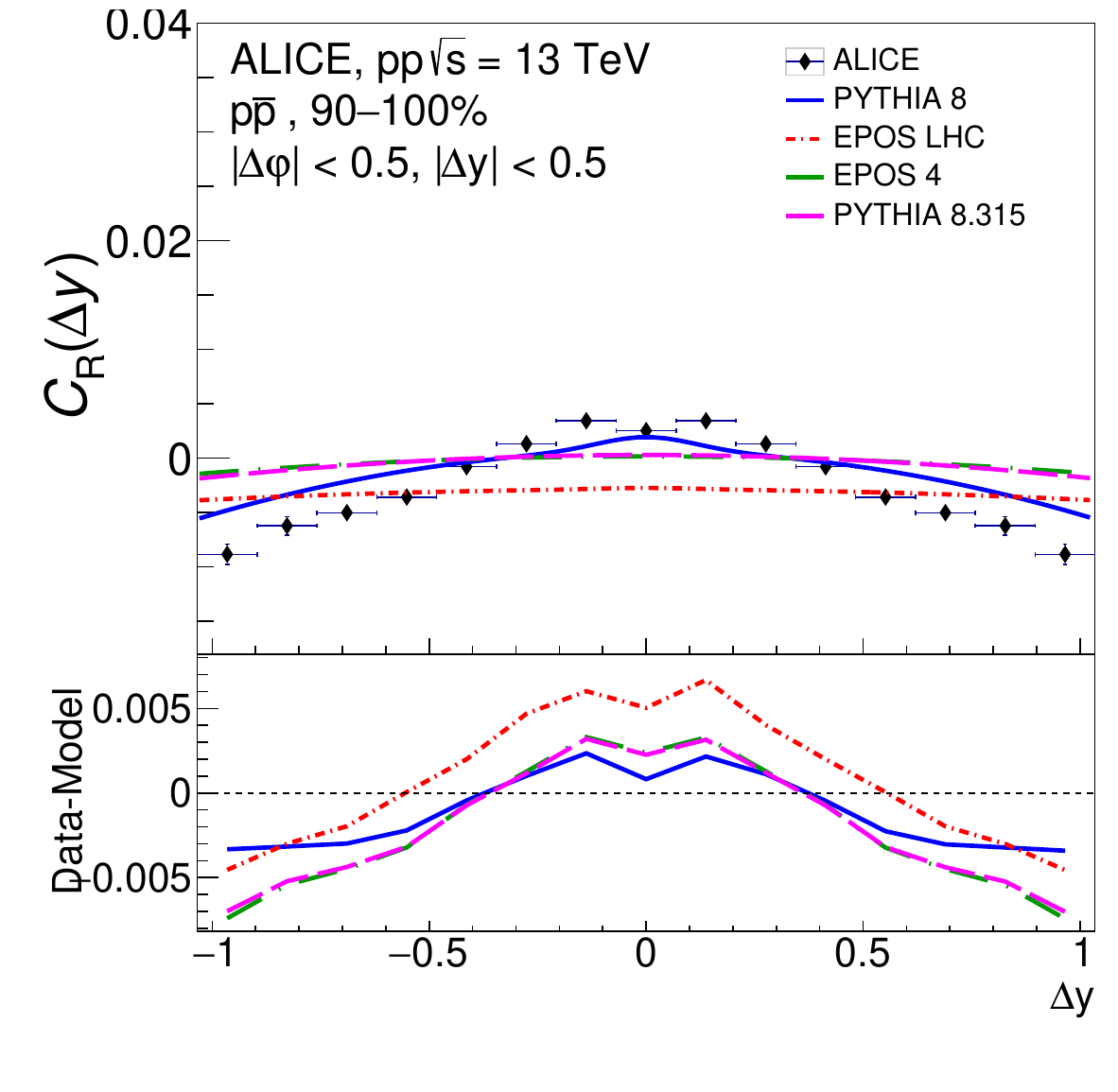}
   \label{fig6-i}}

\caption{$\Delta y$-correlation function distribution compared with MC models (PYTHIA8 Monash 2013 in blue, EPOS LHC in red, EPOS4 in green, and PYTHIA 8.315 in magenta) in pp collision at $\sqrt{s} = 13$ TeV by using the $C_{\rm{R}}$ definition for unlike-sign pions, kaons, and protons (top-to-bottom) for the highest (0--1\%), middle (30--40\%), and lowest (90--100\%) multiplicity classes (left-to-right).}
\label{comparisonMCmodel_rapidity_unlike-sign}
\end{figure}

The $\Delta y$ projections of the near-side region ($\abs{\Delta\varphi} < $ 0.5) of the measured correlation function were also studied and compared with model predictions from PYTHIA8 Monash 2013, PYTHIA 8.315, EPOS LHC, and EPOS 4 as shown in~\Cref{comparisonMCmodel_rapidity_like-sign,comparisonMCmodel_rapidity_unlike-sign}. This comparison is performed separately for like-sign and unlike-sign pairs across three multiplicity classes: the highest (0--1\%), the middle (30--40\%), and the lowest (90--100\%).

In comparison with the results presented in~\Cref{comparisonMCmodel_like-sign,comparisonMCmodel_unlike-sign}, which demonstrate the successful description of the correlations observed in the $\Delta\varphi$ analysis, the $\Delta y$ projection for like- and unlike-sign pions and kaons shown in~\Cref{comparisonMCmodel_rapidity_like-sign,comparisonMCmodel_rapidity_unlike-sign} reveals discrepancies. For like-sign pions, none of the models achieve a satisfactory qualitative or quantitative description of the data. For unlike-sign pions, as well as like- and unlike-sign kaons, PYTHIA8 shows the best agreement, while the other models reproduce only the general correlation trends without reaching quantitative consistency with the measurements.\\
\begin{figure}[!h]
   \centering
 \subfigure[]{
   \includegraphics[width=6cm]{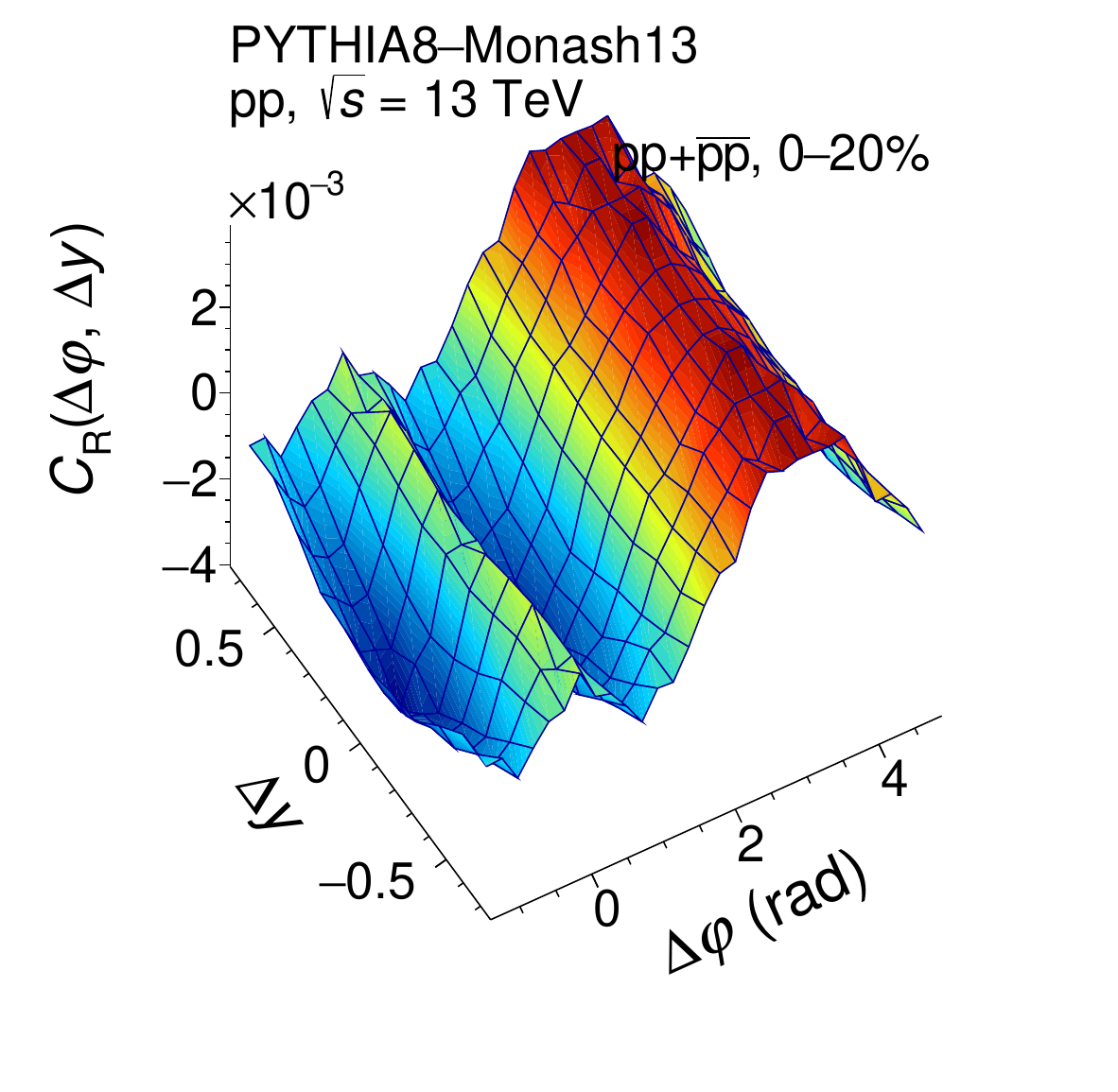} 
   \label{fig7-a}}
    \subfigure[]{
   \includegraphics[width=6cm]{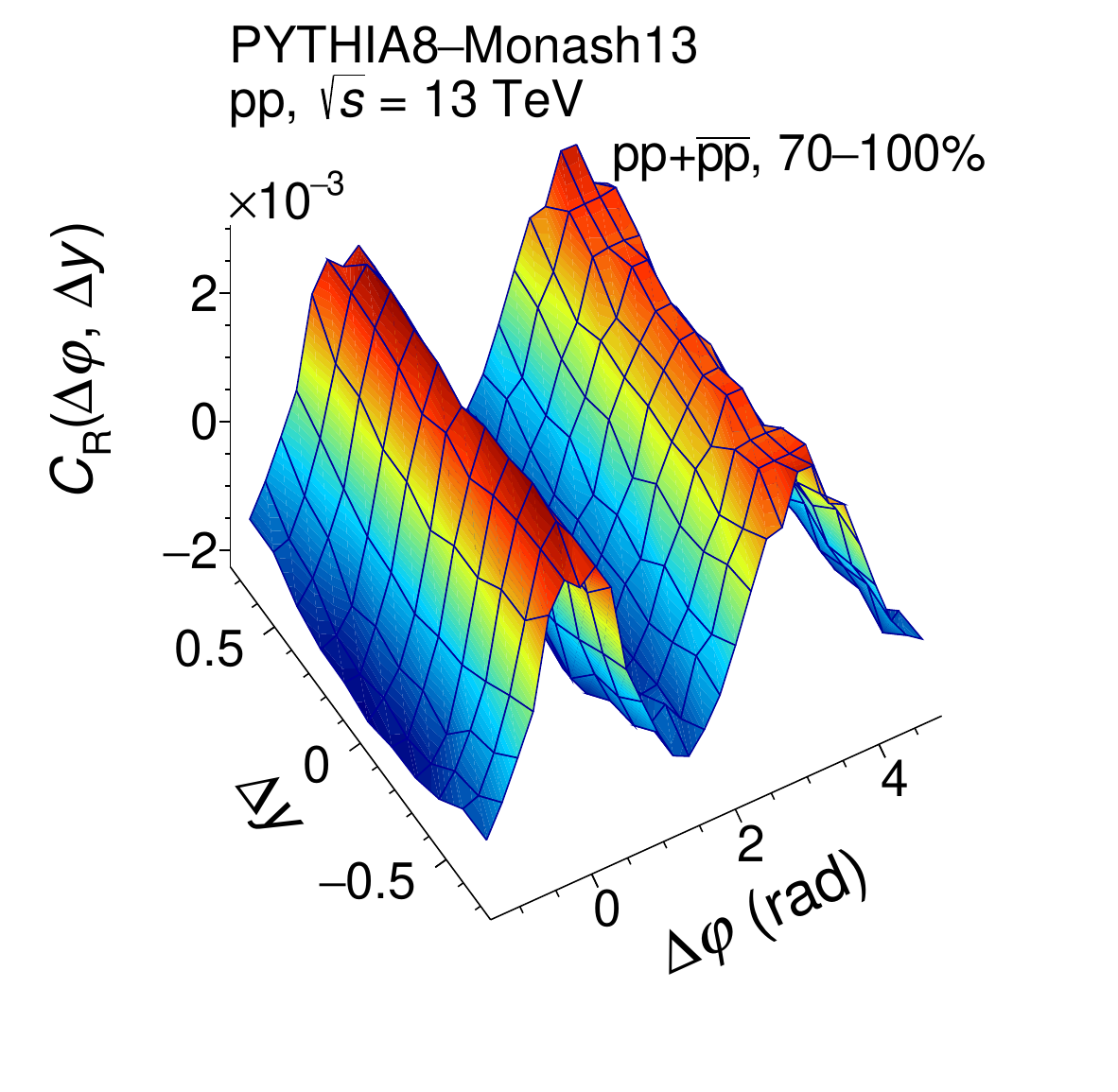} 
   \label{fig7-b}}   
   \subfigure[]{
   \includegraphics[width=6cm]{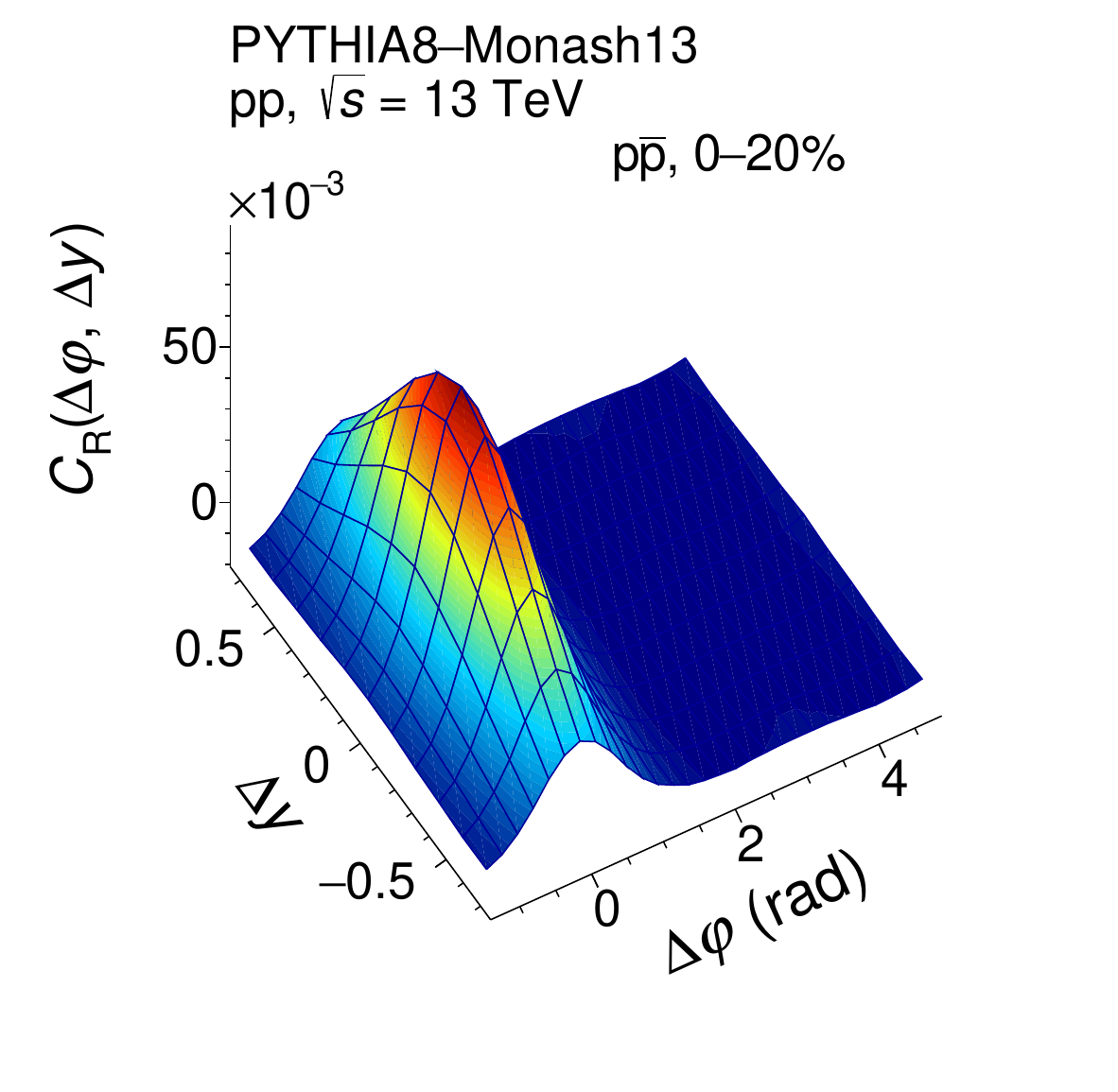}
   \label{fig7-c}}
 \subfigure[]{
   \includegraphics[width=6cm]{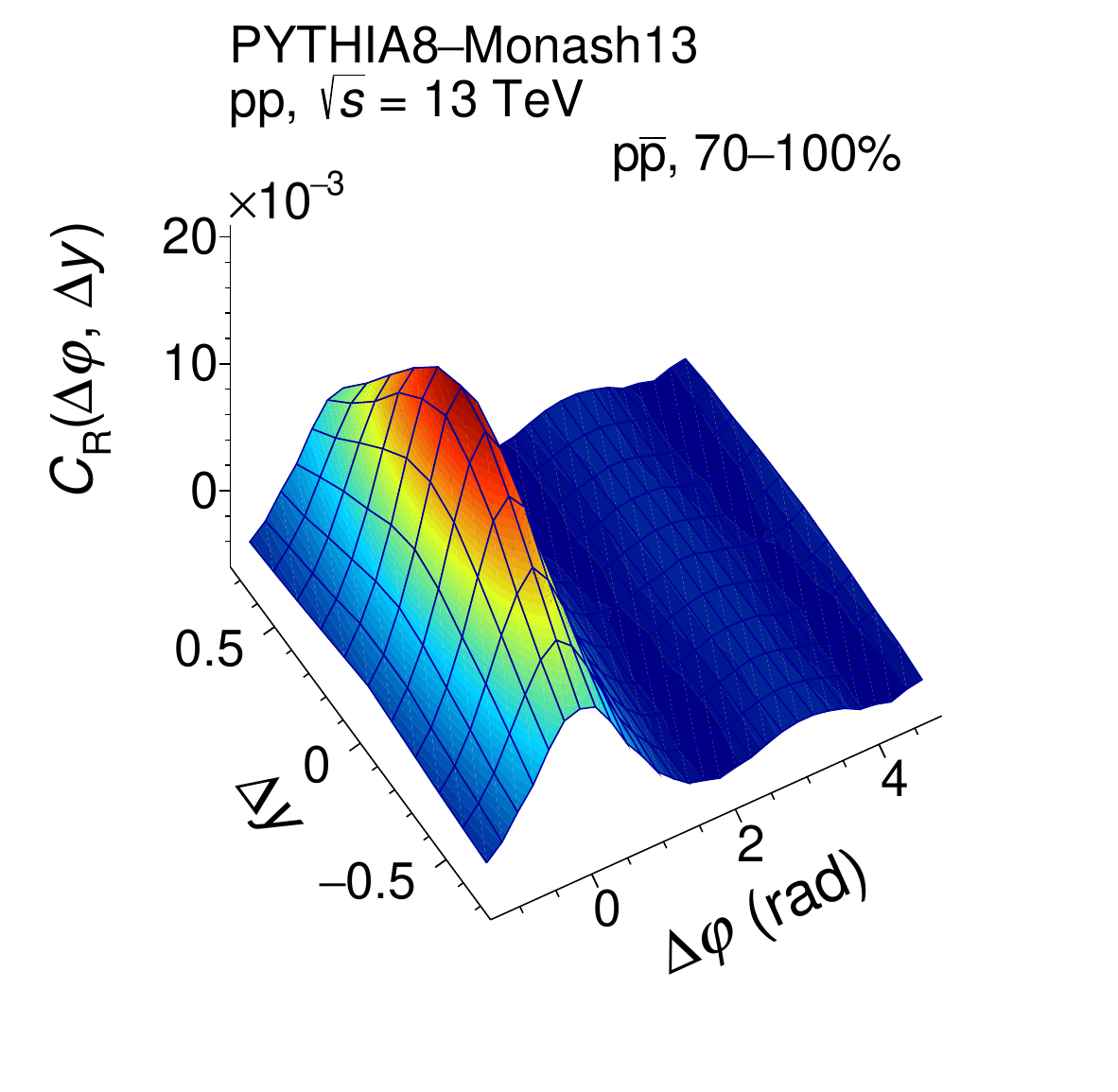}
   \label{fig7-d}}
\caption{The ($\Delta y,\Delta\varphi$)-correlation functions for PYTHIA8--Monash2013 for like- and unlike-sign protons pairs in the 0--20\% and 70--100\% multiplicity classes in pp collisions at $\sqrt{s} = 13$ TeV using the $C_{\rm{R}}$ definition.}
\label{corrfunc2D_pythia}
\end{figure}
\begin{figure}[!h]
   \centering
 \subfigure[]{
   \includegraphics[width=6cm]{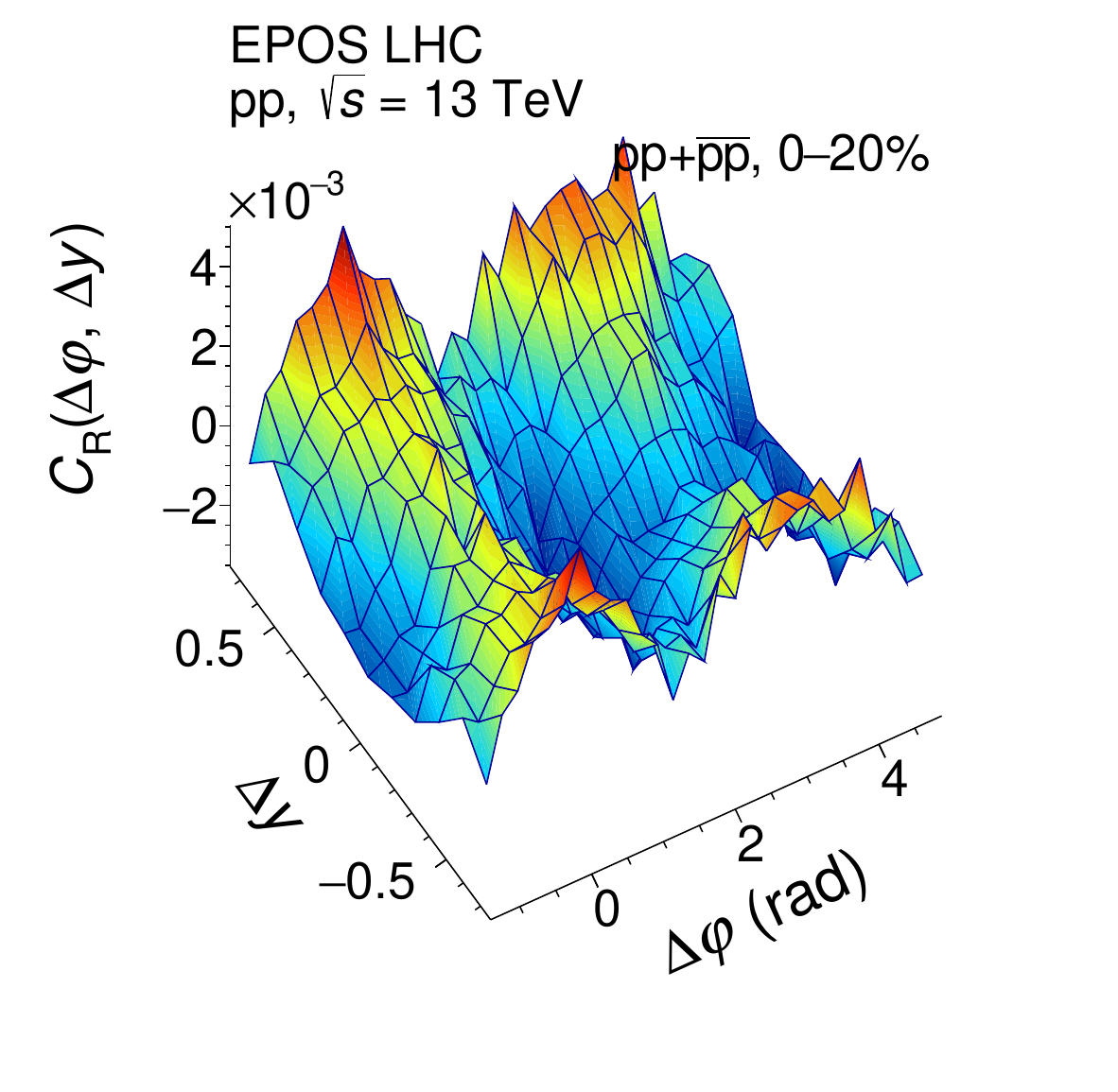} 
   \label{fig7-a}}
    \subfigure[]{
   \includegraphics[width=6cm]{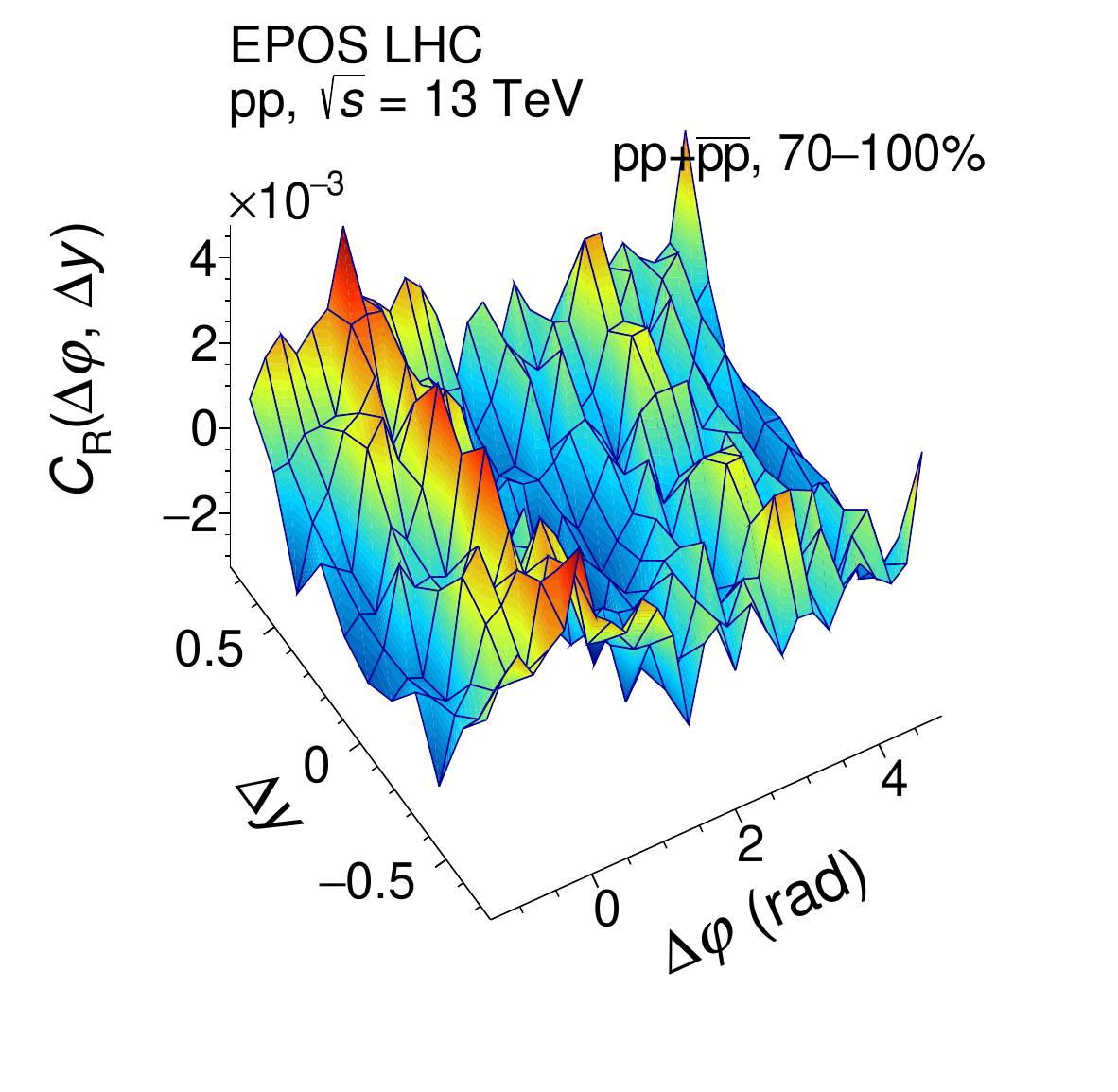} 
   \label{fig7-b}}   
   \subfigure[]{
   \includegraphics[width=6cm]{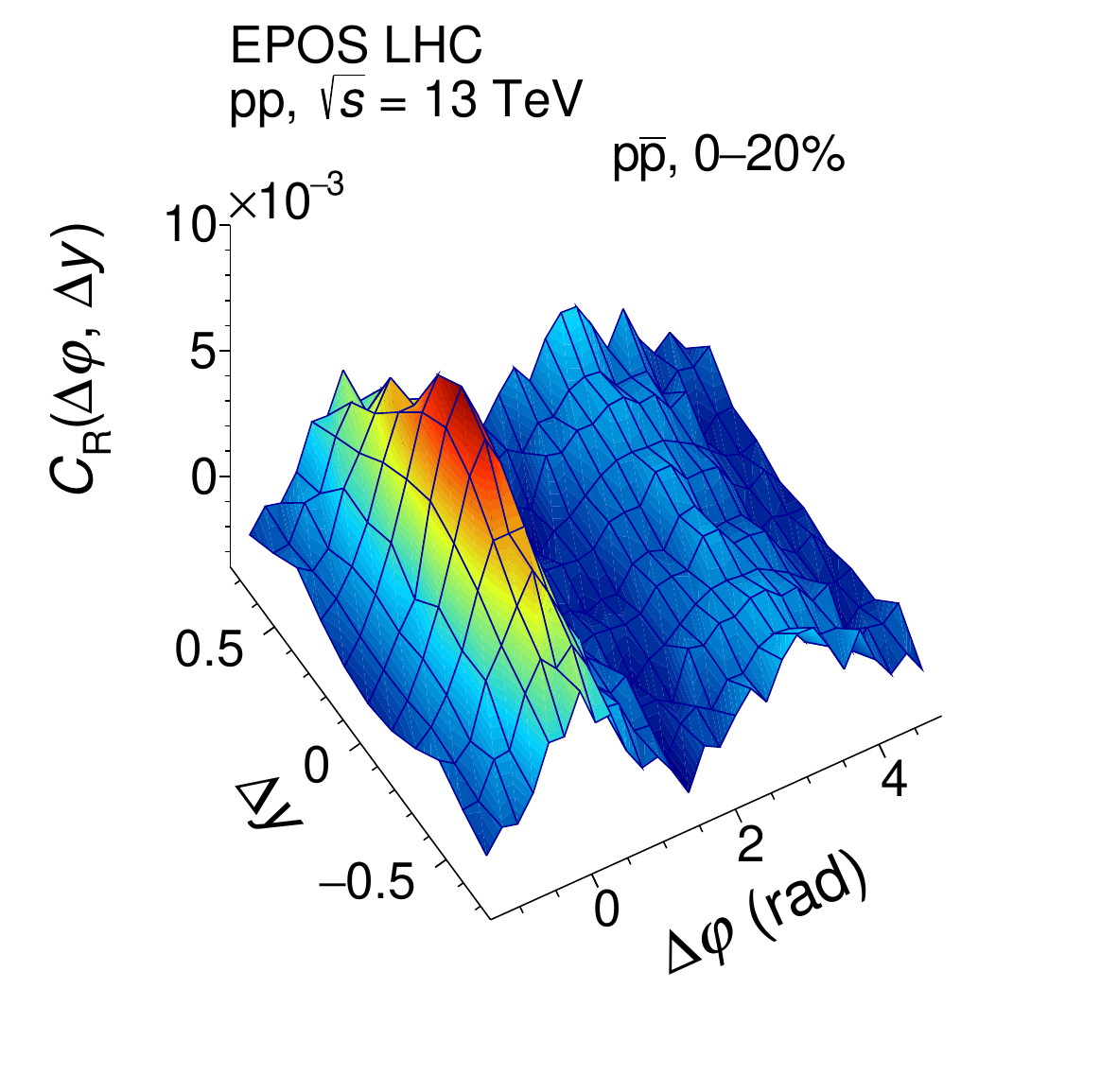}
   \label{fig7-c}}
 \subfigure[]{
   \includegraphics[width=6cm]{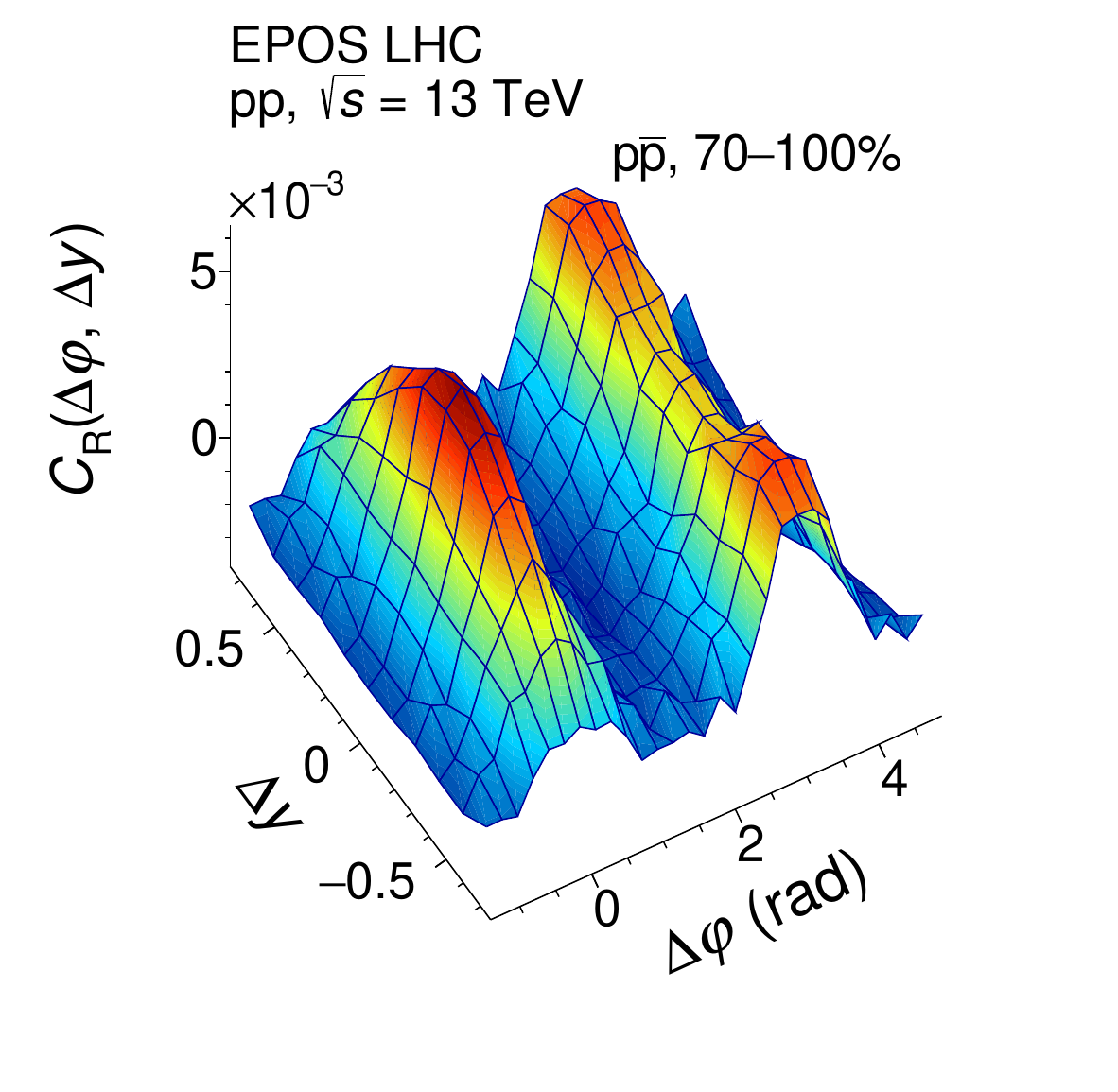}
   \label{fig7-d}}
\caption{The ($\Delta y,\Delta\varphi$)-correlation functions for EPOS LHC for like- and unlike-sign protons pairs in 0--20\% and 70--100\% multiplicity classes in pp collisions at $\sqrt{s} = 13$ TeV using the $C_{\rm{R}}$ definition.}
\label{corrfunc2D_epos}
\end{figure}

For protons, the situation is reversed: no agreement is observed in the $\Delta\varphi$ projections, yet in the $\Delta y$ projections PYTHIA 8.315 describes both like- and unlike-sign correlations within the systematic uncertainties.
An analysis of the 2D correlation functions in \Cref{corrfunc2D_pythia,corrfunc2D_epos}, generated using PYTHIA8 (Monash 2013 tune) and EPOS LHC for 0--20\% and 70--100\% multiplicity classes, highlights that neither model reliably reproduces baryonic correlations, both qualitatively and quantitatively. These multiplicity ranges were chosen to capture the full spectrum of event activity, offering a clear and comprehensive view of the correlation functions in both central and peripheral collisions. While EPOS LHC exhibits a typical evolution of correlation patterns across multiplicity classes, PYTHIA8 displays distinctive behavior. At higher multiplicities, the correlation structure resembles that of lower multiplicities, featuring a dominant away-side ridge and a smaller near-side ridge, indicative of momentum conservation effects. In contrast, at lower multiplicities, the near-side ridge becomes more pronounced. Finally, these distributions show that relying solely on $\Delta y$ projections to compare models with experimental data in angular correlation functions can lead to misleading conclusions about their apparent agreement.

This comparison provides valuable insights into the limitations of current theoretical models in capturing the observed phenomena, emphasizing the need for further development and refinement of these models to better reproduce the experimental data.
\clearpage
\section{Summary}
This study focused on the two-particle angular correlation of identified particles in $\Delta y,\Delta\varphi$ space, extending previous research conducted at $\sqrt{s} = 7$ TeV across various multiplicity classes. To address the trivial multiplicity scaling observed with the probability-ratio definition of the correlation function, this study employed the rescaled two-particle correlation function as well. 

In pp collisions at $\sqrt{s} = 13$ TeV, the anticorrelation observed previously persists and becomes more pronounced at higher multiplicities when using the rescaled two-particle correlation function definition. While this analysis was primarily conducted to investigate the origin of these anticorrelations in $\Delta y \Delta \varphi$, it has also uncovered intriguing trends across different particle species and multiplicity classes. 
In contrast, like-sign particles show a large correlation in the away-side region typical of back-to-back jets, with kaons and protons exhibiting a magnitude similar to the near-side peak. Projection on the azimuthal direction shows that the probability ratio increases with lower multiplicities. However, the rescaled two-particle correlation function exhibits an opposite trend, due to renormalization that removes the trivial multiplicity scaling. Notably, like-sign kaons show no dependence on multiplicity after rescaling.

The results were compared with the predictions of PYTHIA8 and EPOS models in both the $\Delta\varphi$ and $\Delta y$ directions. 
Projections of the experimental data on the azimuthal direction showed a high level of agreement with meson correlations. However, the models exhibited a notable inability to accurately reproduce the baryon--baryon correlations.
In contrast, the simulated correlations for the baryon--antibaryon pairs are qualitatively similar to those observed in the experimental data, but not quantitatively. Regarding projections on the $\Delta y$ axis, the PYTHIA8 and EPOS models for meson pairs show qualitative but not quantitative agreement. These models also fail to reproduce the two-dimensional $\Delta y,\Delta\varphi$ anticorrelation for baryon--baryon pairs. These findings strongly suggest that essential ingredients are missing in the current PYTHIA and EPOS model descriptions, with the most critical deficiency being the inadequate treatment of baryon production. The measured multiplicity dependence of baryon--baryon anticorrelations can help to further explore the possible role of the partonic phase in models such as AMPT and EPOS.

Future experimental studies using the larger LHC Run 3 datasets would enable investigations into how particle mass, strangeness, and charm content affect baryon–baryon anticorrelations.

\newenvironment{acknowledgement}{\relax}{\relax}
\begin{acknowledgement}
\section*{Acknowledgements}

The ALICE Collaboration would like to thank all its engineers and technicians for their invaluable contributions to the construction of the experiment and the CERN accelerator teams for the outstanding performance of the LHC complex.
The ALICE Collaboration gratefully acknowledges the resources and support provided by all Grid centres and the Worldwide LHC Computing Grid (WLCG) collaboration.
The ALICE Collaboration acknowledges the following funding agencies for their support in building and running the ALICE detector:
A. I. Alikhanyan National Science Laboratory (Yerevan Physics Institute) Foundation (ANSL), State Committee of Science and World Federation of Scientists (WFS), Armenia;
Austrian Academy of Sciences, Austrian Science Fund (FWF): [M 2467-N36] and Nationalstiftung f\"{u}r Forschung, Technologie und Entwicklung, Austria;
Ministry of Communications and High Technologies, National Nuclear Research Center, Azerbaijan;
Rede Nacional de Física de Altas Energias (Renafae), Financiadora de Estudos e Projetos (Finep), Funda\c{c}\~{a}o de Amparo \`{a} Pesquisa do Estado de S\~{a}o Paulo (FAPESP) and The Sao Paulo Research Foundation  (FAPESP), Brazil;
Bulgarian Ministry of Education and Science, within the National Roadmap for Research Infrastructures 2020-2027 (object CERN), Bulgaria;
Ministry of Education of China (MOEC) , Ministry of Science \& Technology of China (MSTC) and National Natural Science Foundation of China (NSFC), China;
Ministry of Science and Education and Croatian Science Foundation, Croatia;
Centro de Aplicaciones Tecnol\'{o}gicas y Desarrollo Nuclear (CEADEN), Cubaenerg\'{\i}a, Cuba;
Ministry of Education, Youth and Sports of the Czech Republic, Czech Republic;
The Danish Council for Independent Research | Natural Sciences, the VILLUM FONDEN and Danish National Research Foundation (DNRF), Denmark;
Helsinki Institute of Physics (HIP), Finland;
Commissariat \`{a} l'Energie Atomique (CEA) and Institut National de Physique Nucl\'{e}aire et de Physique des Particules (IN2P3) and Centre National de la Recherche Scientifique (CNRS), France;
Bundesministerium f\"{u}r Forschung, Technologie und Raumfahrt (BMFTR) and GSI Helmholtzzentrum f\"{u}r Schwerionenforschung GmbH, Germany;
General Secretariat for Research and Technology, Ministry of Education, Research and Religions, Greece;
National Research, Development and Innovation Office, Hungary;
Department of Atomic Energy Government of India (DAE), Department of Science and Technology, Government of India (DST), University Grants Commission, Government of India (UGC) and Council of Scientific and Industrial Research (CSIR), India;
National Research and Innovation Agency - BRIN, Indonesia;
Istituto Nazionale di Fisica Nucleare (INFN), Italy;
Japanese Ministry of Education, Culture, Sports, Science and Technology (MEXT) and Japan Society for the Promotion of Science (JSPS) KAKENHI, Japan;
Consejo Nacional de Ciencia (CONACYT) y Tecnolog\'{i}a, through Fondo de Cooperaci\'{o}n Internacional en Ciencia y Tecnolog\'{i}a (FONCICYT) and Direcci\'{o}n General de Asuntos del Personal Academico (DGAPA), Mexico;
Nederlandse Organisatie voor Wetenschappelijk Onderzoek (NWO), Netherlands;
The Research Council of Norway, Norway;
Pontificia Universidad Cat\'{o}lica del Per\'{u}, Peru;
Ministry of Science and Higher Education, National Science Centre and WUT ID-UB, Poland;
Korea Institute of Science and Technology Information and National Research Foundation of Korea (NRF), Republic of Korea;
Ministry of Education and Scientific Research, Institute of Atomic Physics, Ministry of Research and Innovation and Institute of Atomic Physics and Universitatea Nationala de Stiinta si Tehnologie Politehnica Bucuresti, Romania;
Ministerstvo skolstva, vyskumu, vyvoja a mladeze SR, Slovakia;
National Research Foundation of South Africa, South Africa;
Swedish Research Council (VR) and Knut \& Alice Wallenberg Foundation (KAW), Sweden;
European Organization for Nuclear Research, Switzerland;
Suranaree University of Technology (SUT), National Science and Technology Development Agency (NSTDA) and National Science, Research and Innovation Fund (NSRF via PMU-B B05F650021), Thailand;
Turkish Energy, Nuclear and Mineral Research Agency (TENMAK), Turkey;
National Academy of  Sciences of Ukraine, Ukraine;
Science and Technology Facilities Council (STFC), United Kingdom;
National Science Foundation of the United States of America (NSF) and United States Department of Energy, Office of Nuclear Physics (DOE NP), United States of America.
In addition, individual groups or members have received support from:
Czech Science Foundation (grant no. 23-07499S), Czech Republic;
FORTE project, reg.\ no.\ CZ.02.01.01/00/22\_008/0004632, Czech Republic, co-funded by the European Union, Czech Republic;
European Research Council (grant no. 950692), European Union;
Deutsche Forschungs Gemeinschaft (DFG, German Research Foundation) ``Neutrinos and Dark Matter in Astro- and Particle Physics'' (grant no. SFB 1258), Germany;
CONVECS project, CUP C97H23001700002 FESR 2021-2027 program, Italy;
FAIR - Future Artificial Intelligence Research, funded by the NextGenerationEU program (Italy).

\end{acknowledgement}

\bibliographystyle{utphys}   
\bibliography{bibliography}

\newpage
\appendix

%
%

\section{The ALICE Collaboration}
\label{app:collab}
\begin{flushleft} 
\small

I.J.~Abualrob\,\orcidlink{0009-0005-3519-5631}\,$^{\rm 113}$, 
S.~Acharya\,\orcidlink{0000-0002-9213-5329}\,$^{\rm 49}$, 
G.~Aglieri Rinella\,\orcidlink{0000-0002-9611-3696}\,$^{\rm 32}$, 
L.~Aglietta\,\orcidlink{0009-0003-0763-6802}\,$^{\rm 24}$, 
N.~Agrawal\,\orcidlink{0000-0003-0348-9836}\,$^{\rm 25}$, 
Z.~Ahammed\,\orcidlink{0000-0001-5241-7412}\,$^{\rm 133}$, 
S.~Ahmad\,\orcidlink{0000-0003-0497-5705}\,$^{\rm 15}$, 
I.~Ahuja\,\orcidlink{0000-0002-4417-1392}\,$^{\rm 36}$, 
Z.~Akbar$^{\rm 80}$, 
A.~Akindinov\,\orcidlink{0000-0002-7388-3022}\,$^{\rm 139}$
% ...

\section*{Affiliation Notes}

$^{\rm I}$ Deceased\\
$^{\rm II}$ Also at: Max-Planck-Institut fur Physik, Munich, Germany\\
$^{\rm III}$ Also at: Czech Technical University in Prague (CZ)\\
$^{\rm IV}$ Also at: Instituto de Fisica da Universidade de Sao Paulo\\
$^{\rm V}$ Also at: Dipartimento DET del Politecnico di Torino, Turin, Italy\\
$^{\rm VI}$ Also at: Department of Applied Physics, Aligarh Muslim University, Aligarh, India\\
$^{\rm VII}$ Also at: Institute of Theoretical Physics, University of Wroclaw, Poland\\
$^{\rm VIII}$ Also at: Facultad de Ciencias, Universidad Nacional Aut\'{o}noma de M\'{e}xico, Mexico City, Mexico\\

\section*{Collaboration Institutes}

$^{1}$ A.I. Alikhanyan National Science Laboratory (Yerevan Physics Institute) Foundation, Yerevan, Armenia\\
$^{2}$ AGH University of Krakow, Cracow, Poland\\
$^{3}$ Bogolyubov Institute for Theoretical Physics, National Academy of Sciences of Ukraine, Kyiv, Ukraine\\
$^{4}$ Bose Institute, Department of Physics  and Centre for Astroparticle Physics and Space Science (CAPSS), Kolkata, India\\
$^{5}$ California Polytechnic State University, San Luis Obispo, California, United States\\
$^{6}$ Central China Normal University, Wuhan, China\\
$^{7}$ Centro de Aplicaciones Tecnol\'{o}gicas y Desarrollo Nuclear (CEADEN), Havana, Cuba\\
$^{8}$ Centro de Investigaci\'{o}n y de Estudios Avanzados (CINVESTAV), Mexico City and M\'{e}rida, Mexico\\
$^{9}$ Chicago State University, Chicago, Illinois, United States\\
$^{10}$ China Nuclear Data Center, China Institute of Atomic Energy, Beijing, China\\
$^{11}$ China University of Geosciences, Wuhan, China\\
$^{12}$ Chungbuk National University, Cheongju, Republic of Korea\\
$^{13}$ Comenius University Bratislava, Faculty of Mathematics, Physics and Informatics, Bratislava, Slovak Republic\\
$^{14}$ Creighton University, Omaha, Nebraska, United States\\
$^{15}$ Department of Physics, Aligarh Muslim University, Aligarh, India\\
$^{16}$ Department of Physics, Pusan National University, Pusan, Republic of Korea\\
$^{17}$ Department of Physics, Sejong University, Seoul, Republic of Korea\\
$^{18}$ Department of Physics, University of California, Berkeley, California, United States\\
$^{19}$ Department of Physics, University of Oslo, Oslo, Norway\\
$^{20}$ Department of Physics and Technology, University of Bergen, Bergen, Norway\\
$^{21}$ Dipartimento di Fisica, Universit\`{a} di Pavia, Pavia, Italy\\
$^{22}$ Dipartimento di Fisica dell'Universit\`{a} and Sezione INFN, Cagliari, Italy\\
$^{23}$ Dipartimento di Fisica dell'Universit\`{a} and Sezione INFN, Trieste, Italy\\
$^{24}$ Dipartimento di Fisica dell'Universit\`{a} and Sezione INFN, Turin, Italy\\
$^{25}$ Dipartimento di Fisica e Astronomia dell'Universit\`{a} and Sezione INFN, Bologna, Italy\\
$^{26}$ Dipartimento di Fisica e Astronomia dell'Universit\`{a} and Sezione INFN, Catania, Italy\\
$^{27}$ Dipartimento di Fisica e Astronomia dell'Universit\`{a} and Sezione INFN, Padova, Italy\\
$^{28}$ Dipartimento di Fisica `E.R.~Caianiello' dell'Universit\`{a} and Gruppo Collegato INFN, Salerno, Italy\\
$^{29}$ Dipartimento DISAT del Politecnico and Sezione INFN, Turin, Italy\\
$^{30}$ Dipartimento di Scienze MIFT, Universit\`{a} di Messina, Messina, Italy\\
$^{31}$ Dipartimento Interateneo di Fisica `M.~Merlin' and Sezione INFN, Bari, Italy\\
$^{32}$ European Organization for Nuclear Research (CERN), Geneva, Switzerland\\
$^{33}$ Faculty of Electrical Engineering, Mechanical Engineering and Naval Architecture, University of Split, Split, Croatia\\
$^{34}$ Faculty of Nuclear Sciences and Physical Engineering, Czech Technical University in Prague, Prague, Czech Republic\\
$^{35}$ Faculty of Physics, Sofia University, Sofia, Bulgaria\\
$^{36}$ Faculty of Science, P.J.~\v{S}af\'{a}rik University, Ko\v{s}ice, Slovak Republic\\
$^{37}$ Faculty of Technology, Environmental and Social Sciences, Bergen, Norway\\
$^{38}$ Frankfurt Institute for Advanced Studies, Johann Wolfgang Goethe-Universit\"{a}t Frankfurt, Frankfurt, Germany\\
$^{39}$ Fudan University, Shanghai, China\\
$^{40}$ Gauhati University, Department of Physics, Guwahati, India\\
$^{41}$ Helmholtz-Institut f\"{u}r Strahlen- und Kernphysik, Rheinische Friedrich-Wilhelms-Universit\"{a}t Bonn, Bonn, Germany\\
$^{42}$ Helsinki Institute of Physics (HIP), Helsinki, Finland\\
$^{43}$ High Energy Physics Group,  Universidad Aut\'{o}noma de Puebla, Puebla, Mexico\\
$^{44}$ Horia Hulubei National Institute of Physics and Nuclear Engineering, Bucharest, Romania\\
$^{45}$ HUN-REN Wigner Research Centre for Physics, Budapest, Hungary\\
$^{46}$ Indian Institute of Technology Bombay (IIT), Mumbai, India\\
$^{47}$ Indian Institute of Technology Indore, Indore, India\\
$^{48}$ INFN, Laboratori Nazionali di Frascati, Frascati, Italy\\
$^{49}$ INFN, Sezione di Bari, Bari, Italy\\
$^{50}$ INFN, Sezione di Bologna, Bologna, Italy\\
$^{51}$ INFN, Sezione di Cagliari, Cagliari, Italy\\
$^{52}$ INFN, Sezione di Catania, Catania, Italy\\
$^{53}$ INFN, Sezione di Padova, Padova, Italy\\
$^{54}$ INFN, Sezione di Pavia, Pavia, Italy\\
$^{55}$ INFN, Sezione di Torino, Turin, Italy\\
$^{56}$ INFN, Sezione di Trieste, Trieste, Italy\\
$^{57}$ Inha University, Incheon, Republic of Korea\\
$^{58}$ Institute for Gravitational and Subatomic Physics (GRASP), Utrecht University/Nikhef, Utrecht, Netherlands\\
$^{59}$ Institute of Experimental Physics, Slovak Academy of Sciences, Ko\v{s}ice, Slovak Republic\\
$^{60}$ Institute of Physics, Homi Bhabha National Institute, Bhubaneswar, India\\
$^{61}$ Institute of Physics of the Czech Academy of Sciences, Prague, Czech Republic\\
$^{62}$ Institute of Space Science (ISS), Bucharest, Romania\\
$^{63}$ Institut f\"{u}r Kernphysik, Johann Wolfgang Goethe-Universit\"{a}t Frankfurt, Frankfurt, Germany\\
$^{64}$ Instituto de Ciencias Nucleares, Universidad Nacional Aut\'{o}noma de M\'{e}xico, Mexico City, Mexico\\
$^{65}$ Instituto de F\'{i}sica, Universidade Federal do Rio Grande do Sul (UFRGS), Porto Alegre, Brazil\\
$^{66}$ Instituto de F\'{\i}sica, Universidad Nacional Aut\'{o}noma de M\'{e}xico, Mexico City, Mexico\\
$^{67}$ iThemba LABS, National Research Foundation, Somerset West, South Africa\\
$^{68}$ Jeonbuk National University, Jeonju, Republic of Korea\\
$^{69}$ Korea Institute of Science and Technology Information, Daejeon, Republic of Korea\\
$^{70}$ Laboratoire de Physique Subatomique et de Cosmologie, Universit\'{e} Grenoble-Alpes, CNRS-IN2P3, Grenoble, France\\
$^{71}$ Lawrence Berkeley National Laboratory, Berkeley, California, United States\\
$^{72}$ Lund University Department of Physics, Division of Particle Physics, Lund, Sweden\\
$^{73}$ Marietta Blau Institute, Vienna, Austria\\
$^{74}$ Nagasaki Institute of Applied Science, Nagasaki, Japan\\
$^{75}$ Nara Women{'}s University (NWU), Nara, Japan\\
$^{76}$ National and Kapodistrian University of Athens, School of Science, Department of Physics , Athens, Greece\\
$^{77}$ National Centre for Nuclear Research, Warsaw, Poland\\
$^{78}$ National Institute of Science Education and Research, Homi Bhabha National Institute, Jatni, India\\
$^{79}$ National Nuclear Research Center, Baku, Azerbaijan\\
$^{80}$ National Research and Innovation Agency - BRIN, Jakarta, Indonesia\\
$^{81}$ Niels Bohr Institute, University of Copenhagen, Copenhagen, Denmark\\
$^{82}$ Nikhef, National institute for subatomic physics, Amsterdam, Netherlands\\
$^{83}$ Nuclear Physics Group, STFC Daresbury Laboratory, Daresbury, United Kingdom\\
$^{84}$ Nuclear Physics Institute of the Czech Academy of Sciences, Husinec-\v{R}e\v{z}, Czech Republic\\
$^{85}$ Oak Ridge National Laboratory, Oak Ridge, Tennessee, United States\\
$^{86}$ Ohio State University, Columbus, Ohio, United States\\
$^{87}$ Physics department, Faculty of science, University of Zagreb, Zagreb, Croatia\\
$^{88}$ Physics Department, Panjab University, Chandigarh, India\\
$^{89}$ Physics Department, University of Jammu, Jammu, India\\
$^{90}$ Physics Program and International Institute for Sustainability with Knotted Chiral Meta Matter (WPI-SKCM$^{2}$), Hiroshima University, Hiroshima, Japan\\
$^{91}$ Physikalisches Institut, Eberhard-Karls-Universit\"{a}t T\"{u}bingen, T\"{u}bingen, Germany\\
$^{92}$ Physikalisches Institut, Ruprecht-Karls-Universit\"{a}t Heidelberg, Heidelberg, Germany\\
$^{93}$ Physik Department, Technische Universit\"{a}t M\"{u}nchen, Munich, Germany\\
$^{94}$ Politecnico di Bari and Sezione INFN, Bari, Italy\\
$^{95}$ Research Division and ExtreMe Matter Institute EMMI, GSI Helmholtzzentrum f\"ur Schwerionenforschung GmbH, Darmstadt, Germany\\
$^{96}$ Saga University, Saga, Japan\\
$^{97}$ Saha Institute of Nuclear Physics, Homi Bhabha National Institute, Kolkata, India\\
$^{98}$ School of Physics and Astronomy, University of Birmingham, Birmingham, United Kingdom\\
$^{99}$ Secci\'{o}n F\'{\i}sica, Departamento de Ciencias, Pontificia Universidad Cat\'{o}lica del Per\'{u}, Lima, Peru\\
$^{100}$ SUBATECH, IMT Atlantique, Nantes Universit\'{e}, CNRS-IN2P3, Nantes, France\\
$^{101}$ Sungkyunkwan University, Suwon City, Republic of Korea\\
$^{102}$ Suranaree University of Technology, Nakhon Ratchasima, Thailand\\
$^{103}$ Technical University of Ko\v{s}ice, Ko\v{s}ice, Slovak Republic\\
$^{104}$ The Henryk Niewodniczanski Institute of Nuclear Physics, Polish Academy of Sciences, Cracow, Poland\\
$^{105}$ The University of Texas at Austin, Austin, Texas, United States\\
$^{106}$ Universidad Aut\'{o}noma de Sinaloa, Culiac\'{a}n, Mexico\\
$^{107}$ Universidade de S\~{a}o Paulo (USP), S\~{a}o Paulo, Brazil\\
$^{108}$ Universidade Estadual de Campinas (UNICAMP), Campinas, Brazil\\
$^{109}$ Universidade Federal do ABC, Santo Andre, Brazil\\
$^{110}$ Universitatea Nationala de Stiinta si Tehnologie Politehnica Bucuresti, Bucharest, Romania\\
$^{111}$ University of Cape Town, Cape Town, South Africa\\
$^{112}$ University of Derby, Derby, United Kingdom\\
$^{113}$ University of Houston, Houston, Texas, United States\\
$^{114}$ University of Jyv\"{a}skyl\"{a}, Jyv\"{a}skyl\"{a}, Finland\\
$^{115}$ University of Kansas, Lawrence, Kansas, United States\\
$^{116}$ University of Liverpool, Liverpool, United Kingdom\\
$^{117}$ University of Science and Technology of China, Hefei, China\\
$^{118}$ University of Silesia in Katowice, Katowice, Poland\\
$^{119}$ University of South-Eastern Norway, Kongsberg, Norway\\
$^{120}$ University of Tennessee, Knoxville, Tennessee, United States\\
$^{121}$ University of the Witwatersrand, Johannesburg, South Africa\\
$^{122}$ University of Tokyo, Tokyo, Japan\\
$^{123}$ University of Tsukuba, Tsukuba, Japan\\
$^{124}$ Universit\"{a}t M\"{u}nster, Institut f\"{u}r Kernphysik, M\"{u}nster, Germany\\
$^{125}$ Universit\'{e} Clermont Auvergne, CNRS/IN2P3, LPC, Clermont-Ferrand, France\\
$^{126}$ Universit\'{e} de Lyon, CNRS/IN2P3, Institut de Physique des 2 Infinis de Lyon, Lyon, France\\
$^{127}$ Universit\'{e} de Strasbourg, CNRS, IPHC UMR 7178, F-67000 Strasbourg, France, Strasbourg, France\\
$^{128}$ Universit\'{e} Paris-Saclay, Centre d'Etudes de Saclay (CEA), IRFU, D\'{e}partment de Physique Nucl\'{e}aire (DPhN), Saclay, France\\
$^{129}$ Universit\'{e}  Paris-Saclay, CNRS/IN2P3, IJCLab, Orsay, France\\
$^{130}$ Universit\`{a} degli Studi di Foggia, Foggia, Italy\\
$^{131}$ Universit\`{a} del Piemonte Orientale, Vercelli, Italy\\
$^{132}$ Universit\`{a} di Brescia, Brescia, Italy\\
$^{133}$ Variable Energy Cyclotron Centre, Homi Bhabha National Institute, Kolkata, India\\
$^{134}$ Warsaw University of Technology, Warsaw, Poland\\
$^{135}$ Wayne State University, Detroit, Michigan, United States\\
$^{136}$ Yale University, New Haven, Connecticut, United States\\
$^{137}$ Yildiz Technical University, Istanbul, Turkey\\
$^{138}$ Yonsei University, Seoul, Republic of Korea\\
$^{139}$ Affiliated with an institute formerly covered by a cooperation agreement with CERN\\
$^{140}$ Affiliated with an international laboratory covered by a cooperation agreement with CERN.\\

\end{flushleft} 
  
\end{document}